# How much has the Sun influenced Northern Hemisphere temperature trends? An ongoing debate.

**Ronan Connolly[1,2], Willie Soon[1], Michael Connolly[2], Sallie Baliunas[3], Johan Berglund[4], C. J. Butler[5], Rodolfo Gustavo Cionco[6,7], Ana G. Elias[8,9], Valery M. Fedorov[10], Hermann Harde[11], Gregory W. Henry[12], Douglas V. Hoyt[13], Ole Humlum[14], David R. Legates[15], Sebastian Lüning[16], Nicola Scafetta[17], Jan-Erik Solheim[18], László Szarka[19], Harry van Loon[20], Víctor M. Velasco Herrera[21], Richard C. Willson[22], Hong Yan[23] and Weijia Zhang[24,25]**

[1] Center for Environmental Research and Earth Science (CERES), Salem, MA 01970, USA
[2] Independent scientists, Dublin, Ireland
[3] Retired, formerly Harvard-Smithsonian Center for Astrophysics, Cambridge, MA 02138, USA
[4] Independent researcher, Malmö, Sweden
[5] Retired, formerly Armagh Observatory, College Hill, Armagh BT61 9DG, Northern Ireland, UK
[6] Comisión de Investigaciones Científicas de la Provincia de Buenos Aires, Argentina
[7] Grupo de Estudios Ambientales, Universidad Tecnológica Nacional, Colón 332, San Nicolás (2900), Buenos Aires, Argentina
[8] Laboratorio de Física de la Atmósfera, Facultad de Ciencias Exactas y Tecnología, Universidad Nacional de Tucumán, Av. Independencia 1800, 4000 Tucumán, Argentina
[9] Instituto de Física del Noroeste Argentino (Consejo Nacional de Investigaciones Científicas y Técnicas - Universidad Nacional de Tucumán), 4000 Tucumán, Argentina
[10] Faculty of Geography, Lomonosov, Moscow State University, Leninskie Gory St. 1, Moscow 119991, Russia
[11] Helmut-Schmidt-University, Hamburg, Germany
[12] Center of Excellence in Information Systems, Tennessee State University, Nashville, TN 37209 USA
[13] Independent scientist, Berkeley Springs, WV, USA
[14] Emeritus Professor in Physical Geography, Department of Geosciences, University of Oslo, Norway
[15] College of Earth, Ocean, and the Environment, University of Delaware, Newark DE 19716-2541, USA
[16] Institute for Hydrography, Geoecology and Climate Sciences, Hauptstraße 47, 6315 Ägeri, Switzerland
[17] Department of Earth Sciences, Environment and Georesources, University of Naples Federico II, Complesso Universitario di Monte S. Angelo, via Cinthia, 21, 80126 Naples, Italy
[18] Retired, formerly Department of Physics and Technology, UiT The Arctic University of Norway, 9037 Tromsø, Norway
[19] CSFK Geodetic and Geophysical Institute, 9400 Sopron, Csatkai utca 6-8, Hungary
[20] Retired, formerly National Center for Atmospheric Research, Boulder, Colorado, USA.
[21] Instituto de Geofisica, Universidad Nacional Autónoma de México, Ciudad Universitaria, Coyoacán, 04510, México D.F., México
[22] Active Cavity Radiometer Irradiance Monitor (ACRIM), Coronado, CA 92118, USA
[23] State Key Laboratory of Loess and Quaternary Geology, Institute of Earth Environment, Chinese Academy of Sciences, Xi'an 710061, China
[24] Department of Mathematics and Physics, Shaoxing University, Shaoxing, China
[25] Department of AOP Physics, University of Oxford, Oxford, UK

E-mail: ronan@ceres-science.com

## Abstract

In order to evaluate how much Total Solar Irradiance (TSI) has influenced Northern Hemisphere surface air temperature trends, it is important to have reliable estimates of both quantities. Sixteen different estimates of the changes in Total Solar Irradiance (TSI) since at least the 19th century were compiled from the literature. Half of these estimates are "low variability" and half are "high variability". Meanwhile, five largely-independent methods for estimating Northern Hemisphere temperature trends were evaluated using: 1) only rural weather stations; 2) all available stations whether urban or rural (the standard approach); 3) only sea surface temperatures; 4) tree-ring widths as temperature proxies; 5) glacier length records as temperature proxies. The standard estimates which use urban as well as rural stations were somewhat anomalous as they implied a much greater warming in recent decades than the other estimates, suggesting that urbanization bias might still be a problem in current global temperature datasets - despite the conclusions of some earlier studies. Nonetheless, all five estimates confirm that it is currently warmer than the late 19th century, i.e., there has been some "global warming" since the 19th century. For each of the five estimates of Northern Hemisphere temperatures, the contribution from direct solar forcing for all sixteen estimates of TSI was evaluated using simple linear least-squares fitting. The role of human activity on recent warming was then calculated by fitting the residuals to the UN IPCC's recommended "anthropogenic forcings" time series. For all five Northern Hemisphere temperature series, different TSI estimates suggest everything from no role for the Sun in recent decades (implying that recent global warming is mostly human-caused) to most of the recent global warming being due to changes in solar activity (that is, that recent global warming is mostly natural). It appears that previous studies (including the most recent IPCC reports) which had prematurely concluded the former, had done so because they failed to adequately consider all the relevant estimates of TSI and/or to satisfactorily address the uncertainties still associated with Northern Hemisphere temperature trend estimates. Therefore, several recommendations on how the scientific community can more satisfactorily resolve these issues are provided.

## 1. Introduction

The UN's Intergovernmental Panel on Climate Change (IPCC)'s Working Group 1 concluded in their most recent (5th) Assessment Report [1] that:

"*Each of the last three decades has been successively warmer at the Earth's surface than any preceding decade since 1850* […] *In the Northern Hemisphere, 1983-2012 was* ***likely*** *the warmest 30-year period of the last 1400 years*" (IPCC Working Group 1's Summary for Policymakers, 2013, p3 – emphasis in original) [2]

And that:

"*It is* ***extremely likely*** *that human influence has been the dominant cause of the observed warming since the mid-20th century* […] *It is* ***extremely likely*** *that more than half of the observed increase in global average surface temperature from 1951 to 2010 was caused by the anthropogenic increase in greenhouse gas concentrations and other anthropogenic forcings together. The best estimate of the human-induced contribution to warming is similar to the observed warming over this period.*" (IPCC Working Group 1's Summary for Policymakers, 2013, p15 – emphasis in original) [2]

In other words, the IPCC 5th Assessment Report essentially answered the question we raised in the title of our article, "*How much has the Sun influenced Northern Hemisphere temperature trends?*", with: 'almost nothing, at least since the mid-20th century' (to paraphrase the above statement). This followed a similar conclusion from the IPCC's 4th Assessment Report (2007):

"*Most of the observed increase in global average temperatures since the mid-20th century is* ***very likely*** *due to the observed increase in anthropogenic greenhouse gas concentrations*" (IPCC Working Group 1's Summary for Policymakers, 2007, p10 – emphasis in original) [3]

This in turn followed a similar conclusion from their 3rd Assessment Report (2001):

"*...most of the observed warming over the last 50 years is likely to have been due to the increase in greenhouse gas concentrations.*" (IPCC Working Group 1's Summary for Policymakers, 2001, p10) [4]

Indeed, over this period, there have also been several well-cited reviews and articles reaching the same conclusion. For example: Crowley (2000) [5]; Stott et al. (2001) [6]; Laut (2003) [7]; Haigh (2003) [8]; Damon and Laut (2004) [9]; Benestad (2005) [10]; Foukal et al. (2006) [11]; Bard and Frank (2006) [12]; Lockwood and Fröhlich (2007) [13]; Hegerl et al. (2007) [14]; Lean and Rind (2008) [15]; Benestad and Schmidt (2009) [16]; Gray et al. (2010) [17]; Lockwood (2012) [18]; Jones et al. (2013) [19]; Sloan and Wolfendale (2013) [20]; Gil-Alana et al. (2014) [21]; Lean (2017) [22].

On the other hand, there have also been many reviews and articles published over the same period that reached the opposite conclusion, i.e., that much of the global warming since the mid-20th century and earlier could be explained in terms of solar variability. For example: Soon et al. (1996) [23]; Hoyt and Schatten (1997) [24]; Svensmark and Friis-Christensen (1997) [25]; Soon et al. (2000a,b) [26,27]; Bond et al. (2001) [28]; Willson and Mordvinov (2003) [29]; Maasch et al. (2005) [30]; Soon (2005) [31]; Scafetta and West (2006a,b, 2008) [32–35]; Svensmark (2007) [36]; Courtillot et al. (2007) [37,38]; Singer and Avery (2008) [39]; Shaviv (2008) [40]; Scafetta (2009,2011) [41,42]; Le Mouël et al. (2008,2010,2011) [43–46]; Humlum et al. (2011) [47]; Ziskin and Shaviv (2012) [48]; Solheim et al. (2012) [49]; Courtillot et al. (2013) [50]; Solheim (2013) [51]; Scafetta and Willson (2014) [52]; Harde (2014) [53]; Lüning and Vahrenholt (2015,2016) [54,55]; Soon et al. (2015) [56]; Svensmark et al. (2016,2017) [57,58]; Harde (2017) [59]; Scafetta et al. (2019) [60]; Le Mouël et al. (2019,2020) [61,62]; Mörner et al. (2020) [63]; Lüdecke et al. (2020) [64].

Meanwhile, other reviews and articles over this period have either been undecided, or else argued for significant but subtle effects of solar variability on climate change. For example: Labitzke and van Loon [65–67]; Beer et al. (2000) [68]; Reid (2000) [69]; Carslaw et al. (2002) [70]; Ruzmaikin et al. [71–75]; Salby and Callaghan (2006) [76–78]; Kirkby (2007) [79]; de Jager et al. (2010) [80]; Tinsley et al. [81–85]; Dobrica et al. [86–89]; Blanter et al. (2012) [90]; van Loon et al. [91–93]; Roy et al. [94–97]; Lopes et al. (2017) [98]; Pan et al. (2020) [99].

Why were these dissenting scientific opinions in the literature not reflected in the various IPCC statements quoted above? There are probably many factors. One factor is probably the fact that climate change and solar variability are both multifaceted concepts. Hence, as Pittock (1983) noted, historically, many of the studies of Sun/climate relationships have provided results that are ambiguous and open to interpretation in either way [100]. Another factor is that many researchers argue that scientific results that might potentially interfere with political goals are unwelcome. For example, Lockwood (2012) argues that, "*The field of Sun-climate relations […] in recent years has been corrupted by unwelcome political and financial influence as climate change sceptics have seized upon putative solar effects as an excuse for inaction on anthropogenic warming*" [18].

At any rate, one factor that we believe is highly relevant is the fact that a primary goal of the IPCC reports is to "*speak with one voice for climate science*" [101,102]. This drive to present a single "scientific consensus" on issues has given the IPCC a remarkable international "*reputation as the epistemic authority in matters of climate policy*" (Beck et al., 2014 [101]). However, many researchers have noted that this has been achieved by suppressing dissenting views on any issues where there is still scientific disagreement [101–106]. As a result, an accurate knowledge of those issues where there is ongoing scientific *dissensus* (and why) is often missing from the IPCC reports. This is concerning for policy makers relying on the IPCC reports because, as van der Sluijs et al. (2010) note, "*The consensus approach deprives policy makers of a full view of the plurality of scientific opinions within and between the various scientific disciplines that study the climate problem*" [103]. From our perspective as members of the scientific community, we are also concerned that this suppression of open-minded scientific inquiry may be hindering scientific progress into improving our understanding of these challenging issues.

We argue that the Sun/climate debate is one of these issues where the IPCC's "consensus" statements were prematurely achieved through the suppression of dissenting scientific opinions. Indeed, van der Sluijs et al. (2010) specifically listed it as a prime example: "*Examples of such dissent are disputes over the role of man compared to the role of the sun in the observed and projected climate trends…*" [103].

We agree with Sarewitz (2011)'s argument that "*The very idea that science best expresses its authority through consensus statements is at odds with a vibrant scientific enterprise. Consensus is for textbooks; real science depends for its progress on continual challenges to the current state of always-imperfect knowledge. Science would provide better value to politics if it articulated the broadest set of plausible interpretations, options and perspectives, imagined by the best experts, rather than forcing convergence to an allegedly unified voice*" [105].

The co-authors of this article each have quite different views on the Sun/climate debate, and many of us plan on continuing our research into this challenging topic through independent ways. However, we believe that it is timely to convey to the rest of the scientific community the existence of several unresolved problems, as well as establish those points where there is general agreement. Therefore, while not strictly an "empirical adversarial collaboration" as described by e.g., Refs. [107–109], this review shares some of the same philosophy in that we have agreed not to take the "consensus-driven" approach of the IPCC [101–106], but rather to emphasize where dissenting scientific opinions exist as well as where there is scientific agreement. As Bacon (1605) noted, "*If a man will begin with certainties, he shall end in doubts; but if he will be content to begin with doubts he shall end in certainties.*" - Francis Bacon, The Advancement of Learning, Book 1, Chapter 5, Section 8 (1605).

In Section 2, we will provide a historical review of the Sun/climate debate and a discussion of some of the key ongoing debates. We will attempt to estimate how much of the long-term Northern Hemisphere temperature trends since the 19th century (or earlier) can be explained in terms of solar variability assuming a simple linear relationship between Northern Hemisphere surface air temperatures and Total Solar Irradiance (TSI). We will demonstrate that even this rather simple hypothesis has not yet been satisfactorily addressed.

The IPCC (2013) argued that TSI has been decreasing since the 1950s, and this seems to have been one of the primary reasons why they concluded that the observed warming since the 1950s was "extremely likely" to be due to human-caused greenhouse gas emissions [2]. However, Soon et al. (2015) [56] and Scafetta et al. (2019) [60] have noted that the IPCC (2013) reports had only considered a small subset of the TSI estimates available in the literature, and that other TSI estimates imply different trends. Therefore, we compile and consider a more complete set of 16 different estimates of TSI. This includes the 4 estimates considered by IPCC (2013) [2], as well as the larger set of 8 estimates considered by Soon et al. (2015) [56] and Scafetta et al. (2019) [60]. It also includes the new estimate which Matthes et al. (2017) [110] have recommended for use in the upcoming IPCC 6th Assessment Report.

Aside from these debates over a direct linear relationship between TSI and surface air temperatures, we note that there are many studies arguing that the Sun/climate relationships are probably more subtle than that. For instance, some have argued that the relationship is non-linear, e.g., involving thresholds at which prevailing oceanic or atmospheric circulation patterns might shift [63,111–113]. Others note that the solar effect on the climate should be dampened on short time scales due to thermal inertia [32–34,40,41]. Others suggest that the Sun/climate relationships might be more pronounced in some geographical regions than others [36,40,64,65,71,86,87,91,93,96,114,115]. For simplicity, the primary focus in this paper will be on evaluating the relatively simple hypothesis of a direct linear relationship between TSI and surface air temperatures. However, we encourage readers to follow up on the debates over the possibilities of more subtle sun/climate relationships. With that in mind, in Sections 2.5-2.6, we briefly review some of these ongoing debates.

In Section 3, we will compile and generate several different estimates of Northern Hemisphere temperature trends. We will show that the standard estimates used by IPCC (2013) [2], which include urban as well as rural stations, imply a much greater long-term warming than most other estimates. This suggests that the standard estimates have not adequately corrected for urbanization bias [56,116–118].

Our main analysis involves estimating the maximum solar contribution to Northern Hemisphere temperature trends assuming a linear relationship between TSI and temperature. However, since IPCC (2013) concluded that the most important factor in recent temperature trends is “anthropogenic forcings” (chiefly from greenhouse gas emissions), a useful secondary question we will consider is how much of the trends unexplained by this assumed linear solar relationship can be explained in terms of anthropogenic forcings. Therefore, a second step of our analysis will involve fitting the statistical residuals from the first step using the anthropogenic forcings recommended by IPCC (2013) [1]. In Section 4, we will describe the IPCC’s anthropogenic forcings datasets.

In Section 5, we will calculate the best fits (using linear least-squares fitting) for each of the TSI and Northern Hemisphere temperature reconstructions and then estimate the implied Sun/climate relationship from each combination, along with the implied role of anthropogenic (i.e., human-caused) factors.

Finally, we will offer some concluding remarks and recommendations for future research in Section 6. We emphasise that the main research questions of this paper are based on the debates over the role of the Sun in recent climate change. Although we contrast this with the role of anthropogenic factors, we do not explicitly investigate the possible role of other non-solar driven natural factors such as internal changes in oceanic and/or atmospheric circulation, as this is beyond the scope of the paper. However, we encourage further research into these possible factors, e.g., Refs. [119–123].

## 2. Estimating Total Solar Irradiance changes

### *2.1. Challenges in estimating multi-decadal changes in Total Solar Irradiance*

Because most of the energy that keeps the Earth warmer than space comes from incoming solar radiation, i.e., Total Solar Irradiance (TSI), it stands to reason that a multi-decadal increase in TSI should cause global warming (all else being equal). Similarly, a multi-decadal decrease in TSI should cause global cooling. For this reason, for centuries (and longer), researchers have speculated that changes in solar activity could be a major driver of climate change [7,17,18,24,39,56,124–127]. However, a challenging question associated with this theory is, “How exactly has TSI changed over time?”

One indirect metric on which much research has focused is the examination of historical records of the numbers and types/sizes of “sunspots” that are observed on the Sun’s surface over time [101,102,105–110]. Sunspots are intermittent magnetic phenomena associated with the Sun’s photosphere, that appear as dark blotches or blemishes on the Sun’s surface when the light from the Sun is shone on a card with a telescope (to avoid the observer directly looking at the Sun). These have been observed since the earliest telescopes were invented, and Galileo Galilei and others were recording sunspots as far back as 1610 [125,128–131,134]. The Chinese even have intermittent written records since 165 B.C. of sunspots that were large enough to be seen by the naked eye [135,136] Moreover, an examination of the sunspot records reveals significant changes on sub-decadal to multi-decadal timescales. In particular, a pronounced “Sunspot Cycle” exists over which the number of sunspots rises from zero during the Sunspot Minimum to a Sunspot Maximum where many sunspots occur, before decreasing again to the next Sunspot Minimum. The length of this “Sunspot Cycle” or “Solar Cycle” is typically about 11 years, but it can vary between 8 and 14 years. This 11-year cycle in sunspot behaviour is part of a 22-year cycle in magnetic behaviour known as the Hale Cycle. Additionally, multi-decadal and even centennial trends are observed in the sunspot numbers. During the period from 1645 to 1715, known as the “Maunder Minimum” [125,128–130,134], sunspots were very rarely observed at all.

Clearly, these changes in sunspot activity are capturing some aspect of solar activity, and provide evidence that the Sun is not a *constant* star, but one whose activity shows significant variability on short and long time scales. Therefore, the sunspot records initially seem like an exciting source of information on changes in solar activity. However, as will be discussed in more detail later, it is still unclear how much of the variability in TSI is captured by the sunspot numbers. The fact that sunspot numbers are **not** the only important measure of solar activity (as many researchers often implicitly assume, e.g., Gil-Alana et al. (2014) [21]) can be recognized by the simple realisation that TSI does not fall to zero every ~11 years during sunspot minima, even though the sunspot numbers do. Indeed, satellite measurements confirm that sunspots actually reduce solar luminosity, yet paradoxically the average TSI increases during sunspot maxima and decreases during sunspot minima [137–140].

We will discuss the current explanations for the apparently paradoxical relationship between sunspots and TSI in Sections 2.2 and 2.3. In any case, the fact that there is more to solar activity than sunspot numbers was recognized more than a century ago by Maunder and Maunder (1908) [124] (for whom the “Maunder Minimum” is named) who wrote,

> “*…for sun-spots are but one symptom of the sun’s activity, and, perhaps, not even the most important symptom*” - Maunder and Maunder (1908), pp189-190 [124];

and,

> “*A* [‘great’] *spot like that of February, 1892 is enormous of itself, but it is a very small object compared to the sun; and spots of such size do not occur frequently, and last but a very short time. We have no right to expect, therefore, that* *a time of many sun-spots should mean any appreciable falling off in the light and heat we have from the sun. Indeed, since the surface round the spots is generally bright beyond ordinary, it may well be that a time of many spots means no falling off, but rather the reverse.*" - Maunder and Maunder (1908), p183 [124]

At the start of the 20th century, Langley, Abbott and others at the Smithsonian Astrophysical Observatory (SAO) recognized that a more direct estimate of the variability in TSI was needed [141–143]. From 1902 until 1962, they carried out a fairly continuous series of measurements of the "solar constant", i.e., the average rate per unit area at which energy is received at the Earth's average distance from the Sun, i.e., 1 Astronomical Unit (AU). The fact they explicitly considered the solar constant to understand climate change is apparent from the title of one of the first papers describing this project, Langley (1904) [141], i.e., "*On a possible variation of the solar radiation and its probable effect on terrestrial temperatures*". However, they were also acutely aware of the inherent challenges in trying to estimate changes in solar radiation from the Earth's surface:

"*The determination of the solar radiation towards the Earth, as it might be measured outside the Earth's atmosphere (called the "solar constant"), would be a comparatively easy task were it not for the almost insuperable difficulties introduced by the actual existence of such an atmosphere, above which we cannot rise, though we may attempt to calculate what would be the result if we could.*" (Langley, 1904) [141].

The true extent of this problem of estimating the changes in TSI from beneath the atmosphere became apparent later in the program. Initially, by comparing the first few years of data, it looked like changes in TSI of the order of 10% were occurring. However, it was later realized that, coincidentally, major (stratosphere-reaching) volcanic eruptions occurred near the start of the program: at Mt. Pelée and La Soufrière (1902) and Santa Maria (1903). Hence, the resulting stratospheric dust and aerosols from these eruptions had temporarily reduced the transmission of solar radiation through the atmosphere [143].

## 2.2. The debate over changes in Total Solar Irradiance during the satellite era (1978-present)

It was not until much later in the 20th century that researchers overcame this ground-based limitation through the use of rocket-borne [144], balloon-borne [145] and spacecraft measurements [146]. Ultimately, when Hoyt (1979) systematically reviewed the entire ~60 year-long SAO solar constant project, he found unfortunately that any potential trend in the solar constant over the record was probably less than the accuracy of the measurements (~0.3%) [143]. However, with the launch of the Nimbus 7 Earth Radiation Budget (ERB) satellite mission in 1978 and the Solar Maximum Mission (SMM) Active Cavity Radiometer Irradiance Monitor 1 (ACRIM1) satellite mission in 1980, it finally became possible to continuously and systematically monitor the incoming TSI for long periods from above the Earth's atmosphere [137,140,147,148].

Although each satellite mission typically provides TSI data for only 10 to 15 years, and the data can be affected by gradual long-term orbital drifts and/or instrumental errors that can be hard to identify and quantify [149], there has been an almost continuous series of TSI-monitoring satellite missions since those two initial U.S. missions, including European missions, e.g., SOVAP/Picard [150] and Chinese missions [151,152] as well as international collaborations, e.g., VIRGO/SOHO [153], and further U.S. missions, e.g., ACRIMSAT/ACRIM3 [154] and SORCE/TIM [155]. Therefore, in principle, by rescaling the measurements from different parallel missions so that they have the same values during the periods of overlap, it is possible to construct a continuous time series of TSI from the late-1970s to the present.

Therefore, it might seem reasonable to assume that we should at least have a fairly reliable and objective understanding of the changes in TSI during the satellite era, i.e., 1978 to present. However, even within the satellite era, there is considerable ongoing controversy over what exactly the trends in TSI have been [42,52,56,60,68,156–158]. There are a number of rival composite datasets, each implying different trends in TSI since the late-1970s. All composites agree that TSI exhibits a roughly 11-year cycle that matches well with the sunspot cycle discussed earlier. However, the composites differ in whether additional multidecadal trends are occurring.

The composite of the ACRIM group that was in charge of the three ACRIM satellite missions (ACRIM1, ACRIM2 and ACRIM3) suggests that TSI generally increased during the 1980s and 1990s but has slightly declined since then [52,60,154,159]. The Royal Meteorological Institute of Belgium (RMIB)'s composite implies that, aside from the sunspot cycle, TSI has remained fairly constant since at least the 1980s [160]. Meanwhile, the Physikalisch-Meteorologisches Observatorium Davos (PMOD) composite implies that TSI has been steadily decreasing since at least the late-1970s [157,161]. Additional TSI satellite composites have been produced by Scafetta (2011) [42]; de Wit et al. (2017) [156] and Gueymard (2018) [158].

The two main rival TSI satellite composites are ACRIM and PMOD. As we will discuss in Section 3, global temperatures steadily increased during the 1980s and 1990s but seemed to slow down since the end of the 20th century. Therefore, the debate over these three rival TSI datasets for the satellite era is quite important. If the ACRIM dataset is correct, then it suggests that much of the global temperature trends during the satellite era could have been due to changes in TSI [29,35,41,42,52,60,154,159]. However, if the PMOD dataset is correct, and we assume for simplicity a linear relationship between TSI and global temperatures, then the implied global temperature trends from changes in TSI would exhibit long-term global cooling since at least the late-1970s. Therefore, the PMOD dataset implies that none of the observed warming since the late-1970s could be due to solar variability, and that the warming must be due to other factors, e.g., increasing greenhouse gas concentrations. Moreover, it implies that the changes in TSI have been partially reducing the warming that would have otherwise occurred; if this TSI trend reverses in later decades, it might accelerate "global warming" [161,161,162].

The PMOD dataset is more politically advantageous to justify the ongoing considerable political and social efforts to reduce greenhouse gas emissions under the assumption that the observed global warming since the late-19th century is mostly due to greenhouse gases. Indeed, as discussed in Soon et al. (2015) [56], Dr. Judith Lean (of the PMOD group) acknowledged in a 2003 interview that this was one of the motivations for the PMOD group to develop a rival dataset to the ACRIM one by stating,

> "*The fact that some people could use Willson's* [ACRIM dataset] *results as an excuse to do nothing about greenhouse gas emissions is one reason we felt we needed to look at the data ourselves*" – Dr. Judith Lean, interview for NASA Earth Observatory, August 2003 [163]

Similarly, Zacharias (2014) argued that it was politically important to rule out the possibility of a solar role for any recent global warming,

> "*A conclusive TSI time series is not only desirable from the perspective of the scientific community, but also when considering the rising interest of the public in questions related to climate change issues, thus preventing climate skeptics from taking advantage of these discrepancies within the TSI community by, e.g., putting forth a presumed solar effect as an excuse for inaction on anthropogenic warming.*" – Zacharias (2014) [164]

We appreciate that some readers may share the sentiments of Lean and Zacharias and others and may be tempted to use these political arguments for helping them to decide their opinion on this ongoing scientific debate. In this context, readers will find plenty of articles to use as apparent scientific justification, e.g., Refs. [22,150,156,157,160–162,164–166]. It may also be worth noting that the IPCC appears to have taken the side of the PMOD group in their most recent 5th Assessment Report – see Section 8.4.1 of IPCC (2013) [1] for the key discussions. However, we would encourage all readers to carefully consider the counter-arguments offered by the ACRIM group, e.g., Refs. [29,52,60,154,159]. In our opinion, this was not satisfactorily done by the authors of the relevant section in the influential IPCC reports, i.e. Section 8.4.1 of IPCC (2013) [1]. Matthes et al. (2017)'s recommendation that their new estimate (which will be discussed below) should be the only solar activity dataset considered by the CMIP6 modelling groups [110] for the IPCC's upcoming 6th Assessment Report is even more unwise due to the substantial differences between various published TSI estimates. This is aside from the fact that Scafetta et al. (2019) [60] have argued that the TSI proxy reconstructions preferred in Matthes et al. (2017) (i.e., NRLTSI2 and SATIRE) contradict important features observed in the ACRIM 1 and ACRIM 2 satellite measurements. We would also encourage readers to carefully read the further discussion of this debate in Soon et al. (2015) [56].

## *2.3. Implications of the satellite era debate for pre-satellite era estimates*

The debate over which satellite composite is most accurate also has implications for assessing TSI trends in the pre-satellite era. In particular, there is ongoing debate over how closely the variability in TSI corresponds to the variability in the sunspot records. This is important, because if the match is very close, then it implies the sunspot records can be a reliable solar proxy for the pre-satellite era (after suitable scaling and calibration has been carried out), but if not other solar proxies may need to be considered.

In the 1980s and early 1990s, data from the NIMBUS7/ERB and ACRIMSAT/ACRIM1 satellite missions suggested a cyclical component to the TSI variability that was highly correlated to the sunspot cycle. That is, when sunspot numbers increased, so did TSI, and when sunspot numbers decreased, so did TSI [137–140,147,167,168]. This was not known in advance, and it was also unintuitive because sunspots are "darker", and so it might be expected that more sunspots would make the Sun "less bright" and therefore lead to a lower TSI. Indeed, the first six months of data from the ACRIM1 satellite mission suggested that this might be the case because "*two large decreases in irradiance of up to 0.2 percent lasting about 1 week [were] highly correlated with the development of sunspot groups*" – Willson et al. (1981) [148]. However, coincidentally, it appears that the sunspot cycle is also highly correlated with changes in the number of "faculae" and in the "magnetic network", which are different types of intermittent magnetic phenomena that are also associated with the Sun's photosphere, except that these phenomena appear as "bright" spots and features. It is now recognized that the Sun is currently a "faculae-dominated star". That is, even though sunspots themselves seem to reduce TSI, when sunspot numbers increase, the number of faculae and other bright features also tend to increase, increasing TSI, and so the net result is an increase in TSI. That is, the increase in brightness from the faculae outweighs the decrease from sunspot dimming (i.e., the faculae:sunspot ratio of contributions to TSI is greater than 1). For younger and more active stars, the relative contribution is believed to be usually reversed (i.e., the ratio is less than 1) with the changes in stellar irradiance being "spot-dominated" [169–171].

At any rate, it is now well-established that TSI slightly increases and decreases over the sunspot cycle in tandem with the rise and fall in sunspots (which coincides with a roughly parallel rise and fall in faculae and magnetic network features) [137–140,147,167,168]. Many of the current pre-satellite era TSI reconstructions are based on this observation. That is, a common approach to estimating past TSI trends includes the following three steps:

1. Estimate a function to describe the inter-relationships between sunspots, faculae and TSI during the satellite era.
2. Assume these relationships remained reasonably constant over the last few centuries at least.
3. Apply these relationships to one or more of the sunspot datasets and thereby extend the TSI reconstruction back to 1874 (for sunspot areas [139,172]); 1700 (for sunspot numbers [132,133]; or 1610 (for group sunspot numbers [128,129]).

Although there are sometimes additional calculations and/or short-term solar proxies involved, this is the basic approach adopted by, e.g., Foukal and Lean (1990) [139]; Lean (2000) [173]; Solanki et al. (2000,2002) [174,175]; Wang et al. (2005) [172]; Krivova et al. (2007,2010) [176,177]. Soon et al. (2015) noted that this heavy reliance on the sunspot datasets seems to be a key reason for the similarities between many of the TSI reconstructions published in the literature [56].

However, does the relationship between faculae, sunspots and TSI remain fairly constant over multidecadal and even centennial timescales? Also, would the so-called "quiet" solar region remain perfectly constant despite multidecadal and secular variability observed in the sunspot and faculae cycles? Is it reasonable to assume that there are no other aspects of solar activity that contribute to variability in TSI? If the answers to all these questions are yes, then we could use the sunspot record as a proxy for TSI, scale it accordingly and extend the satellite record back to the 17$^{th}$ century. This would make things much simpler. It would mean that, effectively, even Galileo Galilei could have been able to determine almost as much about the changes in TSI of his time with his early 16$^{th}$-century telescope as a modern (very high budget) Sun-monitoring satellite mission of today. All that he would have been missing was the appropriate scaling functions to apply to the sunspot numbers to determine TSI.

If the PMOD or similar satellite composites are correct, then it does seem that, at least for the satellite era (1978-present), the sunspot cycle is the main variability in TSI and that the relationships between faculae, sunspots and TSI have remained fairly constant. This is because the trends of the PMOD composite are highly correlated to the trends in sunspot numbers over the entire satellite record. However, while the ACRIM composite also has a component that is highly correlated to the sunspot cycle (and the faculae cycle), it implies that there are also additional multidecadal trends in the solar luminosity that are not captured by a linear relation between sunspot and faculae records. Recent modelling by Rempel (2020) is consistent with this in that his analysis suggests even a 10% change in the quiet-Sun field strength between solar cycles could lead to an additional TSI variation comparable in magnitude to that over a solar cycle [178]. Therefore, if the ACRIM composite is correct, then it would be necessary to consider additional proxies of solar activity that are capable of capturing these non-sunspot number-related multidecadal trends.

Over the years, several researchers have identified several time series from the records of solar observers that seem to be capturing different aspects of solar variability than the basic sunspot numbers [24,179–185]. Examples include the average umbral/penumbral ratio of sunspots [181], the length of sunspot cycles [49,68,182,183,186–189], solar rotation rates [184], the "envelope" of sunspot numbers [185], variability in the 10.7cm solar microwave emissions [65,190], solar plage areas (e.g., from Ca II K spectroheliograms) [190–193], polar faculae [61,62,194], and white-light faculae areas [195,196]. Another related sunspot proxy that might be useful is the sunspot decay rate. Hoyt and Schatten (1993) have noted that a fast decay rate suggests an enhanced solar convection, and hence a brighter Sun, while a slower rate indicates the opposite [179]. Indeed, indications suggest that the decay rate during the Maunder Minimum was very slow [179], hence implying a dimmer Sun in the mid-to-late 17$^{th}$ century. Owens et al. (2017) developed a reconstruction of the solar wind back to 1617 that suggests the solar wind speed was lower by a factor of two during the Maunder Minimum [197]. Researchers have also considered records of various aspects of geomagnetic activity, since the Earth's magnetic field appears to be strongly influenced by solar activity [194,198–202].

There are many other solar proxies that might also be capturing different aspects of the long-term solar variability, e.g., see Livingston (1994) [180], Soon et al. (2014) [203] and Soon et al. (2015) [56]. In particular, it is worth highlighting the use of cosmogenic isotope records, such as $^{14}C$ or $^{10}Be$ [204], since they are used by several of the TSI reconstructions we will consider. Cosmogenic isotope records have been used as long-term proxies of solar activity since the 1960s [205–209]. Cosmogenic isotopes such as $^{14}C$ or $^{10}Be$ are produced in the atmosphere via galactic cosmic rays. However, when solar activity increases, the solar wind reaching the Earth also increases. This tends to reduce the flux of incoming cosmic rays, thus reducing the rate of production of these isotopes and their quantity. These isotopes can then get incorporated into various long-term records, such as tree rings through photosynthesis. Therefore, by studying the changes in the relative concentrations of these isotopes over time in e.g., tree rings, it is possible to construct an estimate of multidecadal-to-centennial and even millennial changes in average solar activity. Because the atmosphere is fairly well-mixed, the concentration of these isotopes only slowly changes over several years, and so the 8-14 year sunspot cycle can be partially reduced in these solar proxies. However, Stefani et al. (2020) still found a very good match between $^{14}C$ or $^{10}Be$ solar proxies and Schove's (1955) [131] estimates of solar activity maxima, which were based on historical *aurora borealis* observations back to 240 A.D. [210]. Moreover, the records can cover much longer periods, and so are particularly intriguing for studying multi-decadal, centennial and millennial variability.

We note that several studies have tended to emphasize the similarities between various solar proxies [13,17,18,22,158,190]. We agree that this is important, but we argue that it is also important to contrast as well as compare. To provide some idea of the effects that solar proxy choice, as well as TSI satellite composite choice, can have on the resulting TSI reconstruction, we plot in Figure 1 several different plausible TSI reconstructions taken from the literature and/or adapted from the literature. All 9 reconstructions are provided in the Supplementary Materials.

Foukal (2012) [193] and Foukal (2015) [196] used a similar approach to the Wang et al. (2005) [172] reconstruction but used slightly different solar proxies for the 20$^{th}$ century pre-satellite era. Foukal (2012) used 10.7cm solar microwave emissions for the 1947-1979 period and solar plage areas (from Ca II K spectroheliograms) for the 1916-1946 period. Foukal (2015) used the faculae areas (from white light images) for the 1916-1976 period. In contrast, Wang et al. (2005) predominantly relied on the group sunspot number series for the pre-satellite era (after scaling to account for the sunspot/faculae/TSI relationships during the satellite era). These three different reconstructions are plotted as Figure 1(a), (c) and (i) respectively. All three reconstructions have a lot in common, e.g., they all have a very pronounced ~11 year Solar Cycle component, and they all imply a general increase in TSI from the 19$^{th}$ century to the mid-20$^{th}$ century, followed by a general decline to present. However, there are two key differences between them. First, the Wang et al. (2005) reconstruction implies a slightly larger increase in TSI from the 19$^{th}$ century to the 20$^{th}$ century. On the other hand, while Foukal (2012) and Wang et al. (2005) imply the maximum TSI occurred in 1958, the Foukal (2015) reconstruction implies a relatively low TSI in 1958 and suggested two 20$^{th}$ century peaks in TSI – one in the late 1930s and another in 1979, i.e., the start of the satellite era. All three reconstructions imply that none of the global warming since at least 1979 could be due to increasing TSI, and in the case of the Foukal (2012) and Wang et al. (2005) reconstructions, since at least 1958.

Meanwhile, all three of those reconstructions were based on the PMOD satellite composite rather than the ACRIM composite. Therefore, in Figure 1(b) and (d), we have modified the Foukal (2012) and (2015) reconstructions using the ACRIM series for the 1980-2012 period instead of PMOD. We did this by rescaling the ACRIM time series to have the same mean TSI over the common period of overlap, i.e., 1980-2009.

Because the ACRIM composite implies a general increase in TSI from 1980 to 2000 followed by a general decrease to present, while the PMOD composite implies a general decrease in TSI over the entire period, this significantly alters the long-term trends. The modified version of Foukal (2012) implies that the 1958 peak in TSI was followed by an equivalent second peak in 2000. This suggests that at least some of the global warming from the 1970s to 2000 could have been due to increasing TSI, i.e., contradicting a key implication of Foukal (2012). The modified Foukal (2015) reconstruction is even more distinct. It implies that TSI reached an initial peak in the late 1930s, before declining until 1958, and then increasing to a maximum in 2000. As we will discuss in Section 3, this is broadly similar to many of the Northern Hemisphere temperature estimates. Therefore, the modified Foukal (2015) is at least consistent with the possibility of TSI as a primary driver of global temperatures over the entire 20$^{th}$ century.

That said, while these plausible modifications can alter the relative magnitudes and timings of the various peaks and troughs in TSI, all these reconstructions would still be what Soon et al. (2015) [56] and Scafetta et al. (2019) [60] refer to as "low variability" reconstructions. That is, the multi-decadal trends in TSI appear to be relatively modest compared to the rising and falling over the ~11-year Solar Cycle component. As we will discuss in Section 2.6, many researchers have identified evidence for a significant ~11-year temperature variability in the climate of the mid-troposphere to stratosphere [65–67,76–78,91,211–216], which has been linked to the more pronounced ~11-year variability in incoming solar ultraviolet irradiance [8,217–225]. However, in terms of surface temperatures, the ~11-year component seems to be only of the order of 0.02-0.2°C over the course of a cycle [40,226–230]. Some researchers have argued that the relatively large heat capacity of the oceans could act as a "calorimeter" to integrate the incoming TSI over decadal time scales, implying that the multidecadal trends are more relevant for climate change than annual variability [40,48,69,231–233], and others have argued that these relatively small temperature variations could influence the climate indirectly through e.g., altering atmospheric circulation patterns [92,93,95,96,111,234]. However, this observation appears to have convinced many researchers (including the IPCC reports [1]) relying on "low variability" reconstructions that TSI cannot explain more than a few tenths of a °C of the observed surface warming since the 19$^{th}$ century, e.g., [1,5,7–9,11,12,16–18,21,22]. We will discuss these competing hypotheses and ongoing debates (which several co-authors of this paper are actively involved in) in Section 2.6, as these become particularly important if the true TSI reconstruction is indeed "low variability", i.e., dominated by the ~11-year cycle.

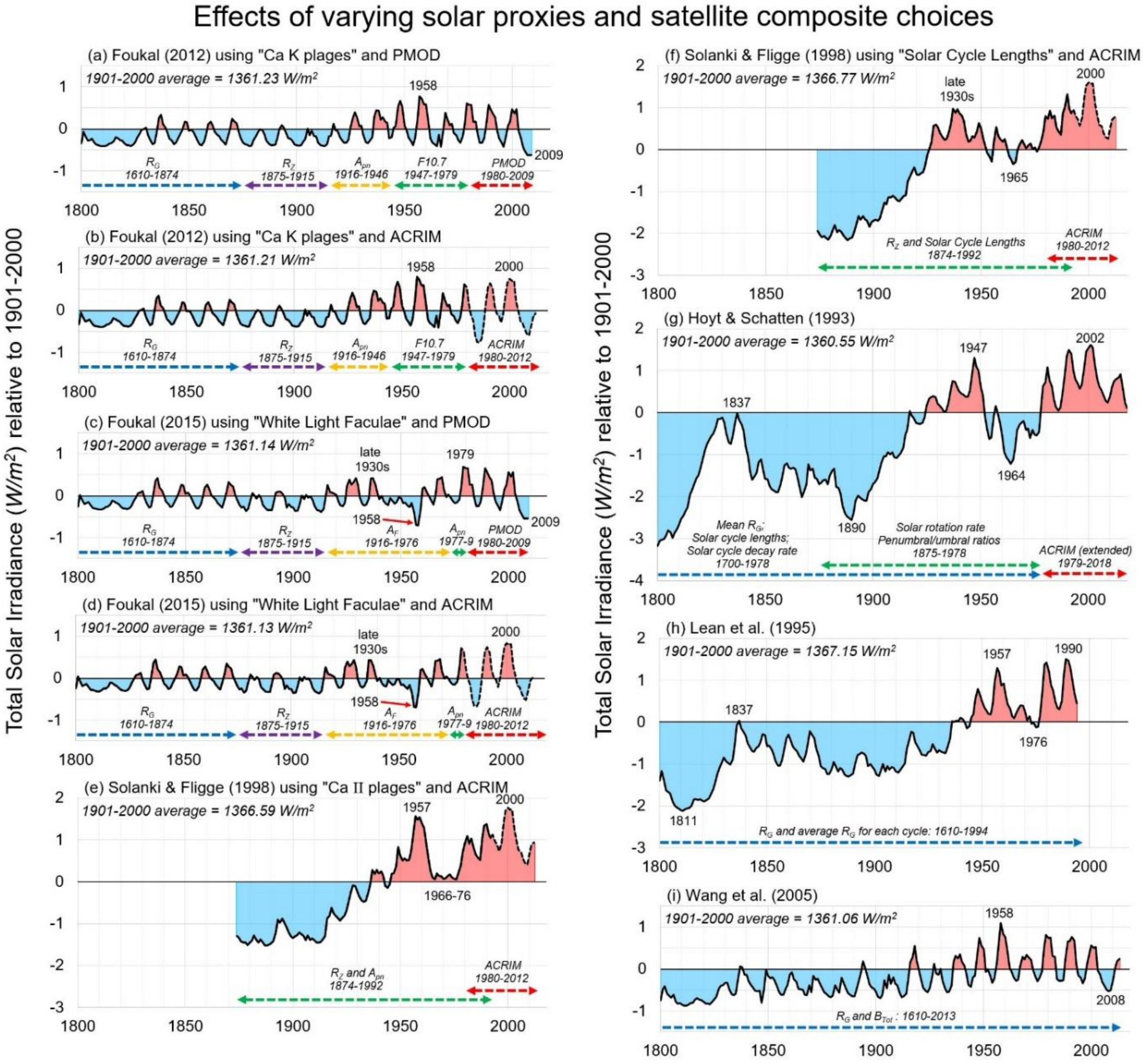

*Figure 1. Examples of different TSI reconstructions that can be created by varying the choice of solar proxies used for the pre-satellite era and the choice of TSI composite used for the satellite era. The Foukal (2012, 2015) [193,196] series using PMOD were downloaded from http://heliophysics.com/solardata.shtml (Accessed 20/06/2020). The equivalent ACRIM series were rescaled using the annual means of the ACRIM TSI composite which was downloaded from http://www.acrim.com/Data%20Products.htm (Accessed 01/07/2020). The two Solanki & Fligge (1998) series were digitized from Figure 3 of that paper [183] and extended up to 2012 with the updated ACRIM annual means. The Hoyt and Schatten (1993) [179] series was updated to 2018 by Scafetta et al. (2019) [60]. The Wang et al. (2005) [172] and Lean et al. (1995) [185] series were taken from the Supplementary Materials of Soon et al. (2015) [56]*

On the other hand, let us consider the possibility that the true TSI reconstruction should be "high variability". In Figure 1(e)-(h), we consider four such "high variability" combinations, and we will discuss more in Section 2.4. All four of these reconstructions include a ~11-year solar cycle component like the "low variability" reconstructions, but they imply that this quasi-cyclical component is accompanied by substantial multidecadal trends. Typically, the ~11-year cycle mostly arises from the solar proxy components derived from the sunspot number datasets (as in the low variability reconstructions), while the multidecadal trends mostly arise from other solar proxy components.

Solanki and Fligge (1998) [183] considered two alternative proxies for their multidecadal component and treated the envelope described by the two individual components as a single reconstruction with error bars. Solanki and Fligge (1999) [235] also suggested that this reconstruction could be extended back to 1610 using the Group Sunspot Number time series of Hoyt and Schatten (1998) [128] as a solar proxy for the pre-1874 period. However, in Figure 1(e) and (f), we treated both components as separate reconstructions, which we digitized from Solanki and Fligge (1998)'s Figure 3, and extended up to 2012 with the updated ACRIM satellite composite. Both reconstructions are quite similar and, unlike the low variability estimates, imply a substantial increase in TSI from the end of the 19th century to the end of the 20th century. They also both imply that this long-term increase was interrupted by a decline in TSI from a mid-20th century peak to the mid-1960s. However, the reconstruction using Ca II K plage areas (Figure 1e) implies that the mid-20th century peak occurred in 1957, while the reconstruction using Solar Cycle Lengths (Figure 1f) implies the mid-20th century peak occurred in the late 1930s and that TSI was declining in the 1940s up to 1965. In terms of the timing of the mid-20th century peak, it is worth noting that Scafetta (2012) found a minimum in mid-latitude aurora frequencies in the mid-1940s, which is indicative of increased solar activity [236].

Figure 1(g) plots the updated Hoyt and Schatten (1993) [179] TSI reconstruction. Although the original Hoyt and Schatten (1993) reconstruction was calibrated to the satellite era using the NIMBUS7/ERB time series as compiled by Hoyt et al. (1992) [147], it has since been updated by Scafetta and Willson (2014) [52] and more recently by Scafetta et al. (2019) [60] using the ACRIM composite until 2013 and the VIRGO and SORCE/TIM records up to present. The Hoyt and Schatten (1993) reconstruction is quite similar to the two Solanki and Fligge (1998) reconstructions, except that it implies a greater decrease in TSI from the mid-20th century to the 1960s, and that the mid-20th century peak occurred in 1947.

We note that there appear to be some misunderstandings in the literature over the Hoyt and Schatten (1993) reconstruction, e.g., Fröhlich and Lean (2002) mistakenly reported that "…*Hoyt and Schatten (1993) is based on solar cycle length whereas the others are using the cycle amplitude*" [162]. Therefore, we should stress that, like Lean et al. (1995) [185], the Hoyt and Schatten (1993) reconstruction did include both the sunspot numbers and the envelope of sunspot numbers, but unlike most of the other reconstructions, they also included multiple additional solar proxies [179]. We should also emphasize that the Hoyt and Schatten (1998) [128] paper describing the widely-used "Group Sunspot Number" dataset is a completely separate analysis, although it was partially motivated by Hoyt and Schatten (1993).

The Lean et al. (1995) [185] reconstruction of Figure 1(h) also implies a long-term increase in TSI since the 19th century and a mid-20th century initial peak – this time at 1957, i.e., similar to Figure 1(e). The Lean et al. (1995) reconstruction was based on Foukal and Lean's (1990) reconstruction [139], and itself evolved into Lean (2000) [173], which evolved into Wang et al. (2005) [172], which in turn has evolved into the Coddington et al. (2016) [237] reconstruction, which as we will discuss in Section 2.4 is a major component of the recent Matthes et al. (2017) [110] reconstruction. However, Soon et al. (2015) [56] noted empirically (in their Figure 9) that the main net effect of the evolution from Lean et al. (1995) [185] to Lean (2000) [173] to Wang et al. (2005) [172] has been to reduce the magnitude of the multi-decadal trends, i.e., to transition towards a "low variability" reconstruction. We note that Coddington et al. (2016) [237] and Matthes et al. (2017) [110] has continued this trend. Another change in this family of reconstructions is that the more recent ones have used the PMOD satellite composite instead of ACRIM (which is perhaps not surprising given that Lean was one of the PMOD team, as mentioned in Section 2.2, as well as being a co-author of all of that family of reconstructions).

Therefore, a lot of the debate over whether the high or low variability reconstructions are more accurate relates to the question of whether or not there are multi-decadal trends that are not captured by the ~11-year solar cycle described by the sunspot numbers. This overlaps somewhat with the ACRIM vs. PMOD debate since the PMOD implies that TSI is very highly correlated to the sunspot number records (via the correlation between sunspots and faculae over the satellite era), whereas the ACRIM composite is consistent with the possibility of additional multidecadal trends between solar cycles [42,52,56,60,156–158].

This has been a surprisingly challenging problem to resolve. As explained earlier, the ~11-year cyclical variations in TSI over the satellite era are clearly well correlated to the trends in the areas of faculae, plages, as well as sunspots over similar timescales [137–140,147,167,168]. However, on shorter timescales, TSI is actually anti-correlated to sunspot area [148,238]. Therefore, the ~11-year rise and fall in TSI in tandem with sunspot numbers cannot be due to the sunspot numbers themselves, but appears to be a consequence of the rise and fall of sunspot numbers being *commensally* correlated to those of faculae and plages. However, Kuhn et al. (1988) argued that, "…*solar cycle variations in the spots and faculae alone cannot account for the total* [TSI] *variability*" and that, "… *a third component is needed to account for the total variability*" [239]. Therefore, while some researchers have assumed, like Lean et al. (1998) that there is "…[no] *need for an additional component other than spots or faculae*" [240], Kuhn et al. [241–245] continued instead to argue that "*sunspots and active region faculae do not* [on their own] *explain the observed irradiance variations over the solar cycle*" [244] and that there is probably a "*third component of the irradiance variation*" that is a "*nonfacular and nonsunspot contribution*" [241]. Work by Li, Xu et al. is consistent with Kuhn et al.'s assessment, e.g., Refs. [246–249], in that they have shown: that TSI variability can be decomposed into multiple frequency components [246]; that the relationships are different between different solar activity indices and TSI [247,249]; and that the relationship between sunspot numbers and TSI varies between cycles [248]. Indeed, in order to accurately reproduce the observed TSI variability over the two most recent solar cycles using solar disk images from ground-based astronomical observatories, Fontenla and Landi (2018) [250] needed to consider *nine different solar features* rather than the simple sunspot and faculae model described earlier.

In summary, there are several key debates ongoing before we can establish which TSI reconstructions are most accurate:

1. Which satellite composite is most accurate? In particular, is PMOD correct in implying that TSI has generally decreased over the satellite era, or is ACRIM correct in implying that TSI increased during the 1980s and 1990s before decreasing?
2. Is it more realistic to use a high variability or low variability reconstruction? Or, alternatively, has the TSI variability been dominated by the ~11-year solar cycle, or have there also been significant multi-decadal trends between cycles?
3. When did the mid-20$^{th}$ century peak occur, and how much and for how long did TSI decline after that peak?

The answers to these questions can substantially alter our understanding of how TSI has varied over time. For instance, Velasco Herrera et al. (2015) used machine learning and four different TSI reconstructions as training sets to extrapolate forward to 2100 AD and backwards to 1000 AD [251]. The results they obtained had much in common, but also depended on whether they used PMOD or ACRIM as well as whether they used a high or low variability reconstruction. As an aside, the forecasts from each of these combinations implied a new solar minimum starting in 2002-2004 and ending in 2063-2075. If these forecasts are correct, then in addition to the potential influence on future climate change, such a deficit in solar energy during the 21$^{st}$ century could have serious implications for food production; health; in the use of solar-dependent resources; and more broadly could affect many human activities [251].

## *2.4. Sixteen different estimates of changes in Total Solar Irradiance since the 19$^{th}$ century and earlier*

Soon et al. (2015) identified eight different TSI reconstructions (see Figure 8 in that paper) [56]. Only four of these reconstructions were used by the CMIP5 modelling groups for the hindcasts that were submitted to the IPCC 5$^{th}$ Assessment Report: Wang et al. (2005) [172] described above, as well as Krivova et al. (2007) [176]; Steinhilber et al. (2009) [252]; and Vieira et al. (2011) [253]. Coincidentally, all four implied very little solar variability (and also a general decrease in TSI since the 1950s). However, Soon et al. (2015) also identified another four TSI reconstructions that were at least as plausible – including the Hoyt and Schatten (1993) [179] and Lean et al. (1995) [185] reconstructions described above. Remarkably, all four implied much greater solar variability. These two sets are the "high solar variability" and "low solar variability" reconstructions discussed in Section 2.3 which both Soon et al. (2015) [56] and more recently Scafetta et al. (2019) [60] have referred to.

Since then, eight additional estimates have been proposed – four low variability and four high variability. Coddington et al. (2016) [237] have developed a new version of the Wang et al. (2005) [172] estimate that has reduced the solar variability even further (it uses a sunspot/faculae model based on PMOD). Recently, Matthes et al. (2017) [110] took the mean of the Coddington et al. (2016) [237] estimate and the (similarly low variability) Krivova et al. (2010) [176,177] estimate, and proposed this as a new estimate. Moreover, Matthes et al. (2017) recommended that their new estimate should be the only solar activity dataset considered by the CMIP6 modelling groups [110]. Clearly, Matthes et al.'s (2017) recommendation to the CMIP6 groups goes against the competing recommendation of Soon et al. (2015) [56] to consider a more comprehensive range of TSI reconstructions.

In Figure 2, we plot the four "low solar variability" reconstructions from Soon et al. (2015) [56] as well as these two new "low variability" estimates along with another two estimates by Dr. Leif Svalgaard (Stanford University, USA), which have not yet been described in the peer-reviewed literature but are available from Svalgaard's website [https://leif.org/research/download-data.htm, last accessed 27/03/2020], and have been the subject of some discussion on internet forums.

Recently, Egorova et al. (2018) [254] proposed four new "high variability" estimates that built on the earlier Shapiro et al. (2011) [255] estimate. The Shapiro et al. (2011) [255] estimate generated some critical discussion [256–258] (see Section 2.5.4). Egorova et al. (2018) [254] have taken this discussion into account and proposed four new estimates using a modified version of the Shapiro et al. (2011) [255] methodology. Therefore, in Figure 3, we plot the four "high solar variability" reconstructions from Soon et al. (2015) [56] as well as these four new "high variability" estimates.

This provides us with a total of 16 different TSI reconstructions. Further details are provided in Table 1 and in the Supplementary Information. For interested readers, we have also provided the four additional TSI reconstructions discussed in Figure 1 in the Supplementary Information.

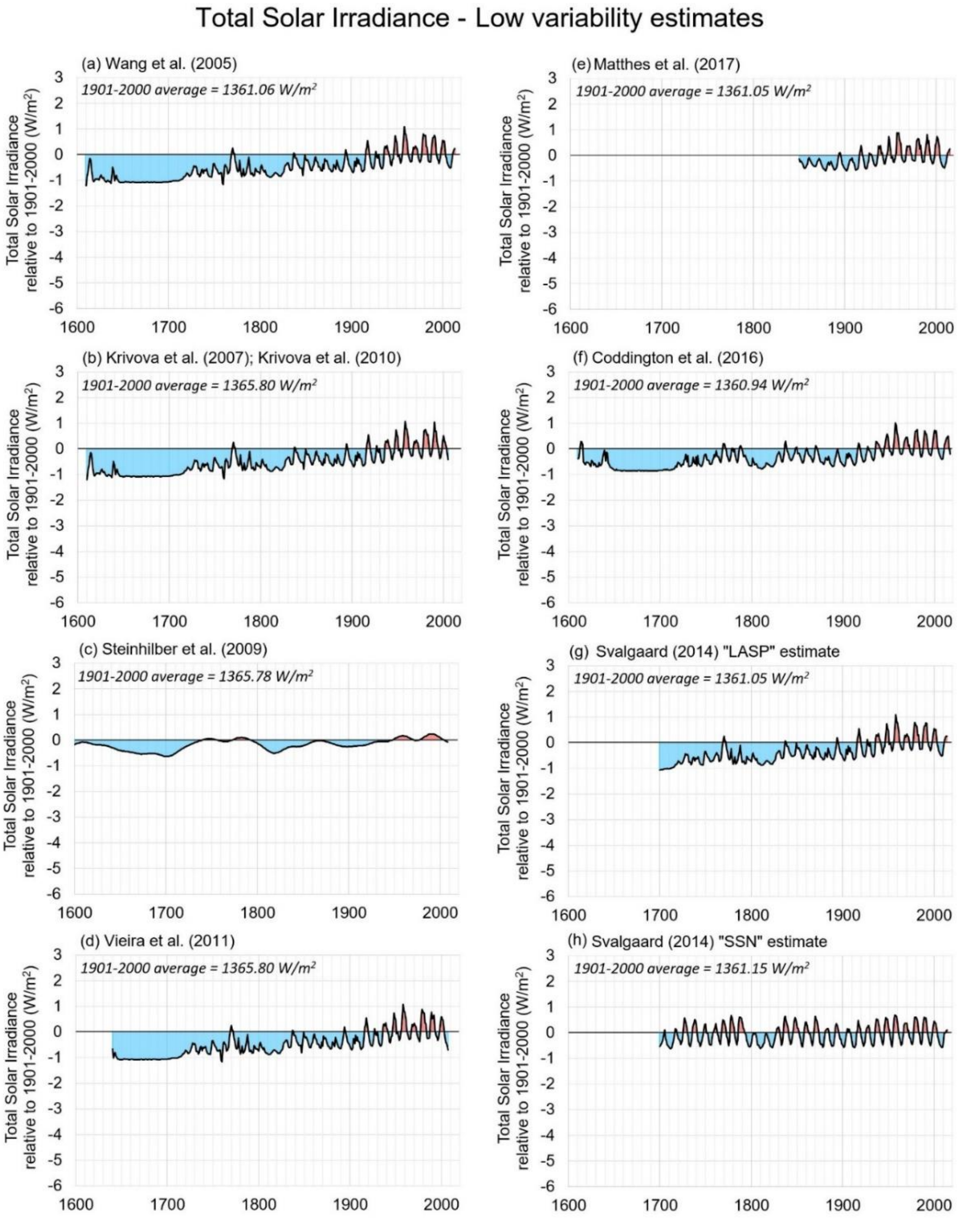

*Figure 2. Eight low variability estimates of Total Solar Irradiance changes relative to 1901-2000 average.*

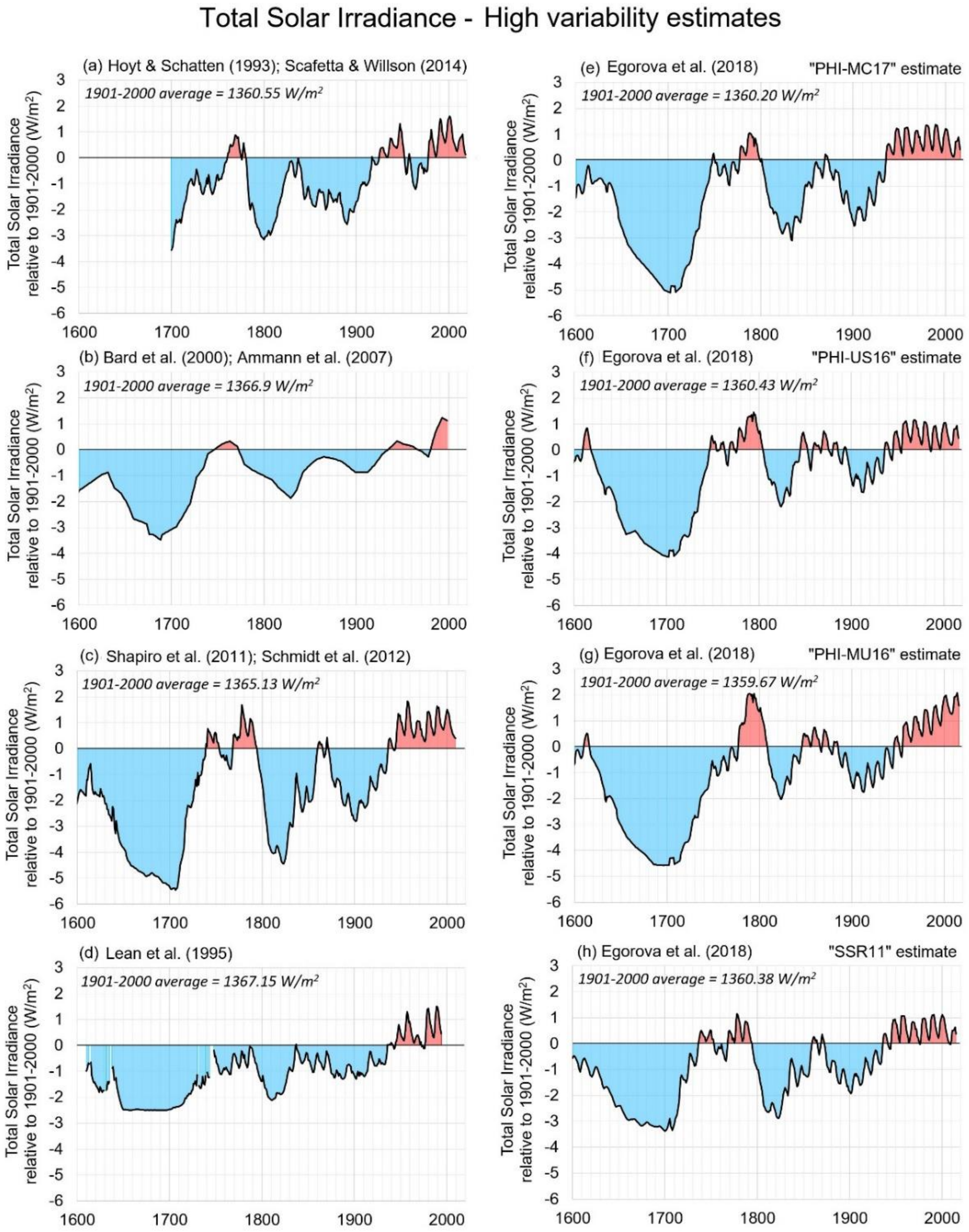

*Figure 3. Eight high variability estimates of the Total Solar Irradiance changes relative to the 1901-2000 average. Note the y-axis scales are the same as in Figure 2.*

*Table 1. The sixteen different estimates of the changes in solar output, i.e., Total Solar Irradiance (TSI), analyzed in this study.*

| IPCC AR5 | Variability | Study | Start | End | 20th Century mean TSI (W/m²) |
|---|---|---|---|---|---|
| Yes | Low | Wang et al. (2005) | 1610 | 2013 | 1361.06 |
| Yes | Low | Krivova et al. (2007); updated by Krivova et al. (2010) | 1610 | 2005 | 1365.8 |
| Yes | Low | Steinhilber et al. (2009) | 7362 BCE | 2007 | 1365.78 |
| Yes | Low | Vieira et al. (2011) | 1640 | 2007 | 1365.8 |
| CMIP6 | Low | Matthes et al. (2017) | 1850 | 2015 | 1361.05 |
| N/A | Low | Coddington et al. (2016) | 1610 | 2017 | 1360.94 |
| N/A | Low | Svalgaard (2014) "LASP" estimate | 1700 | 2014 | 1361.05 |
| N/A | Low | Svalgaard (2014) "SSN" estimate | 1700 | 2014 | 1361.15 |
| No | High | Hoyt & Schatten (1993); updated by Scafetta et al. (2019) | 1701 | 2018 | 1360.55 |
| No | High | Bard et al. (2000); updated by Ammann et al. (2007) | 843 | 1998 | 1366.9 |
| No | High | Shapiro et al. (2011); adapted by Schmidt et al. (2012) | 850 | 2009 | 1365.13 |
| No | High | Lean et al. (1995) | 1610 | 1994 | 1367.15 |
| N/A | High | Egorova et al. (2018) "PHI-MC17" estimate | 6000 BCE | 2016 | 1360.20 |
| N/A | High | Egorova et al. (2018) "PHI-US16" estimate | 6000 BCE | 2016 | 1360.43 |
| N/A | High | Egorova et al. (2018) "PHI-MU16" estimate | 16 | 2016 | 1359.67 |
| N/A | High | Egorova et al. (2018) "SSR11" estimate | 1600 | 2015 | 1360.38 |

## *2.5. Arguments for a significant role for solar variability in past climate change*

The primary focus of the new analysis in this paper (Section 5) is on evaluating the simple hypothesis that there is a direct linear relationship between incoming TSI and Northern Hemisphere surface air temperatures. As will be seen, even for this simple hypothesis, a remarkably wide range of answers are still plausible. However, before we discuss in Section 3 what we currently know about Northern Hemisphere surface air temperature trends since the 19th century (and earlier), it may be helpful to briefly review some of the other frameworks within which researchers have been debating potential Sun/climate relationships.

The gamut of scientific literature which encompasses the debates summarised in the following subsections (2.5 and 2.6) can be quite intimidating, especially since many of the articles cited often come to diametrically opposed conclusions that are often stated with striking certainty. With that in mind, in these two subsections, we have merely tried to summarise the main competing hypotheses in the literature, so that readers interested in one particular aspect can use this as a starting point for further research. Also, several of the co-authors of this paper have been active participants in each of the debates we will be reviewing. Hence, there is a risk that our personal assessments of these debates might be subjective. Therefore, we have especially endeavoured to avoid forming definitive conclusions, although many of us have strong opinions on several of the debates we will discuss here.

The various debates that we consider in this subsection (2.5) can be broadly summarised as being over whether variations in solar activity have been a major climatic driver in the past. We stress that a positive answer does not in itself tell us how much of a role solar activity has played in recent climate change. For instance, several researchers have argued that solar activity was a major climatic driver until relatively recently, but that anthropogenic factors (chiefly anthropogenic $CO_2$ emissions) have come to dominate in recent decades [5,13–17,22,68,80,185]. However, others counter that if solar activity was a major climatic driver in the past, then it is plausible that it has also been a major climatic driver in recent climate change. Moreover, if the role of solar activity in past climate change has been substantially underestimated, then it follows that its role in recent climate change may also have been underestimated [28,33,36–39,54,55,63,182,186,187,259–267].

### *2.5.1. Evidence for long-term variability in both solar activity and climate*

Over the years, numerous studies have reported on the similarities between the timings and magnitudes of the peaks and troughs of various climate proxy records and equivalent solar proxy records [28,30,37–39,54,55,80,182,186,189,209,262,268–277]. Most climate proxy records are taken to be representative of regional climates, and so these studies are often criticised for only representing regionalised trends and/or that there may be reliability issues with the records in question [12,18,100,278] (see also Section 2.6.3). However, others note that similar relationships can be found at multiple sites around the world [30,38,39,54,55,189,262,275,277]. And, it has been argued that some global or hemispheric paleo-temperature reconstructions show similar trends to certain solar reconstructions [39,54,55,262].

These studies are often supplemented by additional studies presenting further evidence for substantial past climatic variability (with the underlying but not explicitly tested assumption that this may have been solar-driven) [30,39,54,55,262]. Other studies present further evidence for substantial past solar variability (with the underlying but not explicitly tested assumption that this contributed to climate changes) [112,261,279,280].

Studies which suggest considerable variability in the past for either solar activity or climate provide evidence that is *consistent with* the idea that there has been a significant role for solar variability in past climate change. However, if the study only considers the variability of one of the two (solar versus climate) in isolation from the other, then this is mostly qualitative in nature.

For that reason, "attribution" studies, which attempt to quantitatively compare specific estimates of past climate change to specific solar activity reconstructions and other potential climatic drivers can often seem more compelling arguments for or against a major solar role. Indeed, this type of analysis will be the primary focus of Section 5. However, in the meantime, we note that the results of these attribution studies can vary substantially depending on which reconstructions are used for past climate change, past TSI, and any other potential climatic drivers that are considered. Indeed, Stott et al. (2001) explicitly noted that the amount of the 20$^{th}$ century warming they were able to simulate in terms of solar variability depended on which TSI reconstruction they used. [6]

For instance, Hoyt and Schatten's (1993) TSI reconstruction was able to "explain ~71% of the [temperature] variance during the past 100 years and ~50% of the variance since 1700" [179]. Soon et al. (1996) confirmed this result using a more comprehensive climate model-based analysis, and added that if increases in greenhouse gases were also included, the percentage of the long-term temperature variance over the period 1880-1993 that could be explained increased from 71% to 92% [23], although Cubasch et al.'s (1997) equivalent climate model-based analysis was only able to explain about 40% of the temperature variability over the same period in terms of solar activity [281]. More recently, Soon et al. (2015) argued that if Northern Hemisphere temperature trends are estimated using mostly rural stations (instead of using both urban and rural stations), then almost all of the long-term warming since 1881 could be explained in terms of solar variability (using Scafetta and Willson (2014)'s update to 2013 of the same TSI reconstruction [52]), and that adding a contribution for increasing greenhouse gases did not substantially improve the statistical fits [56].

On the other hand, using different TSI reconstructions, a number of studies have come to the opposite conclusion, i.e., that solar variability cannot explain much (if any) of the temperature trends since the late-19$^{th}$ century [5,13–16,19,21]. For instance, Lean and Rind (2008) could only explain 10% of the temperature variability over 1889-2006 in terms of solar variability [15], while Benestad and Schmidt (2009) could only explain 7±1% of the global warming over the 20$^{th}$ century in terms of solar forcing [16].

Meanwhile, other studies (again using different TSI reconstructions) obtained intermediate results, suggesting that solar variability could explain about half of the global warming since the 19$^{th}$ century [32,68,199] and earlier [33,185].

### *2.5.2. Similarity in frequencies of solar activity metrics and climate changes*

Another popular approach to evaluating possible Sun/climate relationships has been to use frequency analysis to compare and contrast solar activity metrics with climate records. The rationale of this approach is that if solar activity records show periodic or quasi-periodic patterns and if climate records show similar periodicities, it suggests that the periodic/quasi-periodic climate changes might have a solar origin. Given that the increase in greenhouse gas concentrations since the 19$^{th}$ century has been more continual in nature, and that the contributions from stratospheric volcanic eruptions appear to be more sporadic in nature (and temporary – with aerosol cooling effects typically lasting only 2-3 years), solar variability seems a much more plausible candidate for explaining periodic/quasi-periodic patterns in climate records than either greenhouse gases or volcanic activity.

Hence, much of the literature investigating potential Sun/climate relationships has focused on identifying and comparing periodicities (or quasi-periodicities) in climate, solar activity and/or geomagnetic activity records. For example, Le Mouël et al. [50,61,62,90,98,194,194,201,282,283]; Ruzmaikin and Feynman et al. [73–75]; Scafetta et al. [126,127,259–261,284,285]; White et al. [226,227]; Baliunas et al. (1997)[286]; Lohmann et al. (2004)[287]; Dobrica et al. [86–89]; Mufti and Shah (2011)[288]; Humlum et al. [47,63]; Laurenz, Lüdecke et al. [64,289]; Pan et al. (2020)[99]; Zhao et al. (2020) [267].

Although the exact frequencies of each of the periodicities and their relative dominance vary slightly from dataset to dataset, the authors argue that the periodicities are similar enough (within the uncertainties of the frequency analyses) to suggest a significant role for solar and/or geomagnetic activity in past climate change, albeit without explicitly quantifying the exact magnitude of this role or the exact mechanisms by which this solar influence manifests.

Again, it should be stressed that identifying a significant solar role in past climate change does not in itself rule out the possibility of other climate drivers and therefore does not necessarily imply that *recent* climate change is mostly solar. Indeed, the authors often explicitly state that the *relative* contributions of solar, anthropogenic factors as well as other natural factors in recent climate change may need to be separately assessed [47,62,126,259,287]. However, they typically add that the solar role is probably larger than otherwise assumed [47,62,126,259]. In particular, Scafetta (2013) notes that current climate models appear to be unable to satisfactorily simulate the periodicities present in the global temperature estimates, suggesting that the current climate models are substantially underestimating the solar contribution in recent climate change [259].

That said, one immediate objection to this approach is that one of the most striking quasi-periodic patterns in many solar activity records is the ~11 year solar cycle (sometimes called the "Schwabe cycle") described in previous sections, yet such ~11 year cycles are either absent or at best modest within most climate records [21]. We will discuss the various debates over this apparent paradox in Section 2.6. However, several researchers have countered that there are multiple periodicities other than the ~11 year Schwabe cycle present in both solar activity and climate datasets [62,73,88,99,126,182,194,259,284]. Moreover, many studies have suggested there are indeed climatic periodicities associated with the ~11 year cycle [62,64,73,86–88,90,95,96,99,126,194,259,284,289].

A common limitation of these analyses is that the longer the period of the proposed frequency being evaluated, the longer a time series is required. The datasets with high resolution typically only cover a relatively short timescale (of the order of decades to centuries), meaning that they cannot be used for evaluating multi-centennial cycles [62,99,126,194], while studies using the longer paleoclimate records tend to be focused on longer periodicities [261,262], although some studies combine the analysis of long paleoclimate records with shorter instrumental records [259]. That said, some records can be used for studying both multi-decadal and centennial timescales. For instance, Ruzmaikin et al. (2006) analysed annual records of the water level of the Nile River spanning the period 622-1470 AD. They found periodicities of ~88 years and one exceeding 200 years and noted that similar timescales were present in contemporaneous auroral records, suggesting a geomagnetic/solar link [73]. Interestingly, although they also detected the 11-year cycle, it was not as pronounced as their two multi-decadal/centennial cycles – this is consistent with the 11-year cycle being less climatically relevant than other cycles [73].

Another criticism is the debate over whether the periodicities identified in each of the datasets are genuine, or merely statistical artefacts of applying frequency analysis techniques to "stochastic" data. One problem is that even with the relatively well-defined ~11 year Schwabe cycle, the cycle is *not* strictly periodic, but quasi-periodic, i.e., the exact period for each "cycle" can vary from 8 to 14 years. Meanwhile, there are clearly non-periodic components to both climate and solar activity datasets.

Indeed, some argue that many of the apparent "periodicities" in these datasets are not actually periodic patterns, but rather arise sporadically through stochastic processes [290,291], e.g., Cameron and Schüssler (2019) argue that all "periodicities beyond 11 years are consistent with random forcing" in the various solar activity datasets [291]. Others argue that we should not be expecting strict periodicities but rather quasi-periodic patterns, and therefore we should use frequency analysis techniques that are designed to distinguish between pseudo-periodic components and genuinely periodic (or quasi-periodic) components [61,62,99]. In any case, in recent years, several groups have begun revisiting an old hypothesis that, if valid, could explain genuine multidecadal-to-centennial periodic patterns in solar activity. We will briefly review this hypothesis in Section 2.5.3.

### *2.5.3. Sun-Planetary Interactions as a plausible mechanism for long-term solar variability*

From studying the variability of sunspot cycles in the sunspot record and apparent similarities to estimates of past climate changes over the last millennium, Dicke (1978) was prompted to ask, "Is there a chronometer hidden deep in the Sun?" [292]. That is, he wondered whether the variability between solar cycles might not be "random", but rather due to various periodic but long-term processes that could lead to various periodicities in solar activity on timescales greater than the ~11 year cycle. If Dicke was right, then this would be very consistent with many of the studies described in the previous section. It would imply that many of the quasi-periodicities identified by those studies could be genuine periodicities (not necessarily linear in nature) and not just statistical artifacts as their critics argued. It would also imply that, *in principle*, it should be possible to reliably predict future solar activity as well as retrospectively determine past solar activity. Over the years, some researchers have even suggested that long-term processes internal to the Sun might be on a long-enough timescale to offer an alternative explanation to the prevailing orbital-driven ice age theory (which we will briefly discuss in Section 2.6.5) [293,294].

Dicke's hypothesis has been disputed by others who argue that the variability in solar activity between solar cycles is strictly due to stochastic processes, i.e., that there are no longer-term cyclical periodicities other than the ~11-year cycle [290,291]. However, in recent years, several groups have begun revisiting an old hypothesis to explain long-term solar variability that Wolf had originally proposed in the mid-19th century, which would prove that Dicke was correct [63,126,210,259–261,295–306]. This is the hypothesis that the gravitational effects of the planets orbiting the Sun can in some manner (various mechanisms have been proposed) interact with some of the mechanisms driving solar activity. Note that we will discuss the related, but distinct, issue of the influence that the other planets have on the Earth's orbit of the Sun [307,308] in Section 2.6.5. Here, we are referring to the possibility that the changes in the orbits of each of the planets over time might have an influence on solar activity, including TSI.

Although these Sun-Planetary Interactions (SPI) theories can initially sound more astrological than scientific in nature, many groups have noted that many of the periodicities in solar activity (and climate) records discussed in the previous section are intriguingly similar to the periodicities with which specific planetary alignments occur. Indeed, even the ~11 year cycle might potentially be related to planetary alignments such as the 11.07-year Venus/Earth/Jupiter alignment cycle [210,304,309,309] or harmonics associated with the interactions between Jupiter, Saturn and the Sun that have periodicities of about 10-12 years [127,259,310].

If any of these SPI theories transpire to be valid, then it could have important implications for our understanding of past solar variability, as well as offering us the potential to predict future solar variability [261,300–302,306]. It could also be a powerful vindication of many of the studies described in the previous section. As a result, it is not surprising that the theories have generated significant interest in recent years. However, studies considering these theories have also generated a lot of criticism [311–314], although these critiques have in turn been addressed [127,285,303,315].

Even among proponents of the theory, there is considerable ongoing debate over which combinations of orbitals are most relevant, e.g., if the 2100-2500 year "Bray-Hallstatt oscillation" is driven by SPI, which is more relevant: the 2318-year periodicity involving Jupiter, Saturn, Uranus and Neptune [260,261] or the 2139-year periodicity involving just Jupiter and Saturn [301]?

At any rate, any discussion of the theory appears to be highly controversial and often moves beyond the realm of purely scientific debate. This can be seen from some of the reactions that have occurred when articles considering the concept are published. We give two examples to illustrate the contentiousness of this theory, and how non-scientific arguments often get invoked in a discussion of this theory. We provide the following examples not because we believe the theory is beyond scientific critique (far from it), but rather to emphasise that readers who are interested in the scientific validity of the theory (or otherwise) should recognise that much of the criticism of the articles promoting SPI often moves beyond the realm of pure scientific debate.

As a first example, in 2014, a special issue dedicated to investigations into SPI theory was published in a new journal, "Pattern Recognition in Physics". In response, the managing director of Copernicus Publications terminated the entire journal for reasons that are still not entirely clear, but apparently included the facts that one of the editors of the journal was a "climate skeptic" and that the concluding article in the special issue criticised some of the interpretations and conclusions of "the IPCC project" regarding future climate change trends. Interested readers can find the managing director's full statement on his decision as well as links to an archive of the journal at https://www.pattern-recognition-in-physics.net/. One of the editors in question, the late Nils-Axel Mörner, has also responded in Mörner (2015) [316]. To clarify, we are not arguing here that the articles in that special issue should somehow have been protected from scientific critique or scrutiny. On the contrary, we are noting that the managing director's decision to terminate the journal did not seem to be based on any of the scientific evidence and arguments for SPI which were presented in the articles. Further, now that the journal has been terminated, it is likely that many of those who might have otherwise debated for or against the scientific arguments presented in any of those articles will simply dismiss the articles out of hand.

As another example, Zharkova et al. (2019) [317] was retracted (despite the objections of three of the four authors [318]) because, in one of the subsections in the paper, the authors appear to have made a mistake in their interpretation of SPI theory. Specifically, in their penultimate subsection, they appear to have mistakenly overlooked the fact that as the barycenter of the solar system moves, the Earth mostly moves in tandem with the Sun, i.e., the Earth-Sun distance does not fluctuate as much as they had assumed. This was indeed a mistake as noted by, e.g., Scafetta (2020) [260]. Also, much of the rest of the article built on earlier analysis that has been separately criticised, e.g., Usoskin (2018) [319] (although defended by the authors [320]). However, given that the mistake in question only really related to a subsection of the paper and one sentence in their conclusions, it is surprising that the reaction of the journal was to retract the article rather than encourage the authors to issue a corrigendum.

Most of the researchers currently publishing works that are considering SPI (which includes some of us) appear to be open to the fact that the field is still somewhat speculative and ongoing, and that the theory that SPI significantly influences solar activity has not yet been satisfactorily proven. In particular, most SPI researchers explicitly acknowledge that the direct vertical tides induced on the Sun by the planets are very small (millimeters), and that a more compelling mechanism by which the planetary motions could significantly influence solar activity (including TSI) needs to be established [63,126,210,259–261,295,298–305]. Nonetheless, several such mechanisms have now been proposed in the literature which seem plausible and worthy of further investigation [260,302]. For instance, perhaps the changes in the strength and spatial distribution of potential energy induced by the planetary orbits could influence solar irradiance [298,300,302]. Abreu et al. have proposed that the time-varying torque exerted by the planets on the a non-spherical tachocline could significantly influence solar activity [299,315]. Scafetta (2012) has proposed that the very modest planetary tidal effects implied by classical physics might be substantially amplified in modern physics by modulating the nuclear fusion rates in the Sun and therefore, TSI. He therefore calculates that planetary tides could theoretically induce an oscillating luminosity increases in TSI of between 0.05 and 1.63 W/m$^2$, i.e., a range consistent with the observed variations in TSI during the satellite era [309]. Meanwhile, Stefani et al. have developed a solar dynamo model in which tidal synchronisation amplifies the weak individual effects during "beat periods" [210,304,306]. Scafetta (2020) notes that the various hypotheses should still be treated speculatively especially since often the proposed mechanisms are at least partially inconsistent with each other [260]. However, often the proposed mechanisms are complementary with each other, e.g., Yndestad & Solheim (2017) proposed a hypothesis that combined features of four different mechanisms [302].

#### *2.5.4. Analogies of solar variability with the variability of other "Sun-like" stars*

Another approach that several researchers have taken to try and estimate the magnitude of past solar variability is by analogy with variability of other stars that are "Sun-like" (a somewhat loose term, as will be discussed). Stellar variability does not directly tell us about the exact timings of historic solar activity trends. However, given that the Sun is itself a star, by comparing the behaviour of other stars to what we know of the Sun, we can provide a better context for how we should expect the Sun to behave, including the range of variability in TSI we should expect to see over multi-decadal to multi-centennial time-scales. Of particular relevance for our discussion is the potential help it could provide in resolving the debate over whether the "low variability" or "high variability" TSI reconstructions (Section 2.3-2.4) are more reliable.

This field of studying "Sun-like stars" was largely pioneered by the astronomer, Olin Wilson (1909-1994), working at the Mount Wilson Observatory (CA, USA – the similarity in names was coincidental). To determine which stars are most Sun-like and properly compare the long-term variability of the Sun with other stars, it is important to systematically record measurements of a large sample of potentially "Sun-like stars" over as long a period as possible. Therefore, in 1966, he began a spectroscopic program of regularly recording the relative fluxes of two frequency bands in the stellar emissions from a sample of 91 main-sequence stars [321]. The two frequency bands were those associated with the Ca II "H" and "K" emission lines, as it was known that the ratio in the emission from these two narrow (i.e., about 1 Å) bands varies with solar magnetic activity. The program became known as "the Mount Wilson HK project" and was continued by Baliunas et al. until funding ran out in 2003 [169,265,322–324]. A later program consisted of a collaboration between Fairborn Observatory (AZ, USA) and Lowell Observatory (AZ, USA) to acquire Strömgren *b* and *y* photometry (a different estimate of stellar activity using very broad wavelength bands) of a large sample of stars to approximate their TSI variability [169,264,323,325–328].

In the context of our paper, one of the first points noted from the Mount Wilson HK project as the records for each star increased to about a decade or longer was that many stars (but not all) appear to undergo cyclical variations in the combined fluxes of the H and K lines on timescales similar to those of the Sun's sunspot cycle [265,321,325]. For some stars, the emission fluxes seemed to be mostly constant, while for others, the fluxes seemed to be undergoing a long-term increase or decrease.

Initially, to compare these stellar measurements to those of the Sun, Wilson (1978) used equivalent lunar measurements of the reflected sunlight from the Moon [321]. Others have used Ca II spectroscopic measurements from the National Solar Observatory Sacramento Peak (NM, USA) of the "Sun-as-a-star" program [329–331]. Egeland et al. (2017) has recently compared both approaches and found good agreement between them [324].

These HK measurements of "the Sun as a star" show cyclical changes that closely correspond to the rise and fall in sunspot numbers over a solar cycle [265,324,326,329–331]. Similarly, the variability in Strömgren *b* and *y* photometry also seems to capture much of the variability in TSI over a solar cycle, although surprisingly there is some controversy over whether $(b+y)/2$ is anti-correlated with TSI [332,333] or correlated [169,264,328,334,335]. The controversy appears to arise because solar observations from the Earth are from the ecliptic plane (where the amplitude of the 11-year variability in TSI is relatively low) whereas stellar observations could be from any angle [328,334–336].

At any rate, several studies have suggested that the variability of solar activity for the Sun during the satellite era has been relatively low compared with other stars [169,258,264,265,323,325,326,336–338]. This would be consistent with high solar variability reconstructions. However, other studies have argued that the low solar variability estimates are more plausible, e.g., Hall and Lockwood (2004) [327]; Judge and Saar (2007) [339]. Meanwhile, Judge et al. (2020) using an analysis of a sample of 72 Sun-like stars calculated *an upper bound* for the solar forcing since 1750 which was much larger than the IPCC's low variability estimate of solar variability, although the IPCC's estimate also fell within the bounds of their analysis [264]. As a result, their analysis is compatible with either low or high variability reconstructions.

A major challenge with using Sun-like star data to evaluate long-term solar variability is the difference in timescales, given that we have hundreds of years of sunspot records and proxies covering millennia of solar activity, while only several decades at most for our Sun-like star data.

One approach has been to compare the ranges of the multi-decadal variability in the HK and/or *b+y* measurements of the stellar data to the equivalent measurements for the Sun during recent decades. Many of these studies have suggested that the solar variability in recent decades has been relatively low compared with other Sun-like stars [169,187,257,258,264–266,323,326,328–330]. This would be consistent with high solar variability reconstructions in that it would imply that the solar variability could be greater over longer time-scales. However, other studies disagree and argue that the solar variability in recent decades overlaps quite well with the range of stellar variability for Sun-like stars [170,327,339,340]. This would be consistent with the low solar variability reconstructions.

A major reason for the conflicting conclusions seems to be due to the relatively small samples of suitable stars with large amounts of data and deciding on which stars are most "Sun-like". For instance, in an early analysis of the data, Baliunas and Jastrow (1990) identified 13 stars with relatively long records that appeared to be suitable Sun-like stars. As part of their analysis, they noted that four of these stars (~30%) were non-cycling and that these stars implied much lower activity [265]. Later studies with larger sample sizes have suggested that "non-cycling" stars only represent 10-15% of the Sun-like stars [169,170,341,342]. Nonetheless, Baliunas and Jastrow speculated that maybe these "non-cycling" stars might correspond to Sun-like stars that had entered a "Maunder Minimum"-like state. Lean et al. combined this hypothesis with measurements from the "Sun-as-a-star" program to estimate that the TSI during the Maunder Minimum had been 0.24% lower than present-day [329,330]. This result was later used for calibrating the Lean et al. (1995) TSI reconstruction of Figure 3(d) [185].

However, since then, several studies have suggested that identifying Sun-like stars in "Maunder Minimum"-like states is probably more challenging [327,339,342–344]. Hall and Lockwood (2004) [327] found that 17% of a larger sample of 57 Sun-like stars were "non-cycling", but the distribution of stellar activities was not as neatly divided as Baliunas and Jastrow's original sample. While some have argued that this is an argument in favour of the low-variability reconstructions, e.g., Schmidt et al. (2012) [256], others have noted that we still do not know whether these "non-cycling" stars were genuinely in a Maunder Minimum state, rather than being not as Sun-like as assumed [342,343]. Therefore, there is some interest [339,344] in using more nuanced methods for identifying genuinely Sun-like stars that are currently in a Maunder Minimum-like state than Baliunas and Jastrow's simple first approximation of dividing stars into "cycling" or "non-cycling".

If Sun-like star monitoring programs like the early Mount Wilson, Lowell and Fairborn Observatory programs could be expanded to include a larger sample of potential Sun-like stars (ideally a minimum of several hundred candidates), and these programs were continued for multiple decades, then it is plausible that we could identify samples of Sun-like stars *transitioning* from a cycling state to a non-cycling state (or vice versa).

In the meantime, other studies have taken different independent approaches to using the Sun-like stars data to distinguish between high and low-variability reconstructions. For instance, Zhang et al. (1994) estimated the relationship between stellar brightness (analogous to TSI) and stellar magnetic activity (analogous to sunspot/faculae activity) by comparing the HK and *b+y* measurements[266]. Importantly, they found a reasonably linear relationship. By extrapolating this relationship to zero magnetic activity, and assuming that this was similar to the Maunder Minimum, they calculated that TSI had probably increased by something between 0.2% and 0.6% since the Maunder Minimum. This would be consistent with the high variability reconstructions. Soon et al. (1994) also noted that, like solar activity, the stellar activity of cyclic stars seemed to be inversely proportional to the cycle length, and this offered another metric for comparing solar activity to that of the Sun-like stars [187].

More recently, controversy over the high solar variability in the TSI reconstruction of Shapiro et al. (2011) [255] in Figure 3(c) has led to some interesting comparisons with the Sun-like star data [254,257,258]. Judge et al. (2012) argued that the model "A" for the irradiance from the quiet Sun's photosphere used for generating the Shapiro et al. (2011) reconstruction led to certain unrealistic results, and that using a replacement "model B" reduced the variability of the reconstruction by a factor of two [257]. This would still make the reconstruction a high variability reconstruction, but obviously less high. However, they also noted that when they split the original reconstruction into a series of 15 year segments (for comparison with the various 10-20 year stellar records), the distribution of trends was actually quite consistent with that implied by the Sun-like star data [257]. This was later confirmed by Shapiro et al. (2013) [258] and Judge et al. (2020) [264], suggesting that perhaps the high variability implied by the original reconstruction was coincidentally correct. Egorova et al. (2018) [254] developed an equivalent "model B" that was able to replicate the results of Judge et al. (2012), but they noted that by varying the choice of which solar modulation potential dataset to use, they could get four different TSI reconstructions – Figure 3(e)-(h). Coincidentally, one of these ("PHI-MU16") implied a similar difference between the Maunder Minimum and present to the original Shapiro et al. (2011) reconstruction [254], suggesting a possible explanation for the apparent contradictions between the two separate analyses of Judge et al. (2012). On the other hand, Yeo et al. (2020) disputes whether any of the models of the quiet solar photosphere considered by Shapiro et al. (2011) [255], Judge et al. (2012) [257] or Egorova et al. (2018) [254] are reliable and argues for a different model which implies a rather modest difference between the Maunder Minimum and present [345]. Although, Rempel (2020) clarifies that Yeo et al.'s model does not completely rule out the high TSI changes implied by these reconstructions, but rather suggests that they would "*require substantial changes in the quiet-Sun field strength (about a 50% reduction)*" between the Maunder Minimum and present [178].

Unfortunately, carrying out multi-decadal monitoring of a large sample of Sun-like stars requires considerable effort and resources, and many of these projects have been discontinued due to lack of funding. However, some recent projects such as the Kepler space mission (2009-2013) or the Chinese ground-based Large Sky Area Multi-Object Fiber Spectroscopic Telescope (LAMOST) surveys (2012-present) have provided important additional data for the short-term variability of Sun-like stars [336–338,346–349]. The relatively short observational timespans of these projects mean that they cannot be used for studying the multi-decadal variability. However, the data can be used for comparing the short-term variability of the Sun to other stars on timescales less than a few years [336–338,346–349]. Additionally, the data can improve our understanding of the relationships between the faculae:starspot ratios which we discussed in Section 2.1-2.3. E.g., why are some stars "faculae-dominated" (like the Sun is currently) and others "spot-dominated" [171,335,347,350]?

### *2.6. The apparent paradoxes from the 11 year "Schwabe" quasi-cyclical component*

If you consider all of the TSI reconstructions among the "low variability estimates" (Figure 2), except for the Steinhilber et al. (2009) reconstruction which is based on cosmogenic isotope proxies, it could appear that the most significant feature is the short-term maximum-minimum Sunspot Cycle fluctuations which occur with a roughly-11 year period (i.e., the "Schwabe cycle"). Therefore, initially, it might be supposed that the influence of solar variability on the Earth's climate should be most obvious over the course of each Sunspot Cycle. This applies even more so if you treat the raw Sunspot Number (SSN) record as a proxy for TSI, since the SSN falls to zero during every cycle [21].

This has been a puzzle for the community since the beginning of modern research into possible sun/climate connections, since the fluctuations in global surface air temperature (for instance) over a Sunspot Cycle are relatively small at best [17,41], and often quite ambiguous [100].

Typically, the peak-to-trough variability in global surface air temperatures over a sunspot cycle is estimated empirically at about 0.1°C [17,41], although Scafetta (2009) notes that the estimates of this "11-year solar cycle signature" in the literature vary from about 0.05°C to about 0.2°C [41]. He also notes that typical climate models are unable to simulate even this modest temperature variability over a solar cycle, with some climate models predicting the solar cycle signature to be as low as 0.02-0.04°C [41]. Partly on this basis, he suggests that there are, "*…reasons to believe that traditional climate models cannot faithfully reconstruct the solar signature on climate and are significantly underestimating it*" [41].

In any case, if you assume that (a) the low variability TSI estimates are more reliable than the high-variability estimates, and (b) there is a linear relationship between TSI and global (or hemispheric) surface air temperatures, these relatively low 11-year solar cycle signature estimates would initially appear to put a very modest upper bound on the maximum contribution of solar variability to the Northern Hemisphere surface temperature trends since the $19^{th}$ century. In Section 5, we will compare and contrast the linear fits using the high and low-variability TSI estimates, i.e., we will be implicitly evaluating the first assumption. However, there is also a considerable body of literature critically evaluating the second assumption from several different avenues. Therefore, in this section, we will briefly review some of the main attempts to resolve this apparent "11-year paradox".

The apparent paradox could indicate that the Sun affects the climate by other covarying aspects of solar variability (changes in the ultraviolet (UV) component, galactic cosmic ray fluxes, etc.) rather than just TSI changes. Indeed, much of the literature over the last few decades has suggested that we should not be only looking for a *direct* linear relationship between TSI and global surface air temperatures, but rather considering the possibility of more indirect and/or subtle Sun/climate relationships. Some of the main hypotheses are summarized schematically in Figure 4.

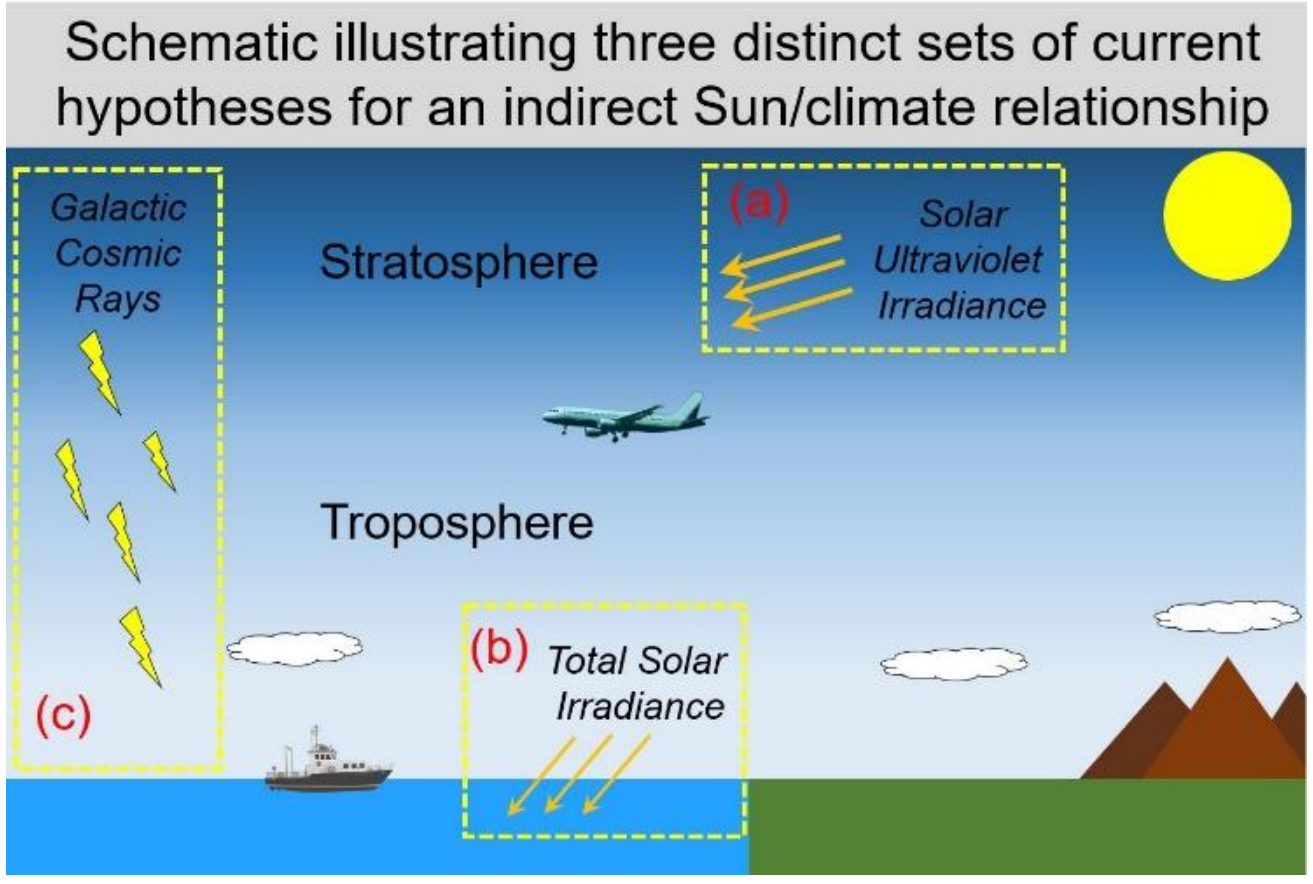

*Figure 4. Schematic illustrating the domains of proposed action for three distinct sets of current hypotheses for how the Sun* ***indirectly*** *influences the Earth's climate. Type (a) notes that there is greater variability in the ultraviolet region of the incoming solar irradiance, but that this is mostly absorbed in the stratosphere. Therefore, it is argued that the main Sun/climate relationships originate in the stratosphere but may be propagated down to the troposphere and surface ("top-down"). Type (b) suggests that there are direct effects within the troposphere from variations in Total Solar Irradiance, but that these are either subtle (e.g., through changes in circulation patterns), or involve heating the oceans which then indirectly alter the tropospheric climate ("bottom-up"). Type (c) notes that solar variability reduces the flux of incoming galactic cosmic rays (GCR) when the solar wind is strong, and that suggests that this GCR flux influences the climate in the troposphere and/or stratosphere.*

### *2.6.1. "Top-down" vs. "bottom-up" mechanisms*

As a surface-dwelling species, we are most interested in the climate at or near ground level, e.g., the surface air temperature. Moreover, most of our climate records similarly describe climate at or near the surface. However, as Dines (1919) noted in the early 20th century from analysis of early weather balloon measurements [351], the variabilities in temperatures and pressures at the surface are somewhat connected to those in the troposphere and stratosphere. Indeed, the temperature variability throughout the troposphere is partially correlated to that at the surface and boundary layer and partially anti-correlated to that in the stratosphere [352]. With that in mind, several researchers looking for Sun/climate relationships have identified potential "top-down" mechanisms whereby a relatively strong ~11-year solar cycle signature in the stratosphere might, in turn, propagate downwards to indirect influence surface climate – perhaps in a subtle and nuanced manner, that could explain the apparent "11-year paradox".

Notably, Labitzke and van Loon (1988) [65] noticed intriguing correlations between Northern Hemisphere winter temperatures (and also the geopotential heights at particular atmospheric pressure levels) and the 11-year solar cycle in the stratosphere, particularly in the polar regions. They found that these correlations were most apparent when they split the data into two halves based on whether the so-called "quasi-biennial oscillation" (QBO) wind was in its west phase or east phase. The QBO is a stratospheric circulation pattern, whereby the prevailing stratospheric winds near the equator appear to alternate from being mostly westerly to mostly easterly roughly every two years. Later work by this group extended these relationships to include the tropics, sub-tropics and both hemispheres as well as other seasons [66,67,211].

Although some of the relationships identified by Labitzke et al. also seem to be partially present within the troposphere, the relationships appear to be most pronounced for the stratosphere. In that context, several researchers noted that most of the ultraviolet (UV) component of the incoming solar irradiance is absorbed within the stratosphere, and the variability of this UV component over the ~11-year cycle seems to be much greater than for TSI [217–219,333,353]. This has led to one of the main sets of current hypotheses for an indirect Sun/climate relationship – illustrated schematically as Figure 4(a). That is, it is argued that the relationships initially identified by Labitzke et al. and built upon by others [76,213,215,222–225] are in some way driven by UV irradiance and therefore originate in the stratosphere, rather than in the troposphere [76,213,217,219,222–225].

These "top-down" mechanisms imply that the Sun/climate relationships identified in the troposphere or at the surface occur *indirectly* through coupling of the stratosphere and troposphere. From this perspective, one solution to the apparent 11-year paradox is that there are Sun/climate relationships, but they are mostly confined to the stratosphere, and, by the time the "solar signal" has reached the surface, only a modest signal remains. Indeed, climate models that attempt to incorporate these "top-down" mechanisms generally simulate a relatively small and diffused "solar signal" at surface level [217,219–221,225]. For example, Haigh and Blackburn's (2006) model simulations suggest that solar heating from increased UV irradiance took at least 50 days to heat the stratosphere but up to 500 days to reach the troposphere [219]. Some studies have found solar signals in the troposphere, but argued that they are less pronounced than in the stratosphere, i.e., consistent with the "top-down" hypothesis [213,215,216,224].

That said, other studies have *also* found evidence for a strong solar signal for temperature variability within the troposphere [27,78,89,91,354]. In particular, Soon et al. (2000) found intriguing correlations between a specific measure of solar activity (the area of the Sun covered by coronal holes) and air temperatures in the lower troposphere (as derived from satellite measurements) [27]. Their results suggested that *most* of the temperature variability within the lower troposphere (over the satellite era at least) could be explained in terms of solar variability, volcanic activity and El Niño/La Niña periods [27]. As can be seen from the schematic in Figure 4, the lower troposphere nominally includes the surface. Therefore, the results of Soon et al. (2000) might *initially* appear to contradict the apparent 11-year paradox [27]. However, we note here an additional nuance in that the satellite-based estimates of "lower troposphere" temperature trends mostly describe the temperatures *above the boundary layer*, i.e., above the first few kilometers. Ongoing work by some of us (RC, MC and WS) suggests that the temperature variability within the regions of the troposphere that are above the boundary layer is more closely related to that of the stratosphere than within the lowest parts of the troposphere closest to the surface. With that in mind, we suggest future research into possible Sun/climate relationships considering the troposphere should distinguish between the boundary layer part of the troposphere and the tropospheric region above the boundary layer (as well as separately considering the "tropopause" transition between the troposphere and stratosphere).

Meanwhile, others have argued for a more nuanced solar signal at the surface/within the troposphere, whereby solar variability directly influences the surface and tropospheric climate, but in more subtle ways that become amplified via positive feedbacks and/or changes in oceanic or atmospheric circulation patterns [214,229]. For instance, van Loon et al. have argued for solar signals that alter circulation patterns associated with, in turn: Hadley and Walker circulations [234,355]; El Niño Southern Oscillation (ENSO) [92,234,356] and the North Atlantic Oscillation (NAO) [93]. Changes in these circulation patterns themselves could alter regional and even hemispheric surface temperatures. Similarly, Ruzmaikin et al. found evidence for a solar signal in the North Annular Mode (NAM), which in turn appears to influence Northern Hemisphere surface temperatures [71,72]. Many other relationships along these lines have now been proposed [111,357–361].

Thereby, these "bottom-up" mechanisms – Figure 4(b) - offer an alternative solution to the apparent 11-year paradox, in which the 11-year cycle only has a modest direct influence on surface temperatures but also indirectly influences the climate (perhaps on multi-decadal timescales) by altering prevailing circulation patterns – especially those associated with key "centers of action" [111].

In terms of Sun/climate relationships at the surface, some researchers have argued that it is difficult to establish whether the "top-down" or "bottom-up" mechanisms are more important [78,212,214]. Others have suggested that both sets of mechanisms are important [95–97,362], with Roy et al. proposing that a complex series of interconnected mechanisms from both sets could be involved [95–97]. Meanwhile, Dima and Voiculescu (2016) suggest that both the "top-down" and "bottom-up" mechanisms might also combine with a third mechanism involving solar-driven variability in cloud cover [362]. They suggest this could be driven by one of the proposed Galactic Cosmic Ray mechanisms, which we will discuss in Section 2.6.4. In that case, their Sun/climate relationships would involve all three of the sets of mechanisms described in Figure 4.

#### *2.6.2. "The ocean as a buffer": Ocean heat capacity as a "filter capacitor"-based buffering mechanism*

Reid (1987; 1991; 2000) [69,231,232] noticed that the global Sea Surface Temperature (SST) time series was intriguingly similar to the multi-decadal trends of the Sunspot Number (SSN) record once the 11-year cycle had been removed by either smoothing both series with an 11-year running mean [231] or by using the "envelope" of the SSN record, i.e., the time series generated by connecting the maxima from each solar cycle [232]. He argued that variability in TSI was influencing ocean temperatures on multi-decadal timescales *but* not as much over the 11-year cycle due to the relatively short time frame and the fact that it was (quasi)-cyclical: "*The cyclical nature of the solar* [11-year cycle] *forcing, however, substantially reduces its climatic impact, since the thermal inertia of the ocean is large enough to dampen an 11-year cycle considerably*" - Reid (2000) [69] .

Hence, Reid's solution for the apparent "11-year paradox" described above was that the heat capacity of the oceans effectively acts as a buffer that filters out much of the short-term cyclical solar variability of the incoming TSI over the 11-year cycle but captures much of the longer multi-decadal to centennial trends and cycles in TSI. If the ocean temperatures are influenced by solar variability on these longer timescales, this could in turn influence oceanic and/or atmospheric circulation patterns, which in turn could influence land surface temperatures.

If this hypothesis is valid, it would imply we need to separately consider TSI variability on multiple timescales (dovetailing with the types of analysis in Section 2.5.2 and 2.5.3) [41]. Indeed, Dima and Lohmann (2009) have proposed that solar variability on millennial time scales could combine with variability in oceanic thermohaline circulations to create "*a possible combined solar-thermohaline circulation origin for the ~1,500-year* [climate change] *cycle*" [112]. Soon and Legates have proposed an analogous solar/thermohaline mechanism that could also operate on multidecadal-to-centennial timescales [113,233].

White et al. (1997; 1998) [226,227] attempted to do such an assessment of the influence of TSI variability on ocean temperatures using the Lean et al. (1995) TSI reconstruction of Figure 3(d). White et al. (1997) used an SST dataset covering the period 1900-1991 and an upper-ocean temperature profile dataset covering the period 1955-1994 [226]. In a second paper, White et al. (1998) repeated the analysis using a time series of the depth-weighted average temperatures (DVT) of the upper oceans for 1955-1996 [227]. Both studies found that the solar influence on ocean temperatures was different on decadal timescales (9-13 years) compared with interdecadal timescales (18-25 years) and for the longer dataset in the first paper, probably multi-decadal to centennial scales (although the dataset was only 92 years long). In principle, this is consistent with Reid's hypothesis that the solar influences on ocean temperatures are different on different timescales. Indeed, the first paper concluded that solar variability could explain between 0.2-0.3°C (i.e., 50-75%) of the 0.4°C global SST warming over the preceding century which at the time had occurred, i.e., largely agreed with Reid. However, in the second paper, they clarified that the long-term increase in TSI in recent decades implied by the Lean et al. (1995) [185] TSI reconstruction was insufficient to explain the rate of warming and argued that a greenhouse gas component was needed. We note Lean was a co-author of both White et al. papers.

More recently, Scafetta (2009) has argued empirically that the global temperature trend estimates of the last 400 years are best fit in terms of solar variability by assuming that solar forcing from changes in TSI act on both fast (less than 0.5 year) and slower, multidecadal time-scales [41]. Although that analysis was empirical in nature, and therefore did not postulate a definitive mechanism for why, this would also be consistent with the "ocean as a buffer" mechanism described above. Indeed, many energy balance climate models (EBMs) compartmentalize the oceans into two or more layers with different time-scales in each layer to explicitly model this buffering mechanism, e.g., Lindzen & Giannitsis (1998) [363]; Held et al. (2010) [364]; Ziskin & Shaviv (2012) [48]; Geoffroy et al. (2013)[365]; Rohrschneider et al. (2019) [366]. Moreover, Wang et al. (2020) suggest that the relationships between TSI and ocean heat content may vary between oceans and on different timescales [367].

The "ocean as a buffer" mechanism on its own could potentially resolve the apparent "11-year cycle paradox" and imply that investigations into a solar influence on the climate should probably prioritise looking at TSI and climate variabilities on timescales longer than the ~11 year cycle, i.e., multidecadal-to-centennial or longer. However, we note that even in terms of the ~11-year component, there is considerable debate over the magnitude of the solar influence. Within the literature, estimates of the solar-induced variability on ocean temperatures over the course of a cycle vary from 0.02-0.2°C [40,226–230,368]. Therefore, further investigation into the role of TSI on the ~11-year timescale does still seem warranted. Shaviv (2008) found evidence for a solar influence over the 11-year cycle on ocean temperatures that was 5-7 times greater than what he would have expected from the changes in TSI alone [40]. He suggested that this indicated that some form of solar amplification mechanism, such as the ones we will review in Section 2.6.4, might be involved. This has support from Solheim (2013) who noted a tight correlation between global annual-averaged sea level changes and annually-averaged sunspot numbers [51]

#### *2.6.3. Sun-climate effects are more pronounced in certain regions*

In Section 2.6.1, we showed that several studies have argued that solar variability could *indirectly* influence *regional* temperature trends, e.g., via altering atmospheric circulation patterns. However, other studies have argued for a more direct relationship between solar variability and *regional* climate trends.

Using one of the high variability TSI estimates (Hoyt and Schatten (1993) [179]), Soon found a striking correlation between TSI and Arctic surface air temperatures since at least 1875, i.e., the entire length of the then-available temperature dataset [31,113]. This suggested that most of the Arctic temperature trends since at least the 19$^{th}$ century (including the Arctic warming since the 1970s) was due to solar variability rather than anthropogenic factors. Interestingly, Callendar (1938) also argued against an anthropogenic role in Arctic temperature trends in his original case for a $CO_2$-driven global warming [369]. Soon et al. (2011) later found a similar result for China [370], while Scafetta et al. found the same for the Central England Temperature dataset from at least 1700 to the present [52,259,284]. Soon et al. (2015) noted that after accounting for urbanization bias in the Northern Hemisphere temperature data, the same TSI estimate could explain most of the long-term temperature trends since at least 1881 for the entire hemisphere (but not for urban areas) [56]. Soon and Legates (2013) also found evidence that the same TSI estimate could explain much of the trends in the so-called Equator-to-Pole Temperature Gradient [233]. In other words, the Hoyt and Schatten (1993) TSI estimate implies a very strong correlation between surface air temperatures and TSI. Indeed, this was already noted by Hoyt and Schatten (1993) [179]. One of us (JES) has noted (manuscript in preparation) this TSI estimate is also well correlated with the 440-year long series of estimated positions of the August ice edge in the Barents Sea described in Mörner et al. (2020) [63].

On the other hand, if researchers use one of the low variability TSI estimates of Figure 2, or even use the raw SSN record as a proxy of solar activity, it is much harder to find a strong relationship. Nonetheless, several researchers have argued that significant correlations can still be identified between solar activity and surface temperature and/or precipitation for certain geographical regions [43–45,49,63,64,86,87,289].

For instance, in a series of papers (independent of their more recent work discussed in Section 2.5.2), Le Mouël and colleagues argued that climatic trends in Europe [43–45,115], the United States [115,371] and possibly Australia [43] were consistent with being at least partially solar-driven. These studies were collectively disputed by Yiou et al. in two papers [372,373], although Le Mouël et al. (2011) defended their analysis [46].

Within the paradigm of the "11-year puzzle", these studies could potentially be interpreted in several distinct ways:

1. They could be case studies representative of much wider global Sun/climate relationships that might be overlooked in global analyses that smooth out subtle relationships through averaging processes, for instance. Also, in some cases, these studies are confined to certain regions simply due to the limited availability of the relevant data for other regions [43,52,56,284]. In other cases, the analysis may be carried out as a case study [86,87,188,370]. This could then potentially contradict the "11-year paradox" if the relationships were later shown to be global in nature. With that in mind, Dobrica et al. (2018) argue that the Sun/climate relationships they identified in their early case studies of Europe [86,87] can now also be extended to much of the Northern Hemisphere, and for different levels of the atmosphere from the surface to the stratosphere [89].
2. On the other hand, it might be argued that these relationships are strictly regional in nature. That is, the studies may have just identified unique geographic regions where the climatic trends have a particularly pronounced solar influence [44,45,49,63,64,86,87,289]. This could then be consistent with the "11-year paradox", which refers to global trends.
3. Another interpretation offers a compromise between the other two – perhaps these regions represent important climatic "centers of action" [111]. In that case, perhaps the solar-induced climatic variability of these regions could in turn lead to shifts in prevailing atmospheric and/or oceanic circulation patterns. Christoforou and Hameed (1997) [111] proposed that, in principle, this could offer potential mechanisms whereby a relatively small variability in TSI over the ~11-year cycle could indirectly lead to multidecadal climatic trends on regional or even global scales. Several examples of such potential mechanisms have been proposed, e.g., Soon (2009) [113]; Mörner et al. (2020) [63]. As an aside, Mörner et al. (2020) [63] proposed that it could be the solar wind and not TSI which is the main climatic driver. They propose that the solar wind interacts with the magnetosphere affecting Earth's rate of rotation (Length-of-Day or LOD [198,202]), and that this alters the Earth's centripetal acceleration and in turn could alter prevailing oceanic circulation patterns.

### *2.6.4. Galactic Cosmic Ray-driven amplification mechanisms*

In Section 2.6.1, we discussed how several researchers have argued for Sun/climate relationships that are driven by the larger variability in the UV component of the solar cycle, rather than the more modest variability over the solar cycle in Total Solar Irradiance (TSI). Since most of the incoming UV irradiance is absorbed in the stratosphere, this has led to various "top-down" mechanisms whereby the Sun/climate relationships begin in the upper atmosphere before being propagated downward, as schematically illustrated in Figure 4(a). However, other researchers have focused on a separate aspect of solar variability that also shows considerable variability over the solar cycle, i.e., changes in the numbers and types of Galactic Cosmic Rays (GCRs) entering the Earth's atmosphere. Because the variability in the GCR fluxes can be different at different altitudes, but some GCRs are absorbed in both the troposphere and the stratosphere, such mechanisms could potentially be relevant throughout the atmosphere [70,374,375] – Figure 4(c). Also, because both the flux and the variability in the incoming GCR fluxes increase with latitude (greatest at the geomagnetic poles [70,374,375]), if such mechanisms transpire to be valid, this might mean that the Sun/climate relationships are more pronounced in some regions than others (Section 2.6.3).

Although some cosmic rays come from the Sun, GCRs are believed to come from other stellar systems, especially from the explosions of nearby supernovae. However, the solar wind appears to reduce the flux of GCRs entering the Earth's atmosphere, and since the solar wind increases with solar activity, the flux of GCRs appears to be inversely proportional to solar activity. Even though the flux of GCRs is much weaker than incoming TSI, GCRs are responsible for much of the ionization that occurs in the atmosphere. Indeed, this is why changes in the ratios of cosmogenic isotopes such as $^{14}C$ or $^{10}Be$ are often used as proxies for solar activity (Section 2.3).

For this reason, Ney (1959) [374] and Dickinson (1975) [375] both hypothesised that changes in the GCR flux might actually be climatically significant through ionization processes and/or interactions with electric fields. For instance, Ney suggested that changes in the GCR flux might lead to changes in storminess (especially thunderstorms) [374]. Dickinson speculated that if GCRs were involved in a significant Sun/climate mechanism, then a plausible candidate mechanism would involve some connection between GCRs and cloud formation. He openly admitted that his hypotheses were strictly speculative and warned, "*I have so piled speculation upon speculation that much further argument does not seem profitable*". However, he hoped, *"…that this discussion has provided some guidance as to fruitful avenues for further research into physical connections between solar activity and the lower atmosphere*" [375].

22 years later, Svensmark and Friis-Christensen (1997) [25] noticed an intriguing result that appeared to have vindicated Dickinson's speculations. They noticed a striking correlation between the GCR flux and satellite estimates of global cloud cover, according to the ISCCP-C2 dataset over the then-available period, 1983-1990. Although this was a relatively short period, it captured a considerable portion of a solar cycle and implied a strong and pronounced Sun/climate mechanism that had not been considered by the climate models. The study was criticised [376,377], but also defended [378].

Kernthaler et al. (1999) reanalysed the ISCCP-C2 dataset to distinguish between "high", "medium" and "low" clouds and also split the global data into latitudinal bands. They argued that taking this more granular approach, the apparent relationship between GCRs and cloud coverage disappeared [376]. When the ISCCP dataset was updated to 1994 and upgraded to version "D2", the dataset providers included similar granular breakdowns. Independently, both Pallé Bagó and Butler (2000;2001) [379,380] and Marsh and Svensmark (2000) [381] confirmed that the original relationship had broken down but that a more nuanced relationship still remained – it appeared that there was a strong correlation between GCR flux and the percentage of low cloud cover, particularly for lower latitudes [379,381]. This appeared counterintuitive, as it had been supposed that any such effect would actually be greatest for high clouds and high latitudes. Nonetheless, the correlation was quite striking and now covered a longer period (1983-1994).

As before, this updated relationship was criticised [7,9,382–384] but also defended [385,386] and gained some support from other researchers [70,387]. However, when the ISCCP-D2 dataset was updated to 2001, Marsh and Svensmark (2003) noticed that the apparent relationship seemed to breakdown again [385]. Yet, they also noted that there was a gap in the available ISCCP calibration satellites between September 1994 and January 1995, and that if a single step calibration adjustment was applied to the data during this gap, the correlation between low cloud cover and GCRs remained for the entire updated 1983-2001 period [385]. Meanwhile, Sun and Bradley (2004) argued that the entire ISCCP D2 time series was unreliable and that ground-based time series (which appeared to contradict the GCR/cloud hypothesis) were preferable [383], while Marsh and Svensmark (2004) argued the opposite [386]. More recently, Agee et al. (2012) argued that the GCR-cloud hypothesis breaks down when the (unadjusted) ISCCP-D2 dataset was updated to 2008, as it implied unusually low cloud cover during a period of unusually high GCR flux [388]. On the other hand, Evan et al. (2007) had already argued that the unusually low cloud cover values of the ISCCP dataset were due to "*satellite viewing geometry artifacts and* […] *not related to physical changes in the atmosphere*" [389].

We sympathise with readers who find these controversies over the reliability of the global cloud cover datasets unsettling. At any rate, Kristjánsson et al. (2004) raised an important additional challenge to the theory by noting that, because GCR fluxes are quite well correlated to other metrics of solar activity, similar correlations could be found between TSI and cloud cover [384]. They also carried out spatial correlations instead of just comparing the global time series and found that some regions had stronger correlations than others [384]. Pallé et al. (2004) also found similar results [390].

In a series of papers, Voiculescu et al. have built on these ideas and carried out regional analyses on the basis that the cloud cover in different regions might be influenced by different factors, including different solar drivers [362,391–393]. Voiculescu et al. found a solar influence on the cloud cover in many regions but, in some regions and for different types of clouds, the correlations were better with changes in UV irradiance. For other regions, the correlations were better with changes in GCR flux, while for others the cloud cover seemed to be influenced by non-solar factors [391,392].

While less exciting than the original Svensmark and Friis-Christensen (1997) result [25], these more nuanced analyses where GCRs are just one of several potential drivers of changes in cloud cover are still consistent with the overall theory that changes in GCR fluxes could be a driver of global temperature changes. However, it suggests that more subtle regional effects need to be considered. It also confirms that it is challenging to separate a specific GCR-driven mechanism from other solar-driven mechanisms [70].

A potentially useful approach for trying to evaluate these more nuanced proposed GCR/cloud connections is to look for any significant cloud changes associated with Forbush decrease (FD) events. These are occasional events (typically a few each year) following a Coronal Mass Ejection (CME) when the solar wind temporarily increases for a few hours, substantially reducing the GCR flux for a few days. Although CMEs also influence other aspects of solar activity, this temporary effect on GCR flux is quite pronounced, and therefore if the GCR/cloud mechanisms are valid, we would expect that evidence for this could be identified by comparing climatic conditions during the event to those of the days immediately before and after the event.

Several studies taking this approach have reported significant climatic changes associated with FD events. For instance, analysing ground-based sunlight measurements at several UK weather stations, Harrison and Stephenson (2006) noted an average reduction in diffuse solar radiation (i.e., cloudier weather) during FD events [387]. Similarly, Svensmark et al. (2009) found that the liquid water content of low clouds could be reduced by up to 7% during FD events [394]. However, again, these studies are typically contested, e.g., Laken et al. (2009) [395] and Calogovic et al. (2010) [396] disputed Svensmark et al.'s (2009) analysis and argued that there was no statistically robust relationship between FDs and cloud cover.

Part of the challenge is that some FD events are stronger than others, and they are so sporadic that the number of strong events over a relatively short period such as the satellite era is quite limited. To overcome this limitation, Dragić et al. (2011)

used records of the "Diurnal Temperature Range" (DTR), i.e., the difference between the daily maximum and daily minimum temperatures from 189 European weather stations, as a proxy for cloud cover [397]. This allowed them to study a much longer time period than the satellite era. They found statistically significant changes in DTR for strong FDs with a GCR reduction of at least 7% [397]. However, Laken et al. (2012) [398] argued that the statistical averaging techniques used by both Dragić et al. (2011) and Svensmark et al. (2009) were inappropriate. On the other hand, after Svensmark et al. (2016) carried out a more robust statistical analysis, they concluded that "there is a real influence of FDs on clouds probably through ions" [57].

An ongoing debate over the relevance of the GCR/cloud theory has been over the exact physical mechanism by which changes in GCRs could influence cloud coverage [70,79]. This has prompted considerable laboratory work to try and replicate in a closed (indoors) system the various steps involved in cloud formation and evaluating the role of GCRs relative to other factors. This includes cloud chamber experiments carried out by Svensmark et al. in the "SKY" project [58,399,400] and an independent CERN-based group as part of the "CLOUD" project by Kirkby et al. [79,401–403]. These experiments have confirmed that ionization by GCRs does appear to increase rates of cloud nucleation under certain circumstances [58,399–401]. However, there is debate over whether there are substantial regions where cloud formation is inhibited by a shortage of GCRs. In particular, computer simulations using global aerosol models that have been calibrated using some of the CLOUD results suggest that GCRs are not a major contributor [402,403]. But, Kirkby et al. (2011) had noted that these models are still not very good at explaining the observed data [401]. Meanwhile, it has been argued that at least for Antarctica, cloud cover appears to be influenced by GCRs [404].

Although much of the discussion on potential links between GCRs and climate have focused on Svensmark et al.'s theory, other groups have argued for more subtle effects, e.g., by influencing stratospheric ozone concentrations [405] or by influencing cyclonic and anti-cyclonic activity [406]. In particular, Tinsley et al. have argued that GCRs interact with the climate by influencing the Global Electrical Circuit [81–83]. This builds on some of Ney's (1959) [374] and Dickinson's (1975) original hypotheses [375]. Carslaw et al. (2002) noted that such mechanisms could themselves influence cloud cover – making it hard to distinguish between Svensmark et al.'s specific theory and other subtler GCR/cloud/climate mechanisms [70]. Tinsley et al. have now published multiple studies suggesting potential links between GCRs and the climate through the intermediary of the Global Electric Circuit, e.g., [81–85]. Others have also provided independent analysis that is somewhat consistent with such mechanisms [70,387,393,407].

Regardless of the exact mechanism by which GCRs might influence climate, Shaviv and Veizer (2003) noted that when they compared estimates of past paleotemperatures and past GCR fluxes over the last 500 million years, they found a much better match than between paleotemperatures and estimates of past $CO_2$ [263]. This study was criticised by both Rahmstorf et al. (2004a) [408] and Royer et al. (2004a) [409], with these criticisms leading to rebuttals and counter-rebuttals from both sides [410–412]. A major challenge is that there is considerable ongoing debate over which estimates of past temperature, $CO_2$ and solar activity on these timescales are most reliable. As a result, some studies argue that Shaviv and Veizer's original analysis was broadly correct [36,413–415], while others disagree [416,417].

Clearly, the evidence for and against a significant influence of GCRs on the climate has been controversial and equivocal, with many proponents [36,114,397,415,418] and critics [20,398,403,417,419] of the theory, while others remain more neutral [362,380,393,407,420]. There is also considerable ongoing debate over what the net effects of changes in GCR fluxes would be on climate. Indeed, it is worth noting that Ney's (1959) original hypothesis implied that increased solar activity should lead to a net cooling effect [374], i.e., the opposite of Svensmark et al.'s theory [25,36,114,385]. A further complication is that the role of clouds appears to depend on whether you are considering short or long timescales, e.g., Young et al. (2019) propose that the net climatic effect of clouds can change from a negative cloud-temperature feedback to a positive cloud-temperature feedback when considering different timescales [421].

Nonetheless, it is acknowledged by proponents of the GCR/cloud/climate theory [36,114,385] that it is not as clear-cut as the intriguing Svensmark and Friis-Christensen (1997) result initially implied [25]. However, it is also acknowledged by critics of the theory that GCRs do seem to have some influence in cloud formation and that our understanding of how and why cloud cover varies is still quite limited [402,403]. It has also been conceded by critics of the theory that interest in the theory has been valuable for the community in that it has prompted more research into these challenging topics, including the SKY and CLOUD projects [402,403].

### *2.6.5. Short-term orbital effects*

#### *2.6.5.1. The difference between the average Earth-Sun distance (1AU) and the daily Earth-Sun distance*

It is also worth briefly distinguishing between changes in the solar irradiance leaving the Sun and changes in the solar irradiance reaching the Earth. Much of the interest in understanding the changes in solar activity have focused on the former. As a result, the TSI is typically described in terms of the output reaching 1 Astronomical Unit (AU), i.e., the average distance of the Earth from the Sun. This applies to most of the TSI reconstructions discussed in this paper. However, as discussed in Soon et al. (2015) [56], because the Earth's orbit of the Sun is elliptical rather than circular, the physical distance of the Earth from the Sun varies quite a bit over the course of the year. Currently, the Earth receives 6.5% more TSI (88 W $m^{-2}$) in January (i.e., the Northern Hemisphere winter) during the *perihelion* (the Earth's closest point to the Sun) than in July (i.e., the Northern Hemisphere summer) during the *aphelion* (the Earth's furthest point from the Sun). Therefore, if we are interested in the effects of changes in the solar activity *on the Earth's climate,* then arguably we are more interested in the changes in TSI actually reaching the Earth, rather than the changes in TSI reaching a distance of 1 AU.

Because the seasonal cycles of the Earth's orbit are almost identical each year, it might initially be supposed that the annual averages of the TSI reaching the Earth and that reaching 1 AU should be perfectly correlated with each other. If so, then this would mean that, when averaged over the year, this difference between the TSI reaching the Earth compared to that at 1 AU would be trivial. In that case, using 1 AU estimates for evaluating the potential effects of varying TSI on the Earth's climate would be easier, since the variations in TSI from year-to-year are easier to see in the TSI data at 1 AU compared to that at Earth's distance (as will be seen below).

This appears to be an implicit assumption within much of the literature evaluating the effects of varying TSI on the Earth's climate. However, we note that there are subtle, but often substantial, differences that can arise between the annual averages at 1 AU versus Earth's distance since most of the trends in TSI (including the ~11-year cycle) are on timescales that are **not** exact integer multiples of the calendar year.

We appreciate that mentally visualising the differences that can arise between the two different estimates is quite tricky, even if you have a high degree of visual spatial intelligence. Therefore, in order to demonstrate that there are indeed subtle but significant differences between the annual averages of TSI at 1 AU versus the TSI reaching the Earth, we compare and contrast the TIM/SORCE TSI datasets for 1 AU and Earth's distance in Figure 5. The TIM/SORCE datasets are the results from a single satellite that operated from 2003 to 2020, and it is particularly relevant because the data is reported for both 1 AU and that at the in-situ Earth's orbit. This period covers roughly 1.5 Solar Cycles – Figure 5 (b).

Before describing the results in Figure 5, we should note some technical points on this analysis. We downloaded the daily-resolved datasets for both versions from http://lasp.colorado.edu/data/sorce/tsi_data/ (accessed 26/06/2020). Unfortunately, the full SORCE TSI data records from February 25, 2003 through February 25, 2020 do not have continuous daily measured values for either the in-situ Earth's orbit or the 1 AU distance adjusted data. On average there are 11 missing days for all years except 2013, 2003 and 2020. 2013 was missing 160 daily values, and discounting 2013, the average for the full 2004-2019 period is 22.5 missing days per year. To fill in the missing daily TSI values, we applied a more sophisticated method than just a linear interpolation. For the 1 AU TSI values, we use an artificial intelligence algorithm that matched not only the amplitude of the measured daily TSI values but also the spectral properties of the measured SORCE TSI record. In addition, PMOD TSI data were used and were calibrated to SORCE's TIM daily series using the method proposed by Soon et al. (2019) [422]. The PMOD data between 2003 and 2017 were standardized with characteristics of the TIM and then added to the TIM records. The data between 2018 and 2020 that were missing in the TIM series were consecutively less than 5 days and therefore were estimated using the method of Radial Basis Function Artificial Neural Networks (RBFANN). Our RBFANN has three layers of neurons: one input (objective data in this case from TIM), one hidden and one output (matching the high- or low- frequency spectral properties). For the daily TSI values at Earth's orbit, another set of RBFANN was constructed to fill in the missing daily values. Therefore, by applying these advanced and more elegant techniques than interpolation per se, we can generate the complete daily SORCE TIM's TSI composite time series for both Earth orbit and at 1 AU perspectives.

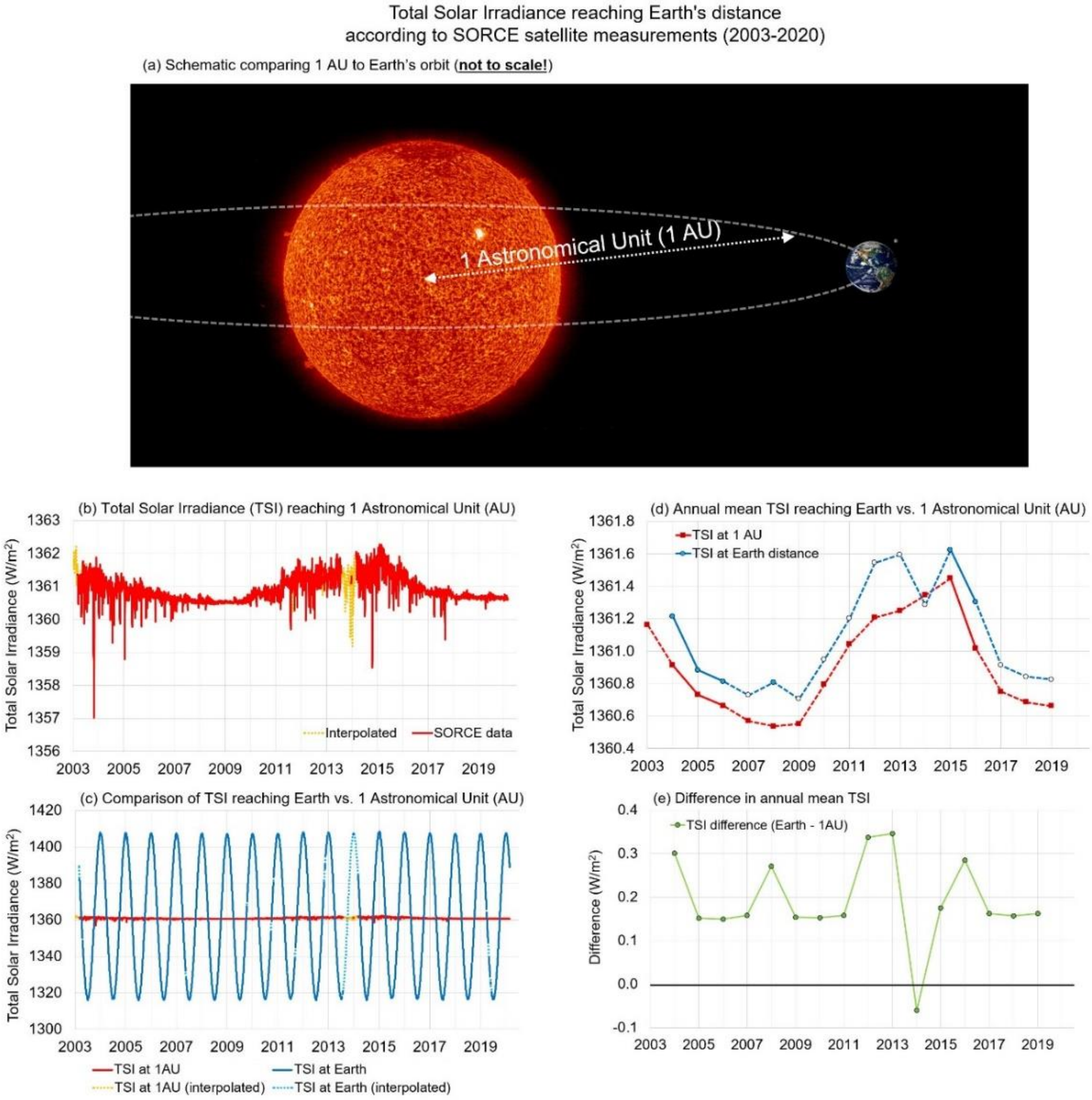

*Figure 5. Comparison of the amount of Total Solar Irradiance (TSI) that reaches the Earth as opposed to 1 Astronomical Unit (AU), i.e., the mean distance of the Earth from the Sun. (a) illustrates schematically (not to scale) how the Earth-Sun distance increases and decreases over the calendar year due to the elliptical nature of the Earth's orbit. The images of the Sun, Earth and Moon are public domain images published by NASA. The Sun image was taken by the SOHO space craft on September 24, 2008 (Credit: SOHO Consortium, EIT, ESA, NASA, https://apod.nasa.gov/apod/ap080924.html). The Earth and Moon image is a composite created by Reto Stöckli, Nazmi El Saleous, and Marit Jentoft-Nilsen, NASA GSFC (https://earthobservatory.nasa.gov/images/885/earth-from-space). (b)-(d) present the results of the TIM/SORCE sun-monitoring satellite program (2003-2020), as downloaded from http://lasp.colorado.edu/data/sorce/tsi_data/ (accessed 26/06/2020). (b) plots the daily averages at 1 AU. (c) compares the daily averages at 1 AU and at the Earth's actual distance from the Sun. (d) compares the annual averages and (e) plots the differences between the annual means. A small number of days had missing data. We interpolated these data points using the method described in the text. In (b) and (c), the interpolated points are indicated using dashed lines. In (d), those years with some interpolated data are indicated by dashed lines.*

The daily results are shown in Figures 5(b) and 5(c), with the interpolated points indicated with dashed lines (and slightly different colors). The annual means of both versions are compared in Figure 5(d). Note that the y-axes each have a different range. This is because, as noted above, the seasonal cycle in the TSI reaching the Earth is of the order of 90 W/m$^2$, while the variability in TSI over the solar cycle is only of the order of a few W/m$^2$. As a result, the pronounced ~11 year solar cycles that can be seen in the 1 AU plot of Figure 5(b) are barely noticeable when viewed on the scale of Figure 5(c).

That said, when the annual averages of both time series are calculated, this seasonal cycle is no longer an issue, and the two time series can be directly compared, as in Figure 5(d). However, as can be seen from Figure 5(e), while the two annual time series are broadly similar, albeit with the 1 AU values being slightly lower than that at Earth distance, the differences between the two time series vary slightly from year-to-year. Over the 16 year period, the differences between the two annual averages varied from +0.35 W/m$^2$ (2013) to -0.06 W/m$^2$ (2014), i.e., a range of 0.41 W/m$^2$. It could be argued that the 2014 estimate is anomalous in that this was the year with the most data interpolation. But, even neglecting that year, the differences between the two averages varied between 0.34 and 0.15 W/m$^2$, i.e., a range of nearly 0.2 W/m$^2$. For comparison, the difference between the maximum and minimum annual TSI at 1AU over the same period was 0.9 $W/m^2$. So, while small as a percentage of the total TSI, these subtle differences are not insignificant.

These differences between the annual averages for each year might initially be surprising. For the analysis in Figure 5, we are assuming that the SORCE datasets at both 1AU and the Earth's distance are reliable. We also are assuming that the interpolations we have carried out (described earlier) are reasonable. However, even if either of those assumptions are problematic, we should stress that the fact that there are differences between the annual means for both versions that vary from year to year is actually to be expected on statistical grounds. The general principle can be understood once we recognise (a) the elliptical shape of the Earth's orbit and (b) that the ~11-year solar cycle does not fall exactly on the calendar year. This means that the times of the year during which a given rise or fall in TSI occurs can make a difference. For instance, if the maximum of a solar cycle occurred during January, then this will lead to a greater mean TSI for the year than if it had occurred during July of the same year. This is because the Earth is currently closer to the Sun in January than in July.

To clarify, if the trends in TSI over a given calendar year are reasonably linear statistically, i.e., it has a constant slope, then this seasonality should not make much difference to the annual mean TSI. This is regardless of the slope itself, i.e., whether the trend is rising, falling or near-zero. However, if the trends for that year are non-constant, then the annual mean may be slightly higher or lower depending on whether the Earth is closer to perihelion or aphelion when the changes in the trends in TSI occur.

More generally, the annual average TSI reaching the Earth depends not just on the changes in TSI, but the times of the year over which those changes occur. The SORCE data in Figure 5 only covers roughly 1.5 Solar Cycles, but in principle, the same could apply to any other multi-decadal trends which might be occurring in addition to the ~11-year cycle.

Since all of the TSI reconstructions discussed in Sections 2.2-2.5 are calculated in terms of the annual averages at 1 AU, for this paper we will limit our analysis to this. However, we encourage researchers who have until now limited their analysis of TSI variability to that at 1 AU to consider this extra complication in future research.

As can be seen from Figure 5(e), the changing differences between the annual average TSI at 1 AU versus that reaching the Earth are subtle, but non-trivial. In addition to this complication in terms of the annual averages, a comprehensive analysis of the effects of TSI on the Earth's climate should consider the seasonal changes in the different latitudinal distributions of the incoming TSI, due to the seasonal orbital of the Earth. There are several different aspects to this, but for simplicity they are collectively referred to as "orbital forcings".

### *2.6.5.2. Comparison with long-term orbital forcing*

The theory that changes in atmospheric $CO_2$ are a primary driver of climate change was originally developed by Arrhenius in the late 19th century as a proposed explanation for the transitions between glacial and interglacial periods during the ice ages [423]. The existence of these dramatic climatic changes on multi-millennial timescales was only established in the 19th century and was one of the great scientific puzzles of the time. [As an aside we note that glaciologically speaking, an "ice age" is usually defined as a period where large permanent ice sheets are present in both hemispheres. These ice sheets can substantially expand during "glacial periods" and retreat during "interglacial periods". As Greenland and Antarctica both currently have large ice sheets, we are currently in an interglacial period (the "Holocene") within an ice age, even though colloquially, the term "ice age" is popularly used just to describe the "glacial periods".]

This $CO_2$ driven explanation for the glacial/interglacial transitions was later criticised by, e.g., Ångström (1901) [424] and Simpson (1929) [425]. However, it was later revived by Callendar (1938) who extended the theory to suggest that anthropogenic $CO_2$ emissions were also the primary driver of the warming from the late-19th century to mid-1930s [369], and Plass (1956) who speculatively proposed (anticipating that this would encourage scientific debate) that atmospheric $CO_2$ was the primary driver of climate change on most timescales [426].

A competing hypothesis that several 19th-century researchers proposed, e.g., Adhémar, and later Croll [308,427–429], was that long-term cyclical changes in the Earth's orbit around the Sun were the driver of the glacial/interglacial transitions. In the early 20th century, Milankovitch carried out an extensive series of calculations that demonstrated that there are several important cyclical variations in the Earth's orbit that vary over tens of thousands of years and that these influence the incoming solar radiation at different latitudes for each of the seasons [430,431].

In the 1970s, relatively high precision estimates of the timings of the glacial/interglacial transitions from ocean sediment cores appeared to vindicate the Milankovitch ice age theory [427,432]. That is, when frequency analyses like those in Sections 2.5.2 and 2.5.3 were carried out on the deep-sea sediment cores, they suggested that the past climate changes of the last few hundred thousand years were dominated by periodicities of ~90,000-120,000 years and to a lesser extent 40,000-42,000 years, and also had peaks at 22,000-24,000 years and 18,000-20,000 years. These peaks were approximately similar to the main astronomical cycles calculated by Milankovitch: 41,000 years; 23,000 years; 19,000 years and to a lesser extent ~100,000 years. As a result, analogous to the arguments described in Sections 2.5.2-2.5.3, it was argued that the glacial/interglacial transitions were indeed driven by orbital forcings [427,432]. Later analysis in terms of ice core measurements also appeared to confirm this theory [433–435].

This appears to have convinced most of the scientific community that the Milankovitch orbital-driven explanation for the glacial/interglacial transitions is correct, and this currently seems to be the dominant paradigm within the literature [430,431,435–440] including the IPCC reports [1]. Ironically, this means that the original $CO_2$-driven theory for the glacial/interglacial transitions as proposed by Arrhenius (1896) [423]; Callendar (1938) [369]; Plass (1956) [426] has been largely discarded even though the current theory that recent climate change has been largely driven by changes in $CO_2$ was developed from that early theory. That said, we emphasise that this is not necessarily a contradiction in that several researchers argue that changes in atmospheric $CO_2$ concentrations caused by orbitally-driven warming or cooling might act as a positive feedback mechanism [1,430,435–440].

We note that there are possible problems with the theory that the Milankovitch orbitals are the primary driver of the glacial/interglacial transitions. To clarify, the Milankovitch orbital variations are clearly climatically significant as discussed above. Also, the idea that some combination of these variations could provide the explanation for the glacial/interglacial transitions seems plausible and intuitive. Indeed, the approximate similarity in the timings of both phenomena is intriguing. However, as will be discussed below, there is still considerable debate over the exact causal mechanisms and over which specific aspects of the Milankovitch orbital variations would drive such dramatic long-term climate changes and why. That said, each of these problems has been countered, and the current consensus among the scientific community is indeed that the glacial/interglacial transitions are driven by orbital forcings.

With that in mind, we will briefly touch on some of the concerns that have been raised over the Milankovitch theory of ice ages and also the corresponding defences that have been made. As for the SPI theories, many of the main concerns with the theory have been highlighted by proponents of the theory. For example, Hays et al. (1976) noted that the dominance of the ~100,000 year periodicities in the climate records was unexpected since the relevant Milankovitch ~100,000 year orbital cycles (associated with changes in eccentricity) had not been expected to have much significance for climate change, and Milankovitch and others had assumed the ~41,000 year obliquity cycle would have been more climatically significant in terms of glacial/interglacials [432].

Imbrie (1982) noted another problem: because there are so many different aspects of the timings of the various orbitals in terms of different latitudes and seasons, there is a danger that researchers could cherry-pick the metric that best fit the data for their hypothesis:

"*There has also been a tendency for investigators to believe they could model the response of the system from a radiation curve representing the input at a single latitude and season […] Since no one could be sure which insolation curve, if any, was the crucial one, investigators had great freedom to choose a curve that resembled a particular set of data. Understandably the resulting ambiguity did much to undermine confidence in the validity of the time domain prediction.*" – Imbrie (1982), p413 [427]

An additional puzzle is that, while the dominant peak on orbital timescales for the last ~1 million years appears to be ~100,000 years, for most of the Pleistocene Ice Age until then, the dominant peak appears to have been the (expected) 41,000 years. This has been dubbed the "mid-Pleistocene transition", and there has been much debate over its explanation [437,441–445].

Another potential concern is the so-called "Devils Hole" record. Winograd et al. (1992) obtained a continuous 500,000 year climate record from a core in Devils Hole, NV (USA), which matched quite well with both the Antarctic ice cores and various ocean sediment cores [446]. However, the Devils Hole record had a key difference from the other estimates – it implied that the previous interglacial period ("Termination II") began several millennia before Milankovitch theory predicted it should have [446–448]. Therefore, if this estimate transpires to be accurate, then it creates a significant problem for the Milankovitch ice age theory. That said, proponents of the Milankovitch explanation have criticised the reliability of the earlier Termination II timing implied by the Devils Hole record [449–453]. However, rebuttals have been offered [454–458].

If the glacial/interglacial transitions are *not* primarily-driven by the Milankovitch orbital cycles, then it raises the question as to why the glacial/interglacial transitions occur. Several different explanations have been proposed [293,294,459–466]. However we note that all of these cited alternative hypotheses are openly speculative, and that several of these studies (and others) argue that Milankovitch orbital cycles are at least partially involved [442,443,445,459,460,465].

At any rate, regardless of the role the Milankovitch orbital cycles play in the glacial/interglacial transitions, we emphasise that the changes in the latitudinal and seasonal variability of the incoming TSI over these cycles are clearly climatically important. Indeed, while much of the research until now has focussed on the gradual variations over millennial timescales or longer, e.g., Huybers and Denton (2008) [467]; Davis and Brewer (2009) [468]; Berger et al. (2010) [469], in recent years, some groups have begun emphasizing the significance of "short-term orbital forcing" (STOF) [307], i.e., secular drifts on multi-decadal to centennial timescales in the average daily insolation at different latitudinal bands for different seasons [307,308,431,470–473].

A challenge that arises when considering the Earth's orbital variability on these shorter timescales, is that it is no longer sufficient to consider the Earth/Sun relationship in isolation. The perturbations of the Earth's orbit around the Sun by the Moon and by the other planets also need to be considered. For this reason, there is some overlap between much of the ongoing research into short-term orbital forcing [307,470–472,472,473] and the calculations within the Sun-Planetary Interactions (SPI) theories described in Section 2.5.3. However, we stress that the two fields of research are distinct. The former field is interested in how the perturbations of the other planetary bodies on the Earth's orbit of the Sun influence the distribution of the incoming TSI reaching the Earth. The latter field is concerned with what effects (if any) the orbits of the various planetary bodies might have on TSI (or more broadly, solar activity) itself. Scafetta et al. (2020) adds a third possible planetary mechanism that could be influencing the Earth's climate, by arguing that planetary configurations could directly modulate the flux of the interplanetary dust reaching the Earth's atmosphere thereby potentially influencing cloud cover [305].

In any case, for simplicity, most discussions on the climatic significance of orbital variations have tended to pick a particular latitudinal band and season, e.g., the average insolation during the Northern Hemisphere summer at 65°N. However, as discussed in the quote above from Imbrie (1982) [427], identifying the particular latitude and season whose curve is "most important" is quite subjective. For instance, Huybers and Denton (2008) argue that "the duration of Southern Hemisphere summer is more likely to control Antarctic climate than the intensity of Northern Hemisphere summer with which it (often misleadingly) covaries" [467]. Indeed, Davis and Brewer (2009) argued that a more climatically relevant factor is the "latitudinal insolation gradient (LIG)", which in turn "… creates the Earth's latitudinal temperature gradient (LTG) that drives the atmospheric and ocean circulation" [468]. Note that this latter LTG concept is equivalent to the "Equator-to-Pole Temperature Gradient (EPTG)" parameter considered by Lindzen (1994) [474] and Soon and Legates (2013)[233] and comparable to the "meridional heat transfer (MHT)" concept of Fedorov (2019) [472]. Another related index is the zonal index pressure gradient, which Mazzarella and Scafetta (2018) argues is linked to the North Atlantic Oscillation, Length-of-day (LOD) and atmospheric temperatures, as well as directly influencing atmospheric circulation patterns [475].

Although Davis and Brewer (2009)'s focus was on timescales of millennia [468], Soon (2009) noted that the variability of the LIG can also be significant on multidecadal-to-centennial timescales [113]. Therefore, recent work into studying the variability of the LIG (or Fedorov (2019)'s related "meridional insolation gradient (MIG)" [472]) on these shorter timescales [470–472,476] could be a particularly important "short-term orbital forcing" for understanding recent climate changes.

## 3. Estimating Northern Hemisphere surface temperature changes

As seen from Table 2 there are many different approaches to estimating the Northern Hemisphere surface temperature changes since the 19th century (or earlier). As will be discussed in this section, most of the estimates in this table share several key similarities. However, there are also subtle differences between the various estimates. For this reason, we analyze the different categories of Northern Hemisphere temperature variability estimates separately. However, although there are subtle differences within a given category (e.g., land-based estimates using both urban and rural stations) – particularly on a year-to-year basis – there are several dozen time series listed in Table 2. Hence, to simplify our analysis, we will construct an upper and lower bound time series for each of our five categories. This then provides us with an uncertainty range for each of our estimates. In this section, we describe each of the five categories in Table 2, and how our upper and lower bounds are calculated.

*Table 2. The different Northern Hemisphere temperature trend datasets analyzed in this study.*

| Region | Type of measurements | Analysis | Start | End | Source |
|---|---|---|---|---|---|
| Land | Rural stations only | SCC2015 weighting | 1881 | 2018 | This study |
| | | Standard weighting | 1841 | 2018 | " |
| | | Equal weighting | 1841 | 2018 | " |
| | | Excluding Arctic | 1841 | 2018 | " |
| | | Excluding U.S. | 1841 | 2018 | " |
| | | Excluding China | 1869 | 2018 | " |
| | | Excluding Ireland | 1841 | 2018 | " |
| Land | Urban and rural stations | Standard weighting | 1800 | 2018 | This study |
| | | CRUTEM4 | 1850 | 2018 | http://www.cru.uea.ac.uk/ |
| | | NOAA NCEI | 1880 | 2018 | https://www.ncdc.noaa.gov/ |
| | | NASA GISS | 1880 | 2018 | https://climexp.knmi.nl |
| | | Berkeley Earth | 1815 | 2018 | " |
| | | Chinese Meteorological Agency | 1900 | 2018 | " |
| | | Cowtan & Way (land mask) | 1850 | 2018 | " |
| Oceans | Sea Surface Temperatures | HadISST | 1870 | 2018 | https://climexp.knmi.nl |
| | | HadSST3 - 100 realisations | 1850 | 2018 | " |
| | | HadSST4 - 100 realisations | 1850 | 2018 | " |
| | | ERSST3 | 1880 | 2011 | https://www.ncdc.noaa.gov/ |
| | | ERSST4 | 1854 | 2018 | " |
| | | ERSST5 | 1854 | 2018 | " |
| Land | Tree-ring based reconstruction | Briffa (2000) | 831 | 1992 | https://www.blogs.uni-mainz.de/fb09climatology/files/2018/06/NH-reconstructions.xlsx |
| | | Esper et al. (2002) | 831 | 1992 | " |
| | | D'Arrigo et al. (2006) | 831 | 1995 | " |
| | | Schneider et al. (2015) | 831 | 2002 | " |
| | | Stoffel et al. (2015) | 831 | 2002 | " |
| | | Wilson et al. (2016) | 831 | 2002 | " |
| Land | Glacier-length based reconstruction | Leclercq & Oerlemans (2012) | 1601 | 2000 | https://www.ncdc.noaa.gov/paleo-search/study/13544 |

### *3.1. Using rural-stations only*

Many have used weather station records to estimate Northern Hemisphere (and global) land surface air temperature trends since at least the late-19th/early-20th century [56,477–484]. The results are typically similar and imply an average warming trend of ~1°C/century since the late-19th century. Thus, most studies assumed these estimates are robust and well-replicated. Recently, however, Soon et al. (2015) [56] constructed a new estimate using only rural (or mostly rural) stations taken from four regions with a high density of rural stations, and they found that this new estimate yielded noticeably different results.

Soon et al.'s (2015)[56] estimate incorporated more than 70% of the available rural stations with records covering both the early and late 20th century. "Rural" was defined in terms of both low night-brightness and low associated population using the urban ratings provided by the dataset. If they are correct that urbanization bias has not been adequately corrected for in the standard estimates, then the Soon et al. (2015) [56] estimate should be more representative of the Northern Hemisphere land surface temperature trends than the standard estimates. McKitrick and Nierenberg (2010) [485], Scafetta and Ouyang (2019) [118] and more recently Scafetta (2021) [486] offers some support for this.

On the other hand, several researchers have claimed that the standard estimates are **not** majorly affected by urbanization bias. Some argue that urbanization bias is only a small problem for global and hemispheric temperature trends, e.g., Jones et al. (1990) [487], Parker (2006) [488], Wickham et al. (2013) [489]. Others concede that urbanization bias is a concern for the raw station data, but argue that after statistical homogenization techniques (usually automated) have been applied to the data, most of the non-climatic biases (including urbanization bias) are removed or substantially reduced, e.g., Peterson et al. (1999) [490], Menne and Williams (2009) [491], Hausfather et al. (2013) [492], Li and Yang (2019) [493], Li et al. (2020) [494]. If either of these two groups is correct, then using all available station records (urban and rural) should be more representative of the Northern Hemisphere land surface temperature trends.

That said, we note that the methodology and/or justification of each of the above studies (and others reaching similar conclusions) has been disputed, e.g., see Connolly & Connolly (2014) for a critique of 13 such studies [495]. We also recommend reading the review comments of McKitrick who was a reviewer of an earlier version of Wickham et al. (2013) [489] and has made his reviews publicly available at: https://www.rossmckitrick.com/temperature-data-quality.html. Also, some of us have argued that these statistical homogenization techniques inadvertently lead to "urban blending" whereby some of the urbanization bias of the urban stations is aliased onto the trends of the rural stations. This means that, counterintuitively, the homogenized station records (whether rural or urban) may often end up less representative of climatic trends than the raw rural station records [116,117].

As in Soon et al. (2015) [56], our analysis in this section is based on version 3 of NOAA's Global Historical Climatology Network (GHCN) monthly temperature dataset [480]. This GHCN dataset was also used by the Hansen et al. (2010) [478] and Lawrimore et al. (2011) [480] estimates, and it became a major component of the Jones et al. (2012) [482], Xu et al. (2018) [484] and Muller et al. (2014) [481] estimates, which will be considered in Section 3.2. However, all these estimates incorporated urban as well as rural stations.

Version 4 of the GHCN monthly temperature dataset [496] has been released since Soon et al. (2015) [56]. This newer version provides a much larger number of stations (~20,000 versus 7,200 in version 3), although most of the new stations have short station records (only a few decades) and ~40% of the stations cover only the contiguous United States. However, unlike version 3, this new dataset does not yet include any estimates of how urbanized each of the stations are. Some of us have begun developing new estimates of the urbanization of the version 4 stations on a regional basis – see Soon et al. (2018) [116] for the results for China. This research is ongoing and has not yet been completed. In the meantime, since it has been claimed that the changes in long-term trends introduced by the switch to version 4 are relatively modest [496], we believe it is useful to update and consider in more detail the Soon et al. (2015) [56] "mostly rural" Northern Hemisphere temperature time series using Version 3.

We downloaded a recent update of the Version 3 dataset from https://www.ncdc.noaa.gov/ghcnm/ [version 3.3.0.20190821, accessed 21/08/2019]. This dataset has two variants – an unadjusted dataset that has only quality control corrections applied and an adjusted dataset that has been homogenized using the automated Menne & Williams (2009) [491] algorithm. A major component of the GHCN dataset is a subset for the contiguous United States called the US Historical Climatology Network (USHCN). As part of our analysis, we use USHCN stations that have been corrected for Time-of-Observation Biases [497] but have not undergone the additional Menne & Williams (2009) [491] homogenization process. We downloaded this intermediate USHCN dataset from https://www.ncdc.noaa.gov/ushcn/introduction [version 2.5.5.20190912, accessed 12/09/2019].

In this study, we are considering the annual average temperature (as opposed to monthly or seasonal averages), and so we introduce the requirement of 12 complete months of data for a given year. Also, to define how the annual average temperatures vary for a given station, we adopt the popular approach of converting each temperature record into an "anomaly time series" relative to a constant baseline period of 1961-1990 (which is the 30 year baseline period with maximum station coverage). After our final hemispheric series have been constructed, they are rescaled from the 1961-1990 baseline period to a 1901-2000 baseline period, i.e., relative to the 20$^{th}$-century average. We require a station to have a minimum of at least 15 complete years of data during this 1961-1990 time period to be incorporated into our analysis. These restrictions, although relatively modest, reduce the total number of stations in the dataset from 7,280 to 4,822.

Version 3 of the GHCN dataset includes two different estimates of how urbanized each station is. Each station is assigned a flag with one of three possible values ("Rural", "Semi-urban" or "Urban") depending on the approximate population associated with the station location. The stations are also assigned a second flag (again with one of three possible values) according to the brightness of the average night-light intensity associated with the station location – see Peterson et al. (1999) [490] for details.

In isolation, both flags are rather crude and somewhat out-dated estimates of the station's degree of urbanization. Nonetheless, stations categorized as "rural" according to *both* flags are usually relatively rural, while those categorized as "urban" according to *both* flags are usually highly urbanized.

Therefore, *by using both flags together*, it is possible to identify most of the highly rural and highly urban stations. As mentioned above, some of us are currently developing more sophisticated and more up-to-date estimates of urbanization for version 4 of the GHCN dataset and the related ISTI dataset, e.g., Soon et al. (2018) [116]. However, for this study (which uses GHCN version 3), we follow the approach proposed in Soon et al. (2015) [56] and divide the stations into three categories:

- "Rural" = rural according to both flags (1278 of the 4822 stations, i.e., 27%)
- "Urban" = urban according to both flags (1129 of the 4822 stations, i.e., 23%)
- "Semi-urban" = the remainder of the stations (2415 of the 4822 stations, i.e., 50%).

In this paper, we are studying the Northern Hemisphere, which has much greater data coverage. However, for context, in Figure 6, we also compare the data availability for both hemispheres. The total number of stations (in either hemisphere) available for each year is plotted in Figure 6(a). The total number of stations that meet the "Rural" requirements are plotted in Figure 6(b). It can be seen that both plots reach a maximum during the 1961-1990 period, but the available data are fewer outside this period, and much lower for the early 20$^{th}$ century and earlier. This has already been noted before, e.g., Lawrimore et al. (2011) [480]. However, as Soon et al. (2015) [56] pointed out, the problem is exacerbated when we consider the rural subset, i.e., Figure 6(b). While 27% of the GHCN stations with at least 15 complete years in the optimum 1961-1990 baseline period are rural, most of these rural stations tend to have relatively short and/or incomplete station records.

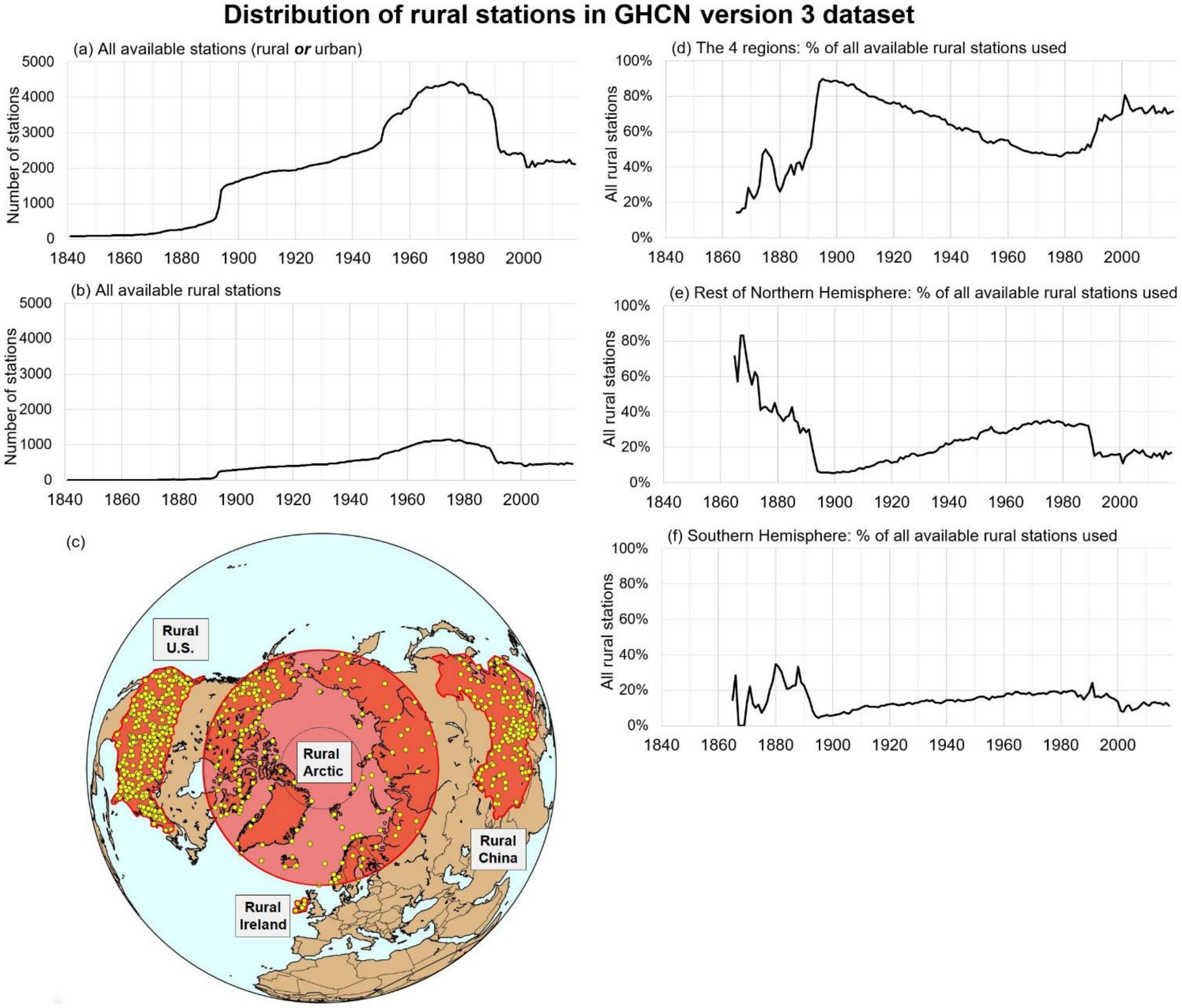

*Figure 6. Distribution of rural stations in the Global Historical Climatology Network (GHCN version 3) dataset used for constructing our rural Northern Hemisphere temperature series. (a) The total number of valid stations (either rural or urban) with data for each year. Valid stations had at least 15 years of data during the 1961-1990 period of maximum station coverage, where a year consists of 12 complete months of data. (b) The total number of valid stations that are considered "rural" in terms of both night-brightness and population density according to the GHCN metadata. (c) The locations of the four regions and the rural stations in those regions used for constructing our rural Northern Hemisphere temperature series. (d) The percentages of the available rural data in those four regions for each year. (e) The percentages of the available rural data in the rest of the Northern Hemisphere. (f) The percentages of the available rural data in the Southern Hemisphere.*

These findings are not surprising since it is more challenging to staff and maintain a long and continuous multidecadal record for an isolated rural location than for a well-populated urban location [56,116,498–500]. This was especially the case before the relatively recent invention of automated weather stations. As a result, the percentage of available stations that are rural is much lower for the earlier periods. For instance, only 454 of the 2163 stations (21%) with data for 1931 and only 300 of the 1665 stations (18%) with data for 1901 are rural, according to our categorisation.

A further difficulty is that many of the most rural stations with relatively long station records do not have complete records. The station records might nominally cover a relatively long period, but there will often be large gaps of several years or even decades, and the average annual temperature of the station will often be quite different before and after the gap [495,498,501]. For instance, using the same categorisation as here, there are 8 rural stations in Version 3 of the GHCN dataset for India and all of them cover a relatively long period (i.e., five have at least some data for the late 19th century, and two of the station records begin in the first decade of the 20th century) [501]. However, most of the records contain substantial data gaps – often accompanied by substantial "jumps" in the average annual temperature that suggest a station move or some other non-climatic bias. None of the records is complete enough to *continuously* describe temperature trends from the late 19th century to present.

If station histories (often called "station metadata") that indicate documented changes in station location, instrumentation, time of observation, etc., exist then, in many cases, it may be possible to correct the station records for these non-climatic biases – especially if parallel measurements associated with the station change are available (as is becoming more common in recent years). However, unfortunately, station histories are not currently provided with the GHCN dataset.

As mentioned earlier, several groups have argued that by applying statistically-based "homogenization" techniques to the dataset, the homogenization algorithms will accurately detect and correct for the main biases [490–494]. However, when applied to rural stations using urban neighbors, these techniques are prone to "urban blending", which has a tendency to "alias" urbanization bias onto the rural station records. That is, the homogenization process can contaminate the station records of nominally "rural" stations even if they had been unaffected by urbanization bias *before homogenization* [56,116,117,495,498,501].

For those reasons, Soon et al. (2015) [56] constructed their estimate, using the *non-homogenized* dataset, but only rural (or mostly rural) stations taken from four regions with a high density of rural stations and/or where they had relevant station history information. Here, we will adopt this approach, but updating and slightly modifying the four regions as follows:

- Rural Arctic. The gridded mean average of all 110 rural GHCN stations north of 60°N. Soon et al. (2015) [56] limited their analysis to the Arctic Circle (i.e., north of 66.7°N), which included fewer stations, and, therefore, they also included some urban and semi-urban stations. However, following Connolly et al. (2017) [502], we include only rural stations and have expanded the "Arctic" region to include any stations north of 60°N.
- Rural Ireland. Soon et al. (2015) [56] demonstrated that the adjustments applied by NOAA to the longest rural record for Ireland, Valentia Observatory, were remarkably inconsistent and also failed to identify the -0.3°C cooling bias introduced by the station's February 2001 move. For this reason, we use the unadjusted rural records for Ireland and manually apply a correction to account for this non-climatic bias. The Soon et al. (2015) [56] composite used this manually homogenized Valentia Observatory record as representative of rural Ireland. However, they noted that the other four rural Irish stations in the GHCN dataset also implied similar trends over their period of overlap. Therefore, in this study, we use the gridded mean of all five rural Irish stations.
- Rural United States. Our rural U.S. series is the same as in Soon et al. (2015) [56], except updated to 2018.
- Rural China. Our Chinese series is also the same as in Soon et al. (2015) [56], except updated to 2018. As there are very few rural stations in China with records covering more than ~70 years, this component of our analysis does include some stations that are currently urban or semi-urban, but whose station records have been explicitly adjusted to correct for urbanization bias as described in Soon et al. (2015) [56].

The locations of the stations in the four regions are shown in Figure 6(c). The regions are distributed at different locations in the Northern Hemisphere and include sub-tropical regions (lower United States and China) as well as polar regions (the Arctic). However, none of the regions are in the Southern Hemisphere; thus, this is purely a Northern Hemisphere estimate.

If accurate estimates of the urbanization bias associated with individual station records could be determined and corrected for (as for the Chinese subset), then in principle, this analysis could be expanded to include more of the available GHCN data. This could be particularly promising for some of the longer station records, especially if they have only been modestly affected by urbanization bias. For instance, Coughlin and Butler (1998) estimated that the total urbanization bias at the long (1796-present) Armagh Observatory station in Northern Ireland (UK) was probably still less than 0.2°C by 1996 [503]. Similarly, Moberg and Bergström (1997) were able to develop urbanization bias corrections for two long records in Sweden (Uppsala and Stockholm) [504]. Given the variability in urbanization biases between stations, we suggest that attempts to correct records for urbanization bias may require case-by-case assessments along these lines, rather than the automated statistical homogenization techniques currently favoured. We encourage more research into expanding the available data in this way, and some of us are currently working on doing so for some regions with relatively high densities of stations. However, in the meantime, we suggest that the approach described in this paper of only using rural data is an important first step in overcoming the urbanization bias problem.

As can be seen from Figure 6(d)-(f), the four regions alone account for more than 80% of the rural data for the early 20th century from either hemisphere. Therefore, this estimate is more likely to be representative of *rural* Northern Hemisphere land surface temperature trends than the standard estimates described in Section 3.2 since, in those estimates, most of the additional data comes from stations that are more urbanized. Nonetheless, because the stations are confined to four regions rather than distributed evenly throughout the hemisphere, it is unclear what is the most suitable method for weighting the data from each station. In Soon et al. (2015), weighting was carried out in two stages. For each region, the stations were assigned to 5°×5° grid boxes according to latitude and longitude. For each year where data for at least one station exist for a grid box, the temperature anomaly for that grid box was the mean of all the temperature anomalies of the stations in that grid box. However, the surface area of a grid box decreases with latitude according to the cosine of the latitude. Therefore, when calculating the regional average for a given year, the grid boxes with data were weighted by the cosine of the mid-latitude for that box.

Soon et al. (2015) [56] argued that each of the four regions sampled a different part of the Northern Hemisphere and therefore the average of all four regions was more representative of hemispheric trends than any of the individual regional estimates. However, since the samples covered different regions of the Northern Hemisphere, they also weighted the four regions according to the cosine of the mid-latitude of the region. They also confined their time series to the period from 1881 onward, i.e., when all four regions had data. The updated version of this weighting approach is shown in Figure 7(a).

Other approaches for weighting the data could be used instead. One approach would be to give all four regions equal weighting. The results from this "equal weighting scheme" are shown in Figure 7(c), and we extend this series back to the earliest year in which we have data, i.e., 1841. Another approach is perhaps more conventional – instead of calculating each regional estimate and averaging them together, the hemispheric averages are calculated directly from all available gridded averages (from all four regions) for each year. This is the "standard weighting scheme" shown in Figure 7(b).

It might be argued that one of our four regions could be unusual for some reason and, thereby, might be unrepresentative of the hemispheric trends. For that reason, we also calculate four additional versions of the "standard weighting scheme" in which we only use three of the four regions. These four estimates are shown in Figure 7(d)-(g).

Slight differences exist between each of the estimates. Therefore, for our analysis in this paper, we use the mean as well as the upper and lower bounds of all seven estimates. These bounds are calculated as the mean of the seven series for each year ±2σ. Nonetheless, all seven estimates are broadly similar to each other, implying that the effects of different weighting schemes are relatively minor. Specifically, all estimates suggest that the Northern Hemisphere warmed from the late-19$^{th}$ century to the mid-1940s; cooled to the mid-1970s; and then warmed until present. According to the longer estimates, a relatively warm period also existed in the mid-19$^{th}$ century. Therefore, according to the rural data, the current warm period is comparable to the earlier warm periods in the mid-1940s and possibly mid-19$^{th}$ century. Although the long-term linear trend is of warming, i.e., there has been global warming since the late-19$^{th}$ century, this seems to be mainly because the late-19$^{th}$ century was relatively cold. This is quite different from the estimates that IPCC (2013) considered, although interestingly it is quite similar to Lansner and Pepke-Pederson (2018)'s analysis based on "sheltered" stations [505]. Such a result might be surprising since many studies have claimed that urbanization bias is not a problem for estimating Northern Hemisphere air temperature trends *and* that all land-based global temperature estimates imply almost identical results [477–484,487–490,492–494,496,506,507]. Therefore, in the next subsection, possible reasons for these differences will be assessed.

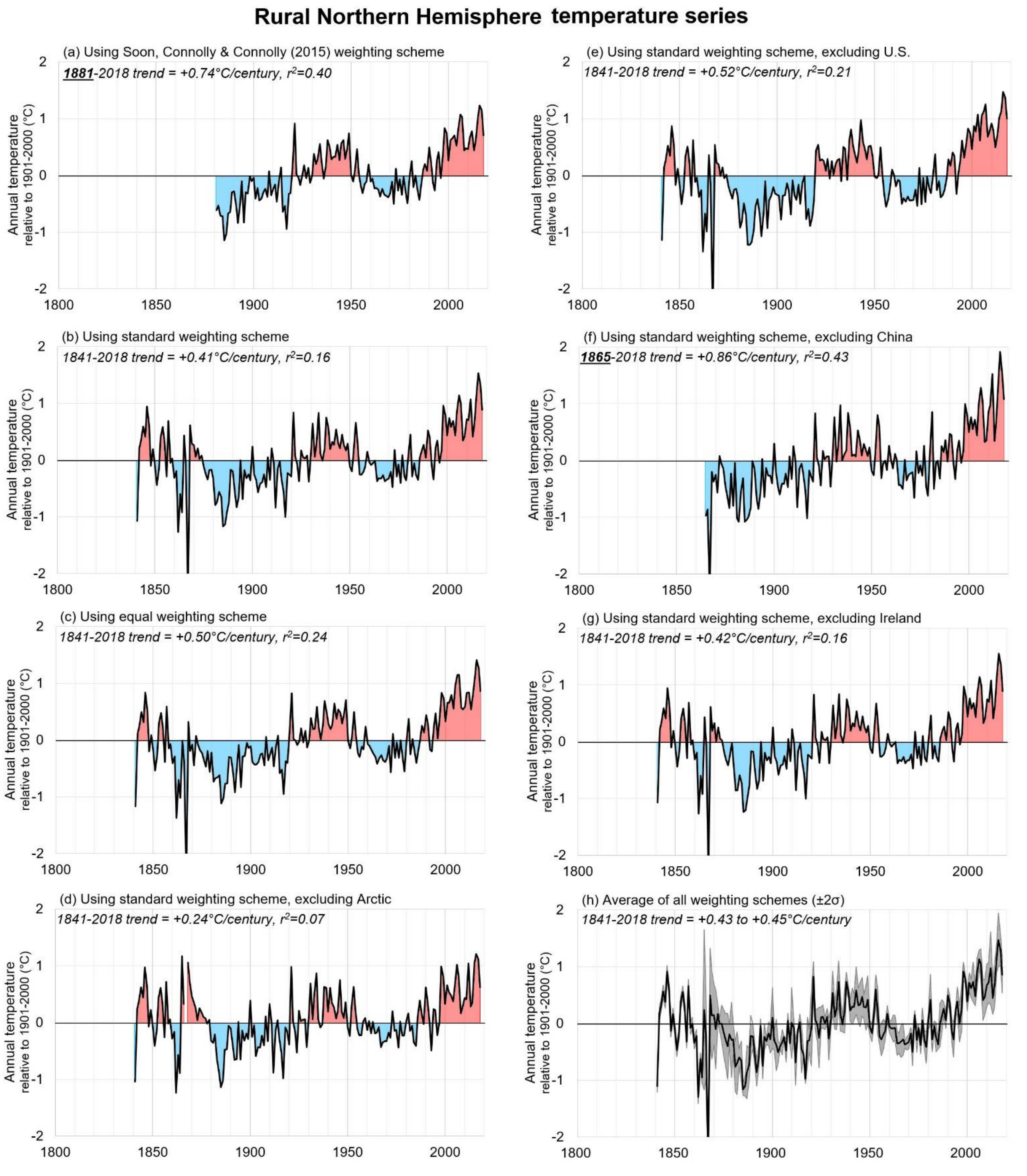

*Figure 7. Rural Northern Hemisphere temperature series constructed using (a)-(c) three different weighting schemes, or (d)-(g) using the standard weighting scheme but excluding one of the four regions. (h) plots the mean as well as upper and lower (±2σ) bounds of the seven alternative versions (a)-(g). Th ±2σ envelope in (h) is the final time series used for the analysis in this paper. For ease of comparison, the y-axes in Figures 7-13, 17 and 18 are all plotted to the same scale, as are the x-axes except for some of the extended plots using paleoclimate estimates.*

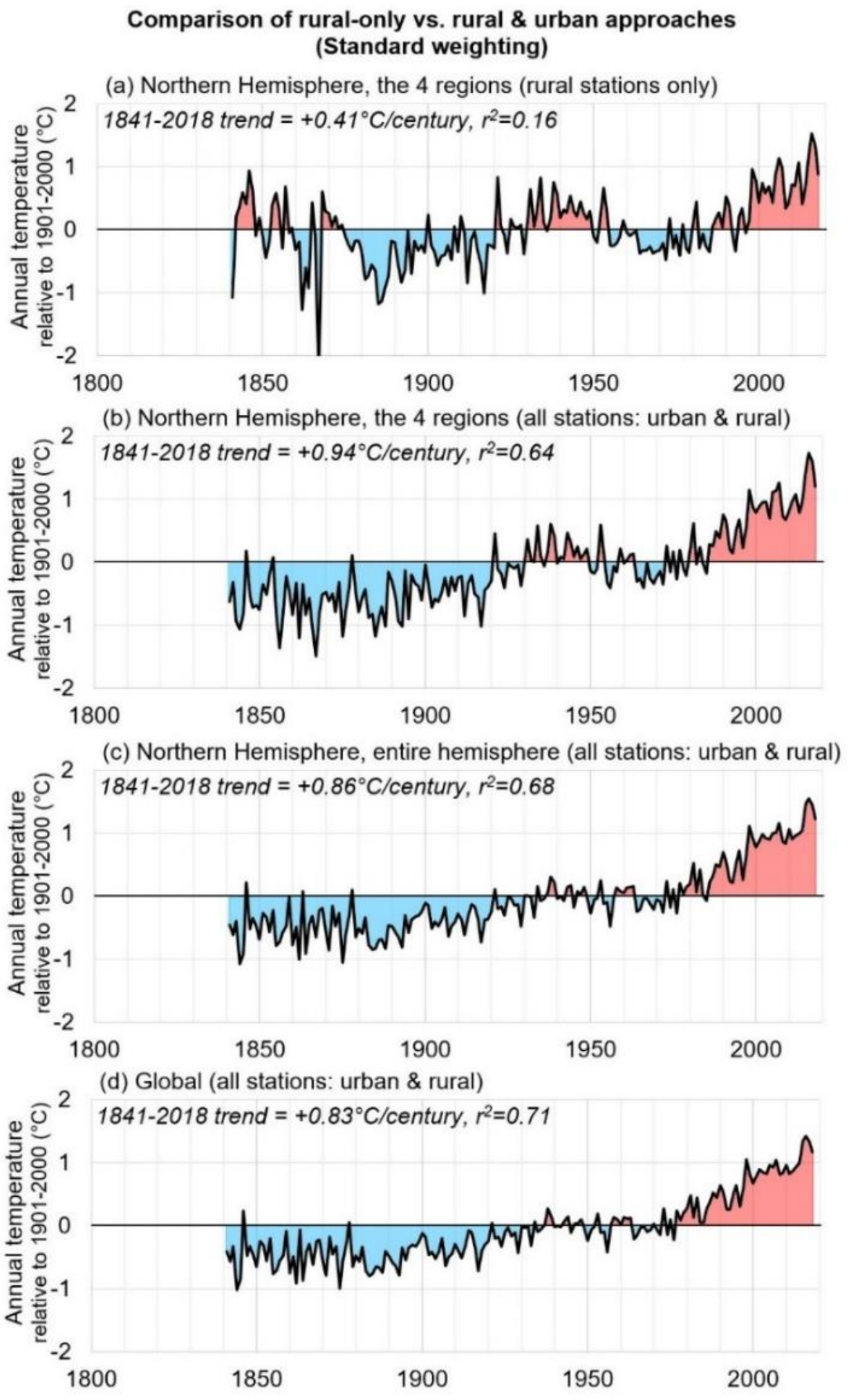

*Figure 8. (a) Our rural Northern Hemisphere temperature series using only rural (or urban-corrected) stations from the four regions. (b) The equivalent temperature series using all stations from the four regions whether urban or rural (but homogenized using the Menne & Williams, 2009 automated homogenization [491]). (c) as for (b) except using all valid Northern Hemisphere stations (urban and rural). (d) as for (b) except using all valid stations from either hemisphere. The 1841-2018 linear trends for each temperature series are shown in the corresponding panel along with the accompanying* $r^2$ *statistic for the linear trend. For ease of comparison, the y-axes in Figures 7-13, 17 and 18 are all plotted to the same scale, as are the x-axes except for some of the extended plots using paleoclimate estimates.*

### *3.1.1. Is our new rural-only estimate better or worse than the standard estimates that include both urban and rural stations?*

Peterson et al. (1999) [490] compared trends from a subset of only rural stations with the full GHCN dataset, but they argued that the trends were equivalent and concluded that urbanization bias was a negligible problem. However, the Peterson et al. (1999) study used a version of the GHCN that had been adjusted using an automated statistically-based homogenization algorithm in an attempt to reduce the effects of non-climatic biases. On the other hand, Soon et al. have argued that not only do the current homogenization algorithms perform poorly when applied to the non-climatic biases in the GHCN, but they also inadvertently lead to a blending of the non-climatic biases in different stations [56,116,117]. Hence, after homogenization, urban blending would have transferred much of the urbanization bias of the neighbouring urban stations into the "rural" stations used by Peterson et al. (1999).

To identify the reasons for the differences between our rural-only estimates and the standard estimates, in Figure 8, we compare our rural-only estimate, Figure 8(a), to three alternative estimates that are comparable to the standard estimates. The estimate in Figure 8(b) was calculated using all stations (rural, semi-urban or urban) for the same four regions and using the version of the GHCN dataset that has been homogenized using the automated Menne & Williams (2009) homogenization algorithm [491]. As in our rural-only estimate, USHCN stations are also corrected for changes in time-of-observation in the homogenized GHCN dataset [497]. However, no additional attempts to correct for non-climatic biases are applied other than NOAA's Menne & Williams (2009) homogenization. Note that this series has slightly more rural stations for the United States component than our rural-only series (295 instead of 272) because there are some non-USHCN stations for the contiguous U.S in the GHCN dataset.

The estimate in Figure 8(c), was calculated using all Northern Hemisphere GHCN stations (rural, semi-urban or urban), and only using the Menne & Williams (2009) homogenized GHCN dataset. Finally, the estimate in Figure 8(d) is calculated as for Figure 8(c), except it is a global estimate using all stations from either hemisphere.

Let us now consider five plausible objections which might be raised as to the reliability of the rural-only series relative to the standard estimates:

1. **Could the four regions be unrepresentative of global trends?** Comparing Figure 8(a) to Figure 8(c) and Figure 8(d) shows that the rural-only estimate implies much less warming than the standard estimates. Specifically, the rural-only estimate implies a long-term 1841-2018 linear trend of only 0.41°C/century for the Northern Hemisphere, which is less than half the equivalent trends using the standard approach, i.e., 0.86°C/century for the Northern Hemisphere, Figure 8(c), and 0.83°C/century globally, Figure 8(d). One possibility might be that these four regions are coincidentally underestimating the warming for the rest of the Northern Hemisphere (and globe). However, as can be seen from Figure 8(b), when the standard approach is applied to those four regions, the linear trend is even greater (0.94°C/century). In other words, if anything, the four regions are ones that slightly *overestimate* the warming of the rest of the world.

2. **Why was an automated statistically-based homogenization algorithm like the standard estimates not used to reduce the effects of non-climatic biases?** Various non-climatic biases in the data need to be corrected. However, as described earlier, Soon et al. have shown that such algorithms perform poorly when applied to the non-climatic biases in the GHCN and frequently result in inappropriate adjustments to the data [56,116,117]. Moreover, they showed that the algorithms inadvertently lead to a blending of the non-climatic biases in different stations. Therefore, we argue that empirically-based homogenization adjustments (ideally using station history information) are more reliable than the statistically-based homogenization adjustments currently used by most groups, e.g., Refs. [478–480,482–484,496,507].
3. **Why were the rural stations from outside these four regions not included?** As can be seen from Figure 6, there are quite a few rural stations in the GHCN dataset that were outside of the four regions. Specifically, there are 503 rural stations in our four regions, but 455 rural stations in the rest of the Northern Hemisphere and 269 in the Southern Hemisphere. Therefore, initially it might appear that our four regions only considered 50% of the total Northern Hemisphere rural data and 39% of the global rural data. However, as can be seen from Figure 6(d)-(f), our four regions comprise the vast majority of the available rural data with relatively long station records. For instance, the four regions account for more than 80% of the rural stations from either hemisphere with early 20$^{th}$ century records. This is not accidental, as Soon et al. (2015) had specifically identified those four regions as being ones that contained a relatively high density of rural stations or for which they had station history information for applying empirically-based homogenization adjustments [56]. Therefore, while it might be surprising to many readers, these four regions account for the vast majority of the *long-term* rural station records.
4. **Were too few stations used?** The rural-only estimate was constructed from a total of 554 stations (this figure is slightly higher than the 503 mentioned above, due to the non-rural Chinese stations used for some of the Chinese component), whereas the Northern Hemisphere estimate in Figure 8(c) used 4176 stations and the global estimate in Figure 8(d) used 4822 stations. Moreover, several of the estimates by other groups purport to use even more stations. Indeed, the Berkeley Earth group claim to use ~40,000 stations [481] (although most of these stations have station records with fewer than 30 years of data). Therefore, some might argue that the rural-only series differs from the standard estimates because the sample size was too small. However, as discussed in Soon et al. (2015) [56] (see in particular their Table 6), and argued by Hawkins and Jones (2013) [507], many of the earlier attempts to estimate Northern Hemisphere temperature trends used even fewer stations, yet obtained fairly similar results to the latest estimates using the standard approach. In particular, Mitchell (1961) [508] used only 119 stations; Callendar (1961) [509] used only ~350; and Lugina et al. (2006) [477] used only 384. Moreover, Jones et al. (1997) [510] calculated that about 50 well-distributed stations should be adequate for determining annual global air temperature trends. Since our composite is for the Northern Hemisphere only, even fewer stations should be needed. Therefore, the differences in trends between the rural-only series and the standard estimates are *not* a result of the smaller number of stations.
5. **Why are the differences most pronounced for the earlier period?** Given that the rate of urbanization has accelerated in recent decades, initially it might be assumed that the differences between urban and rural time series would be greatest for recent decades. However, we remind readers of two factors:
   All the time series in Figure 8 are relative to the 20$^{th}$-century average. This partially reduces the apparent differences between the time series during the 20$^{th}$ century.
   From Figure 6, it can be seen that the GHCN dataset has a relatively high rural composition during the 1951-1990 period. The shortage of rural stations in the standard estimates is most pronounced before (and, to a lesser extent, after) this period. Therefore, counterintuitively, the differences between "rural-only" and "urban and rural" tend to be larger before 1951, due to changes in data availability.

Nonetheless, some readers might still prefer the standard estimates that use both urban and rural stations. We will discuss these in the next section.

### *3.2. Using urban and rural stations*

Figure 9 compares the Northern Hemisphere land-only air temperature trends of seven different estimates calculated using the standard approach, i.e., including both urban and rural stations but applying homogenization procedures such as the Menne and Williams (2009) [491] automated statistical homogenization algorithm or Berkeley Earth's "scalpel" procedure [481]. Although slight differences exist between the approaches taken by each group, the results are remarkably similar. For comparison, the "all Northern Hemisphere stations" estimate of Figure 8(c) is replicated here in Figure 9(a), albeit extended back to 1800, and it is very similar to the other six estimates. The striking similarity of all these estimates has been noted by most of the groups when describing their individual estimates, e.g., CRUTEM [482], i.e., Figure 9(b); Cowtan and Way [483], i.e., Figure 9(c); NOAA NCEI [480], i.e., Figure 9(d); NASA GISS [478,479], i.e., Figure 9(e); Berkeley Earth [481], i.e., Figure 9(f); and the Chinese Meteorological Administration [484], i.e., Figure 9(g).

The fact that all of these estimates are so similar appears to have led to a lot of confidence within the community that these estimates are very reliable and accurate. As discussed in the previous section, we believe this confidence is unjustified and that the new rural-only estimate is more reliable. Nonetheless, for the sake of argument, and because we appreciate that many readers may disagree with us, we will also carry out our analysis for these "urban and rural"-based estimates.

Although the seven estimates are all remarkably similar, as noted above, slight differences still exist between each of the estimates. Therefore, as before, we use the mean with the upper and lower bounds of all seven estimates. These bounds are calculated as the mean of the seven series for each year $\pm 2\sigma$ – see Figure 9(h).

### *3.3. Sea surface temperatures*

One way to potentially bypass the debate over urbanization bias might be to consider ocean surface temperature trends instead of land surface temperature trends. Unfortunately, considerable uncertainties are also associated with the ocean surface temperature data, especially before the 1950s. Before the International Geophysical Year of 1957/1958, available data are quite sparse for the Northern Hemisphere and very sparse for the Southern Hemisphere [511]. Moreover, the methods and instrumentation that were used for measuring ocean surface temperatures varied from ship to ship and are often poorly documented [512]. For this reason, it has been argued that the ocean surface temperature data is less reliable than the land surface temperature data, e.g., Jones (2016) [506]. Nonetheless, the ocean data are largely independent of the land surface temperature data – even if some groups have partially reduced this independence by adjusting the ocean surface temperature data to better match the land surface temperature records, e.g., Cowtan et al. (2018) [513].

Two competing sets of measurements for the ocean surface temperature exist. The Marine Air Temperatures (MAT) are based on the average air temperature recorded on the decks of ships. The Sea Surface Temperatures (SST) are based on the average water temperature near the surface. In some senses, the Marine Air Temperature is more comparable to the Land Surface Air Temperature since they are both measurements of the air temperature near the surface. However, the available Marine Air Temperature records are more limited than the Sea Surface Temperature records, and it has been suggested that the Marine Air Temperatures are more likely to be affected by non-climatic biases, especially the day-time measurements, e.g., Rayner et al. (2003) [514]; Kent et al. (2013) [515].

For this reason, most of the analysis of ocean surface temperature trends has focused on the Sea Surface Temperature data. For brevity, our ocean surface temperature analysis will be confined to the Sea Surface Temperature data, but we encourage further research using both types of dataset and note that they can yield different results [516]. Yet, even with the Sea Surface Temperature data, it is widely acknowledged that non-climatic biases are likely to exist in the data, especially for the pre-1950s period [506,511–513,517,518].

It is likely that if the true magnitudes and signs of these biases could be satisfactorily resolved, as Kent et al. (2016) [512] have called for, this could also help us to resolve the debates over the land surface temperature data. For instance, Davis et al. (2018) [518] noted that if you separately analyze the SST measurements that were taken via "engine-room intake" and those taken by "bucket" measurements, this implies quite different global temperature trends over the period 1950-1975. Specifically, the "engine-room intake" measurements imply a substantial global cooling trend from 1950 to 1975, commensurate with our rural estimates for land surface temperatures in Section 3.1. Interestingly, the "sheltered" stations subset of Lansner and Pepke-Pederson (2018) [505] suggests the same. By contrast, the "bucket" measurements imply a slight increase over this same period – consistent with our "urban and rural" estimates described in Section 3.2 and also the "non-sheltered" subset of Lansner and Pepke-Pederson (2018) [505].

Currently, three different versions of Sea Surface Temperatures are available from the Hadley Centre (HadISST [514], HadSST3 [517,519,520] and HadSST4 [511]) and three from NOAA (ERSST v3 [521], v4 [522] and v5 [523]). The Japanese Meteorological Agency also has a version (COBE SST2 [524]), but here, our analysis will be limited to the more widely-used NOAA and Hadley Centre datasets.

Figure 10 compares the various Northern Hemisphere Sea Surface temperature estimates. In recognition of the considerable uncertainties associated with estimating sea surface temperature trends, the Hadley Centre have taken to providing ensembles of multiple different plausible estimates (100 each) for their two most recent versions, i.e., HadSST3 [517,519,520] and HadSST4 [511]. However, the previous version, HadISST [514], and the three NOAA versions did not take this approach. Therefore, if we were to treat all HadSST3 and HadSST4 ensemble members as separate estimates, then the other four datasets would not make much contribution to our analysis. On the other hand, if we only considered the "median" estimates from the HadSST3 and HadSST4 datasets, then we would be underestimating the uncertainties associated with these datasets by the Hadley Centre. With that in mind, for both of these datasets, we have extracted the “lower realisation” and “upper realisation”, i.e., the lower and upper bounds, and treated these as four separate estimates that are representative of the uncertainties implicit within the combined 200 “ensemble” estimates of the two datasets – see Figure 10(b)-(e). We then treat the Hadley Centre’s other version, HadISST – see Figure 10(a) - and the three NOAA estimates – see Figure 10(f)-(h) - as an additional four distinct estimates.

This provides us with a total of eight different estimates. For our Sea Surface Temperature analysis, we use the mean and the upper and lower bounds of all eight estimates. These bounds are calculated as the mean of the eight series for each year ±2σ. See Figure 10(i).

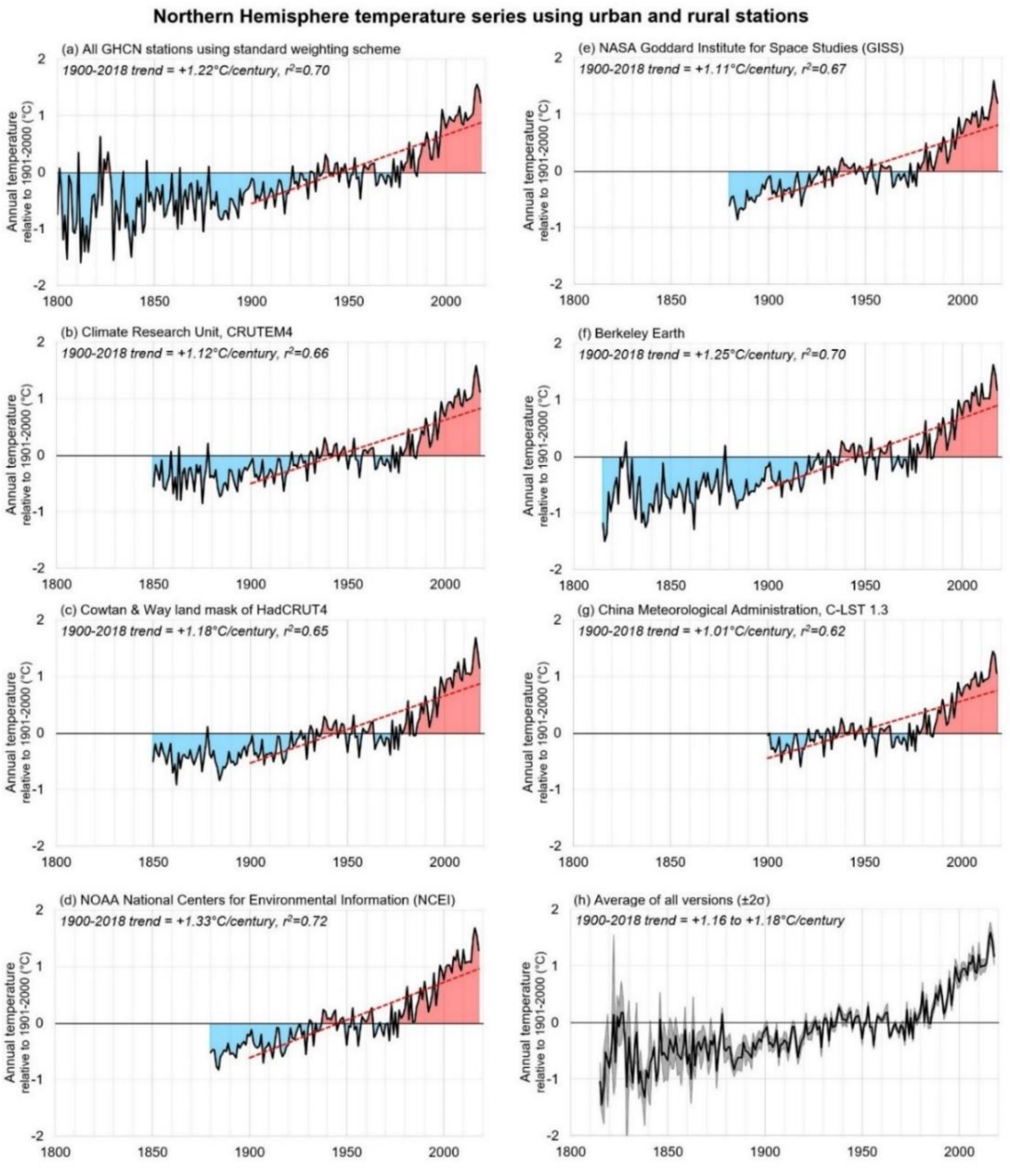

*Figure 9. Different estimates of Northern Hemisphere land surface temperatures constructed using all stations (either urban or rural). (a) Our estimate from Figure 8(c) using the GHCN version 3 dataset. (b) The Climate Research Unit’s CRUTEM4 estimate. (c) Cowtan & Way’s land mask component of HadCRUT4. (d) NOAA National Centers for Environmental Information’s estimate. (e) NASA Goddard Institute for Space Studies’ estimate. (f) Berkeley Earth’s estimate. (g) China Meteorological Administration’s C-LST 1.3 estimate. (h) plots the mean and upper and lower (±2σ) bounds of the seven alternative estimates (a)-(g). The mean and ±2σ envelope in (h) are the final time series used for the analysis in this paper. The linear trend for the common period of overlap for all series, 1900-2018, are shown for all estimates for comparison purposes. Note that this is different from the 1841-2018 linear trends considered in Figure 8. For ease of comparison, the y-axes in Figures 7-13, 17 and 18 are all plotted to the same scale, as are the x-axes except for some of the extended plots using paleoclimate estimates.*

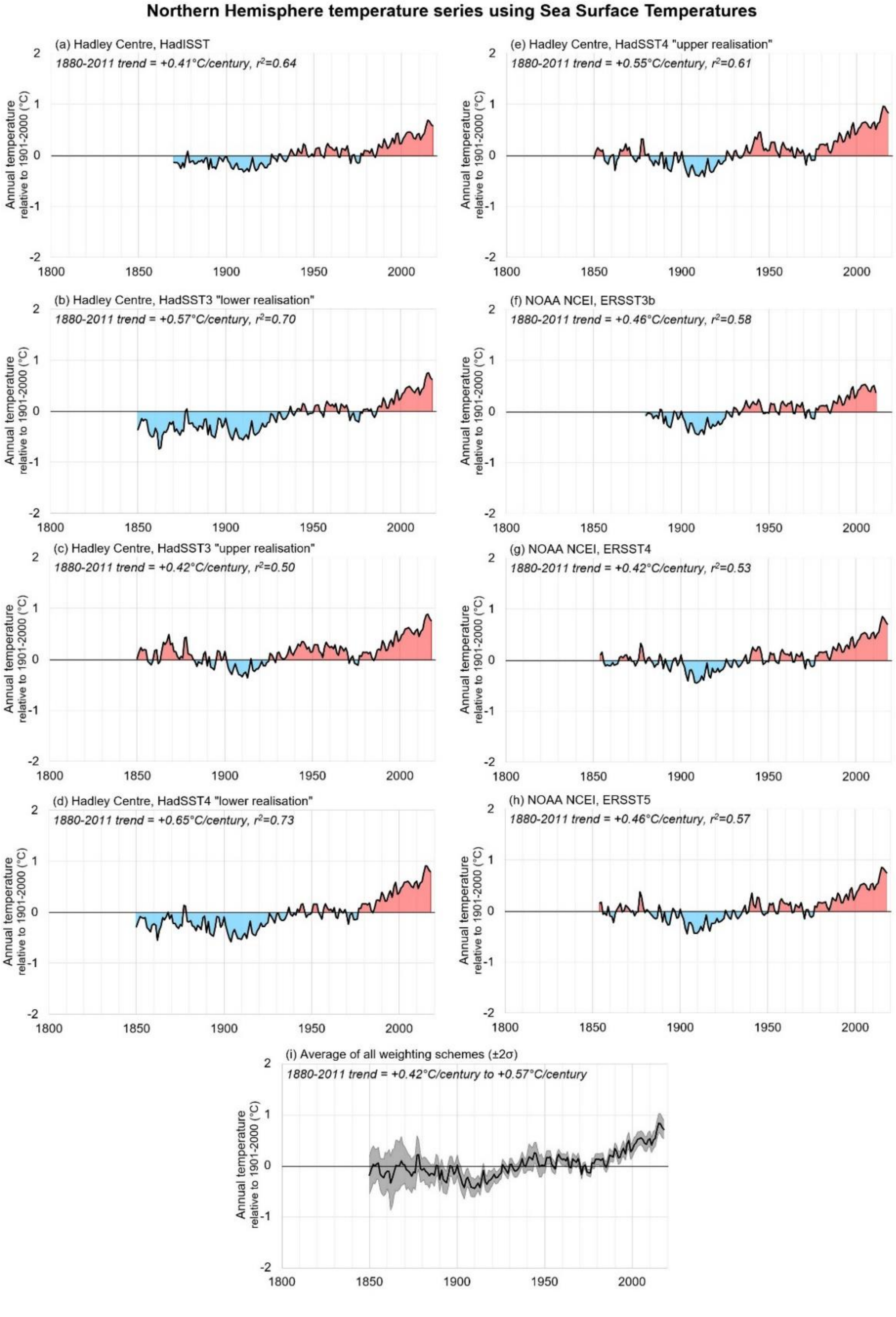

*Figure 10. Different estimates of Northern Hemisphere surface temperatures constructed from Sea Surface Temperature data. (a)-(e) Various estimates developed by the Hadley Centre – see text for details. (f)-(h). Various estimates developed by NOAA National Centers for Environmental Information (NCEI), formerly known as National Climate Data Center (NCDC). (i) plots the mean with upper and lower (±2σ) bounds of the eight alternative estimates (a)-(h). The mean ±2σ envelope in (i) are the final time series used for the analysis in this paper. The linear trends for the common period of overlap of all series, i.e., 1880-2011 are listed for each estimate for comparison purposes. For ease of comparison, the y-axes in Figures 7-13, 17 and 18 are all plotted to the same scale, as are the x-axes except for some of the extended plots using paleoclimate estimates.*

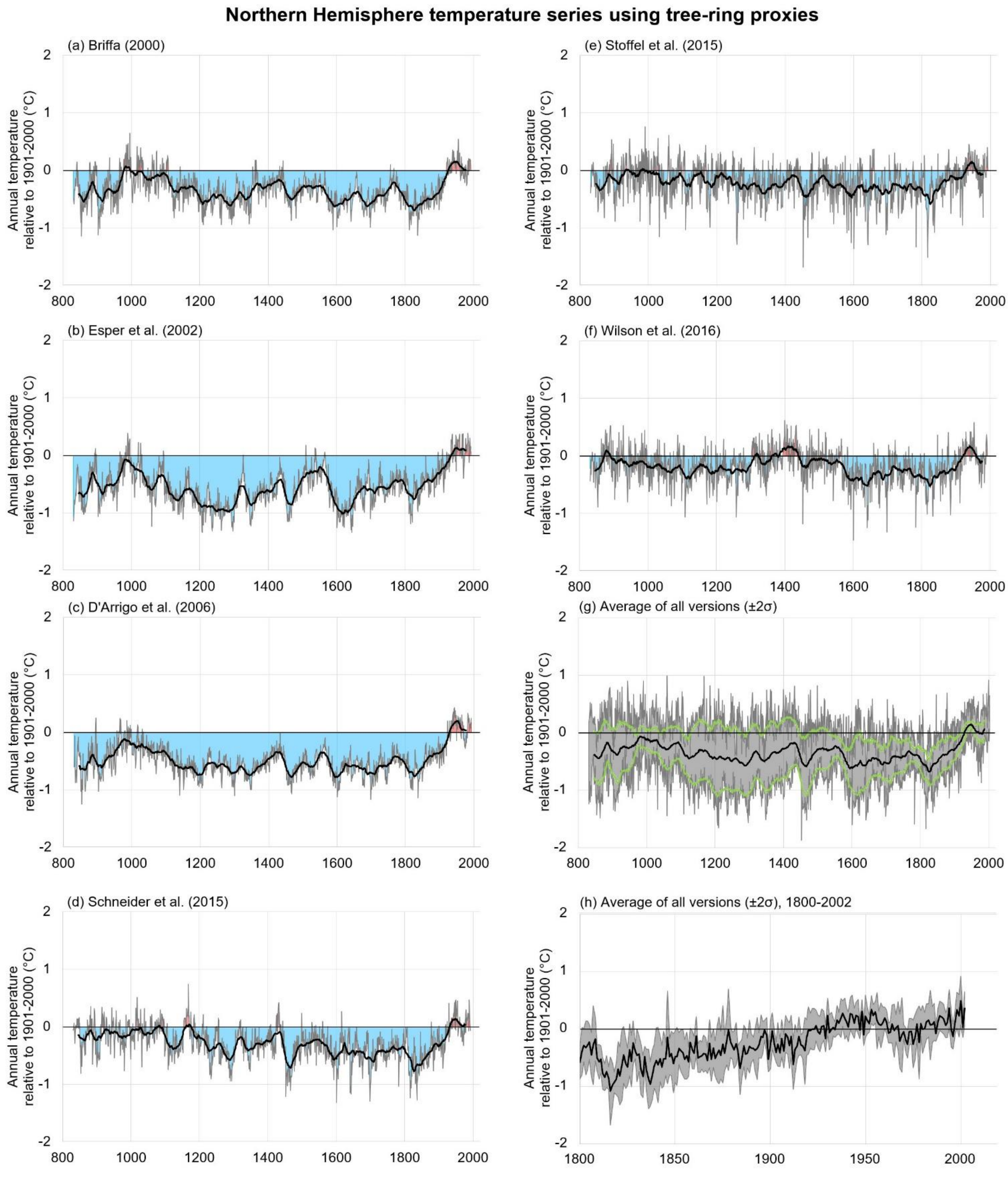

*Figure 11. (a) – (f) Plots of Esper et al. (2018)'s six tree ring-based Northern Hemisphere land surface temperature proxy series. (g) plots the upper and lower (±2σ) bounds of the six proxy series (a)-(f). (h) is the same as (g) except only covering the most recent 1800-2002 period. The mean and ±2σ envelope in (g) and (h) are the final time series used for the analysis in this paper. The solid black line curves in (a) – (g) and the solid green upper and lower bounds in (g) represent the 31-year running averages and are included simply for visual purposes. For ease of comparison, the y-axes in Figures 7-13, 17 and 18 are all plotted to the same scale, as are the x-axes except for some of the extended plots using paleoclimate estimates.*

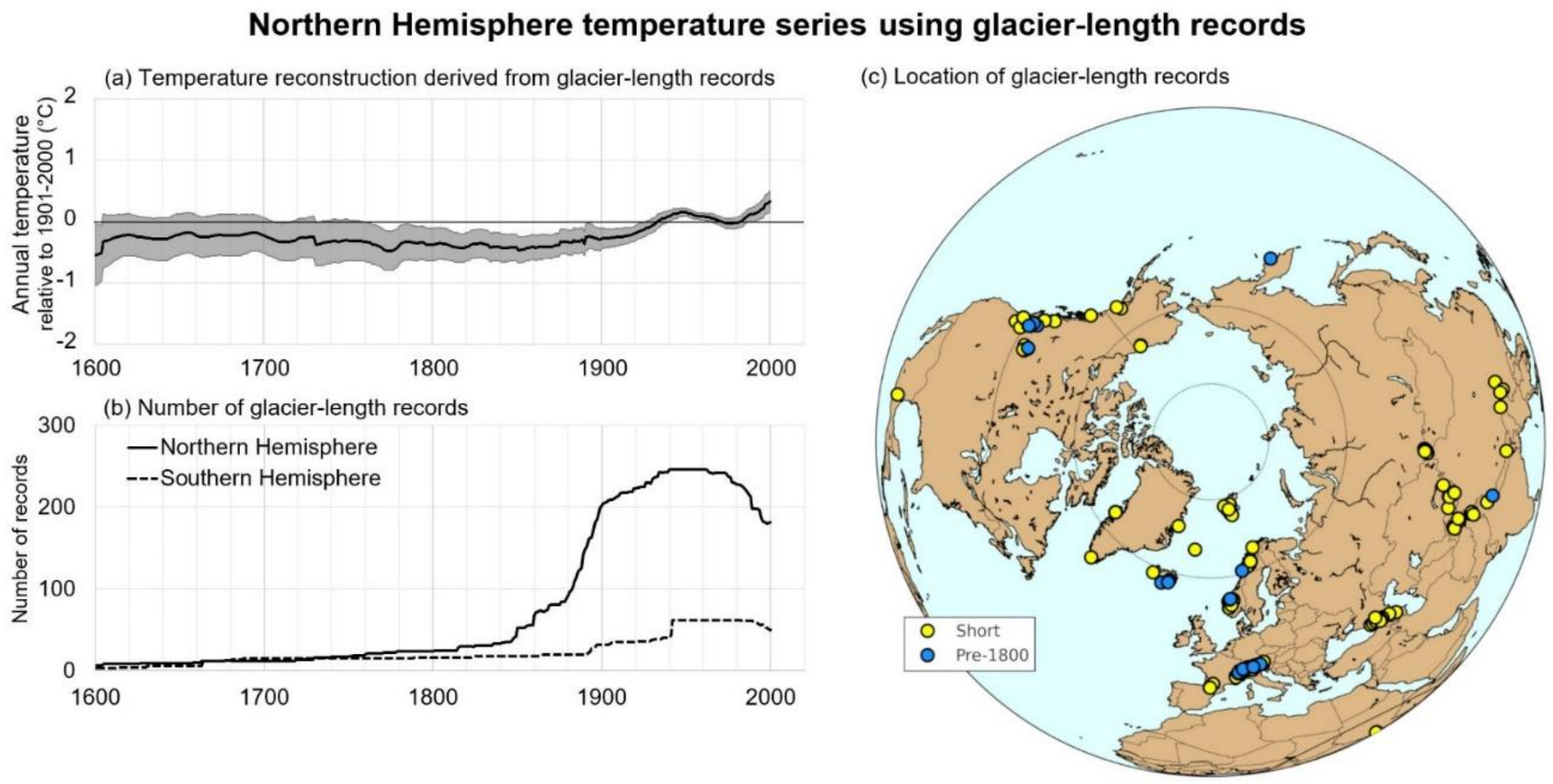

*Figure 12. (a) The mean with upper and lower bounds of Leclercq & Oerlemans' (2012) Northern Hemisphere temperature reconstruction derived from glacier-length records [525]. (b) The number of glacier length records available for either hemisphere over time. (c) Location of the glaciers used by Leclercq & Oerlemans (2012) for their Northern Hemisphere reconstruction. The records with data before 1800 are indicated with a blue fill. For ease of comparison, the y-axes in Figures 7-13, 17 and 18 are all plotted to the same scale, as are the x-axes except for some of the extended plots using paleoclimate estimates.*

### *3.4. Tree-ring proxy based reconstructions*

To assess the role of the Sun in Northern Hemisphere temperature trends before the mid-19th century, analysis must be mostly limited to indirect estimates from paleoclimate reconstructions developed using temperature proxy series, e.g., tree rings [526].

Esper et al. (2018) [526] recently compiled and reviewed six different tree-ring based millennial-length reconstructions for the Northern Hemisphere. Figure 11 compares the six different Northern Hemisphere tree-ring temperature proxy-based reconstructions described by Esper et al. (2018) [526]. Although all these reconstructions share some of the same underlying tree-ring data, and they imply broadly similar long-term temperature trends, Esper et al. (2018) [526] stressed that each of the reconstructions gave slightly different results. They stressed that it was inappropriate to simply average the reconstructions together – a point which St. George and Esper (2019) [527] reiterated. Therefore, here, each of these six reconstructions should be treated as a different estimate. However, for consistency with the rest of our estimates, our analysis will be based on the mean along with the upper and lower bounds, which are calculated as the mean of the six series for each year ±2σ. See Figure 11(g) for the entire period spanned by these estimates, which begin in the 9th century, and Figure 11(h) for a close-up of the estimates since the start of the 19th century.

### *3.5. Glacier-length based reconstruction*

Leclercq & Oerlemans (2012) [525] constructed a dataset of all the available multi-decadal glacier length records for both hemispheres. They then used the changes in glacier length at each of these glaciers as a proxy for local temperature. Leclercq & Oerlemans nominally provided separate estimates for global temperatures as well as each hemisphere. However, as can be seen from Figure 12(b), most of the data was for the Northern Hemisphere, which is the hemisphere our analysis focuses on. It can also be seen that most of the glacier length records began in the 20th century.

Nonetheless, they had some data for both hemispheres to cover the entire period 1600-2000, although the error bars provided by Leclercq & Oerlemans (2012) [525] are quite substantial, especially for the pre-20th century period. We will treat the mean with upper bound and the lower bound from the Leclercq & Oerlemans (2012) [525] Northern Hemisphere reconstruction as our estimates - see Figure 12(a).

### *3.6. Comparison of all five types of estimate*

In Figure 13, we compare and contrast the five different estimates of Northern Hemisphere temperature trends since the 19th century (we only present the post-18th century data for our two longer proxy-based estimates in this figure for clarity). Most of the estimates share key similarities, specifically,

- They all imply that current temperatures are higher than at the end of the 19th century, i.e., that some warming has occurred since the end of the 19th century.
- There was an "Early 20th Century Warm Period" (ECWP) in the mid-20th century – peaking sometime in the 1930s-1950s – followed by several decades of *either* cooling *or* a lack of warming.
- There was a definite period of warming from the late-1970s to the end of the 20th century. Some estimates suggest that this warming has continued to the present day, although there is some debate in the literature over whether there has been a "warming hiatus" after the end of the 20th century and if so whether this hiatus has ended.

However, subtle differences exist between the various estimates, including,

- The timings of the various warming and cooling periods, e.g., most of the estimates imply the Early 20th Century Warm Period peaked during the 1940s, while the tree ring-based proxy estimates imply a later peak during the 1950s.
- The exact magnitudes of the various warming and cooling periods, e.g., the estimates using both urban and rural stations imply a larger long-term warming trend and that the warming has been almost entirely continuous, while the other estimates imply a more nuanced alternation between multi-decadal warming and cooling periods.
- The two proxy-based estimates imply a more dampened temperature multidecadal variability than the instrumental estimates. This is not surprising considering that they are based on indirect estimates of temperature variability, rather than direct temperature measurements. For instance, while glacier lengths are certainly influenced by the local air temperature during melt season (i.e., summer), they are also influenced by winter precipitation (i.e., how much snow accumulates during the growth season). Indeed, Roe and O'Neal (2009) note that glaciers can, in principle, advance or retreat by kilometers without any significant climate change, but simply due to year-to-year variability in local weather [528]. Similarly, García-Suárez et al. (2009) argue that tree-ring proxies are often proxies for multiple different climatic variables as well as temperature, e.g., sunshine, precipitation, soil moisture, etc., and also mostly reflects conditions during the growing season [529]. Therefore it can be difficult to extract a pure "temperature" signal [529]. Meanwhile, Loehle (2009) notes that even with tree-ring proxies from "temperature-limited regions" (usually either high latitude or high elevation sites), the growth rates may still be influenced by other factors, and that the relationship between growing season temperatures and tree-ring growth may be non-linear [530].

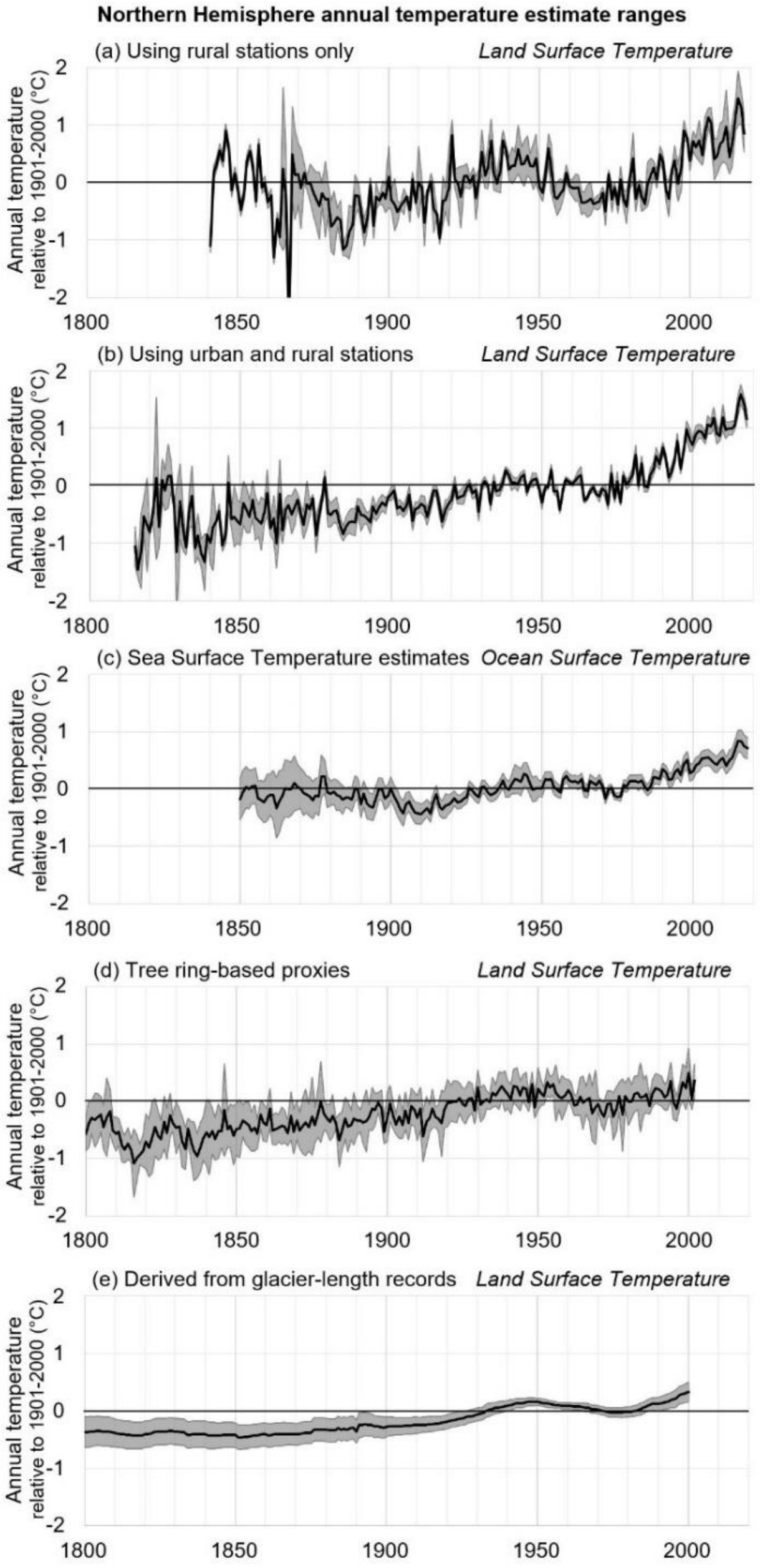

*Figure 13. Different estimates of Northern Hemisphere surface air temperature trends since the 19th century. For ease of comparison, the y-axes in Figures 7-13, 17 and 18 are all plotted to the same scale, as are the x-axes except for some of the extended plots using paleoclimate estimates.*

It is plausible that some (or even all) of these differences arise from non-climatic biases in one or more (and perhaps all) of the time series. That said, some (or even all) may arise because each time series is capturing a slightly different aspect of the true climatic variability, e.g., the oceanic surface temperature variability may be slightly different from the land surface temperature variability.

At any rate, these nuanced differences are important to consider for our analysis, because we will be estimating the influence of the Sun's variability on Northern Hemisphere temperature trends using a relatively simple linear least-squares fit between a given temperature estimate and each of our 16 estimates of solar variability, which we discussed in Section 2.4. The successes of the fits will depend largely on the linear correlations between the timings and magnitudes of the various rises and falls in the time series being fitted. For this reason, we propose to analyze the different categories of Northern Hemisphere temperature variability estimates separately. Our fits will be repeated separately for each of the 16 solar variability estimates to both the upper and lower bounds. This then provides us with an uncertainty range for each of our estimates.

## 4. Changes in "anthropogenic forcings"

The primary focus of this paper is to consider the role of solar variability in Northern Hemisphere temperature trends since the 19th century. However, as discussed in the Introduction, the IPCC 5th Assessment Report argued that the dominant driver of global (and hemispheric) temperature trends since at least the mid-20th century has been "anthropogenic forcings" [531].

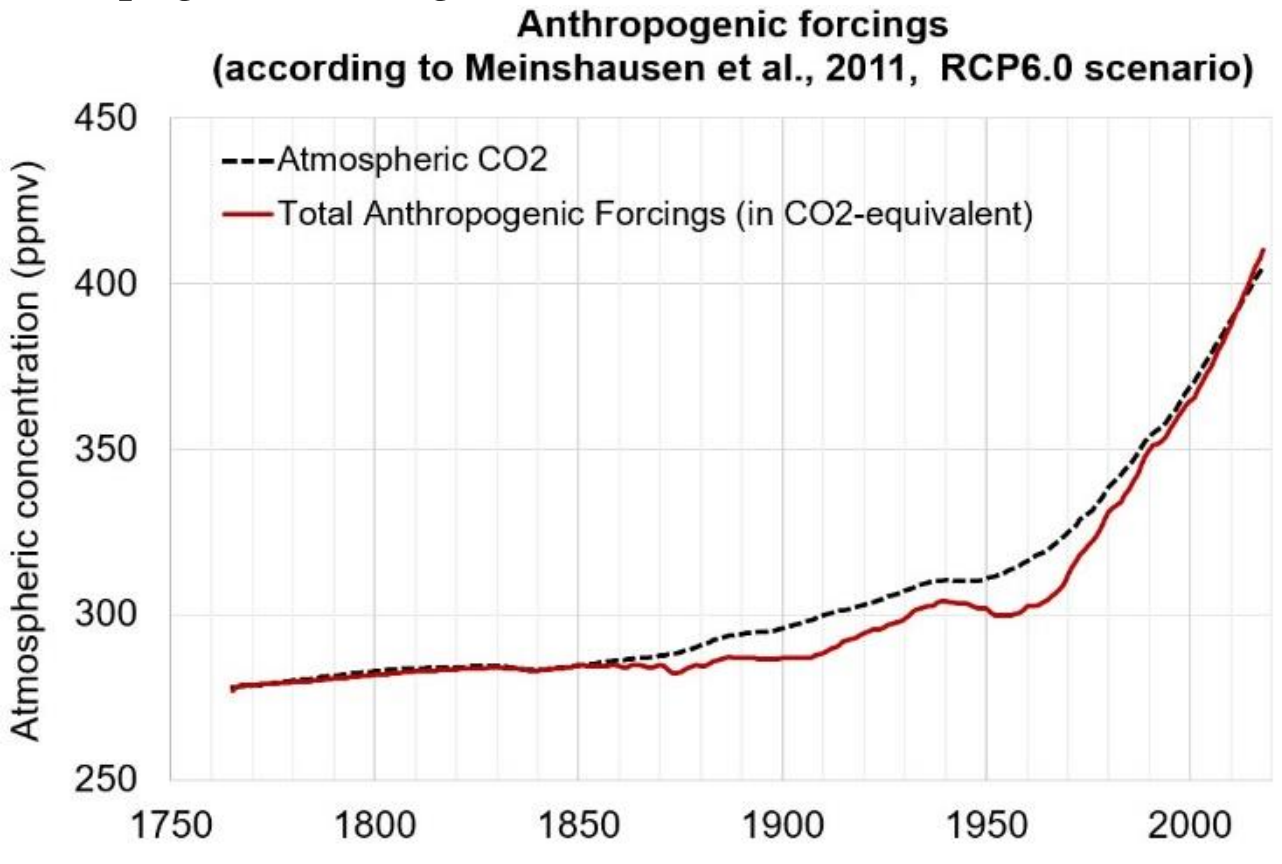

*Figure 14. Comparison of the Total Anthropogenic Forcings dataset used in this study (red solid line) in units of CO2-equivalent atmospheric concentrations vs. atmospheric CO2 concentrations (dashed black line) (ppmv). Both time series are taken from Meinshausen el.'s (2011) Representative Concentration Pathways (RCP) datasets. 1765-2005 corresponds to the "historical forcing" data, while the 2006-2018 extension is taken from the "RCP6.0" scenario, but similar results are found for the other two scenarios that assume a continuous increase in anthropogenic forcing over the 21st century, i.e., RCP4.5 and RCP8.5.*

Therefore, we should also compare the role of "anthropogenic forcings" to that of "solar forcing" (i.e., changes in TSI). However, since we have identified 16 different estimates of TSI (Section 2.4) and 5 different sets of estimates of Northern Hemisphere temperature trends (Section 3), this already gives us 80 (5×16) different combinations to consider. With this in mind, for simplicity, we have taken the approach of selecting a single time series to be representative of "anthropogenic forcings".

According to the Representative Concentration Pathways (RCP) scenarios [532], which were used by the IPCC 5th Assessment Report [2] and Annex II of the 5th Assessment Report, the main "anthropogenic forcings" since the 19th century are: (1) a warming forcing from increasing atmospheric $CO_2$, and (2) a cooling forcing from increasing stratospheric aerosol concentrations. However, they also include several other "anthropogenic forcings", e.g., changes in other greenhouse gases such as methane and nitrous oxide. Recently, in Connolly et al. (2020) [533], three of us have carried out a breakdown of the relationship between changing concentrations of these three greenhouse gases, including a comparison of the RCP scenarios to other scenarios. But, for simplicity, in this paper, we define "anthropogenic forcings" for each year as being the annual sum of all the individual anthropogenic forcings in the historical forcings of the RCP dataset from 1765 to 2005, and we extend this time series up to 2018 using the "RCP 6.0" scenario. In one version of the historical forcings and RCP scenarios, all of the anthropogenic forcings have been converted into $CO_2$ equivalent concentrations. Therefore, we use this "$CO_2$ equivalent" time series. The time series is shown in Figure 14 as a solid red line, along with the equivalent trends of atmospheric $CO_2$ alone as a black dashed line. It can be seen that both time series are broadly similar, but the "total anthropogenic forcings" slightly deviates from the $CO_2$-only time series for parts of the late 19th and 20th century. We note that unlike anthropogenic forcings datasets used for earlier IPCC reports, the Meinshausen et al. (2011) "total anthropogenic forcings" appears to include a slight "bump" in the 1940s followed by a slight valley in the 1950s [532].

## 5. Estimating the role of the Sun in Northern Hemisphere surface temperature trends since the 19th century and earlier

As indicated by the title of this article, in this study, we are attempting to estimate how much of a role solar variability has had on Northern Hemisphere temperature trends. However, to do this, we need to know how solar output (i.e., TSI) and Northern Hemisphere temperatures have changed. In Section 2.4, we compiled sixteen different estimates of Total Solar Irradiance (TSI) trends covering the same period – summarised graphically in Figures 2 and 3. In Section 3, we compiled five different estimates of Northern Hemisphere surface temperature trends since the 19th century (or earlier) – summarised graphically in Figure 13. Therefore, we now have 80 (i.e., 16 × 5) different, but plausible, combinations of our various estimates of the trends in TSI and of Northern Hemisphere temperature trends. Meanwhile, in Section 4, we summarised the trends of "total anthropogenic forcings" in terms of a single time series derived from the sum of the various individual “anthropogenic forcings” considered in the “RCP” scenarios [532] as used for the IPCC’s 5th Assessment Report [1] – summarised graphically in Figure 14.

Given the huge range of plausible combinations to be considered, we have chosen to take a relatively simple statistical approach to answer this question. Specifically,

1. For each of the 80 combinations, the linear least-squares best fits between Northern Hemisphere temperatures and TSI are evaluated for the maximum common period of overlap between the time series. In the case of the two longest Northern Hemisphere temperature series (i.e., the glacier-derived and tree-ring-derived proxy series), 1765 is used as the starting point since this marks the beginning of the “anthropogenic forcings” time series.
2. The “solar contribution” to Northern Hemisphere temperature trends will be defined as the percentage of the long-term temperature trend which can be explained in terms of this linear least-squares solar fitting. Fits are carried out for the upper and lower bounds of each Northern Hemisphere temperature series as well as for the mean, and the lowest and highest values from these two fits are presented as the lower and upper bounds of our “solar contribution” estimates.
3. We then calculate the statistical residuals remaining after subtracting the solar fits from the temperature series (i.e., the unexplained variability of the temperature series).
4. We then calculate the linear least-squares best fits between these residuals and the “total anthropogenic forcings” time series (i.e., that illustrated in Figure 14).
5. We define the “anthropogenic contribution” to Northern Hemisphere temperature trends as the percentage of the long-term temperature trend that can be explained in terms of this linear least-squares fitting to the residuals.
6. The sum of these two contributions to the trend of the original Northern Hemisphere temperature series will then be evaluated. If the sum is less than 100%, then this suggests that some extra component other than solar and anthropogenic forcing is needed to explain the rest of the trend. However, as we will see, often the sum is greater than 100%. This would suggest that either or both components are overestimated and/or there is a missing unexplained “global cooling” component that is neither solar nor part of the anthropogenic forcings.

We stress that this is a relatively simplistic approach to estimating the relative contributions of “solar” and “anthropogenic forcings” to Northern Hemisphere temperature trends. For this analysis, we are explicitly assuming that there is a direct linear relationship between incoming TSI and Northern Hemisphere surface temperatures. However, as discussed in Sections 2.5 and 2.6, there is a lot of evidence to suggest that the relationships between solar activity and the Earth’s climate are non-linear and a lot more subtle. In particular, this simplistic assumption of a direct linear relationship does not take into account: any “top-down” or “bottom-up” mechanisms (Section 2.6.1); ocean buffering (Section 2.6.2); the possibility that Sun/climate effects might vary regionally (Section 2.6.3); Galactic Cosmic Ray-driven amplification mechanisms (Section 2.6.4); or short-term orbital variability (Section 2.6.5). The goal of this analysis is *not* to dismiss these more nuanced approaches to investigating the Sun/climate relationships. Indeed, many of us have contributed to the literature reviewed in Sections 2.5-2.6, and we plan on pursuing further research along these avenues. Rather, we want to emphasise that, as will be seen shortly, even with this approach, a surprisingly wide range of results can be found. As researchers actively interested in resolving these issues, we find this wide range of plausible results disquieting.

We also want to emphasise that by fitting the anthropogenic forcings to the residuals *after* fitting the data to TSI, we are implicitly maximising the solar contribution relative to the anthropogenic contribution. This can be seen by, for instance, comparing and contrasting the two separate approaches to fitting carried out by Soon et al. (2015) [56] in their Section 5.1. That is, they found a larger solar contribution when they fitted the data to TSI first and a larger anthropogenic contribution when they fitted the data to anthropogenic forcings first. With that in mind, it might be argued that the various contributions should be estimated simultaneously, e.g., via the use of a multilinear regression analysis or an energy balance model or a general circulation model. Indeed, several of us have carried out such analyses in the past [23,26,41,59] and are also planning similar approaches for future research. However, we caution that there is a distinction between the TSI estimates that are calibrated against empirical measurements (i.e., satellite measurements) and anthropogenic forcings that are usually calculated from theoretical modelling. That is, they are not necessarily directly comparable “forcings”.

With those caveats in mind, in Figures 15 and 16, we present the best fits for the eight “low variability” TSI estimates (Figure 15) and the eight “high variability” TSI estimates (Figure 16) in terms of the mean estimates of the five Northern Hemisphere temperature datasets. The corresponding fits to the upper and lower bounds can be found in the Microsoft Excel dataset in the Supplementary Materials as well as some extra details on the statistical fittings, etc. However, for brevity, here we will focus on some of the main findings.

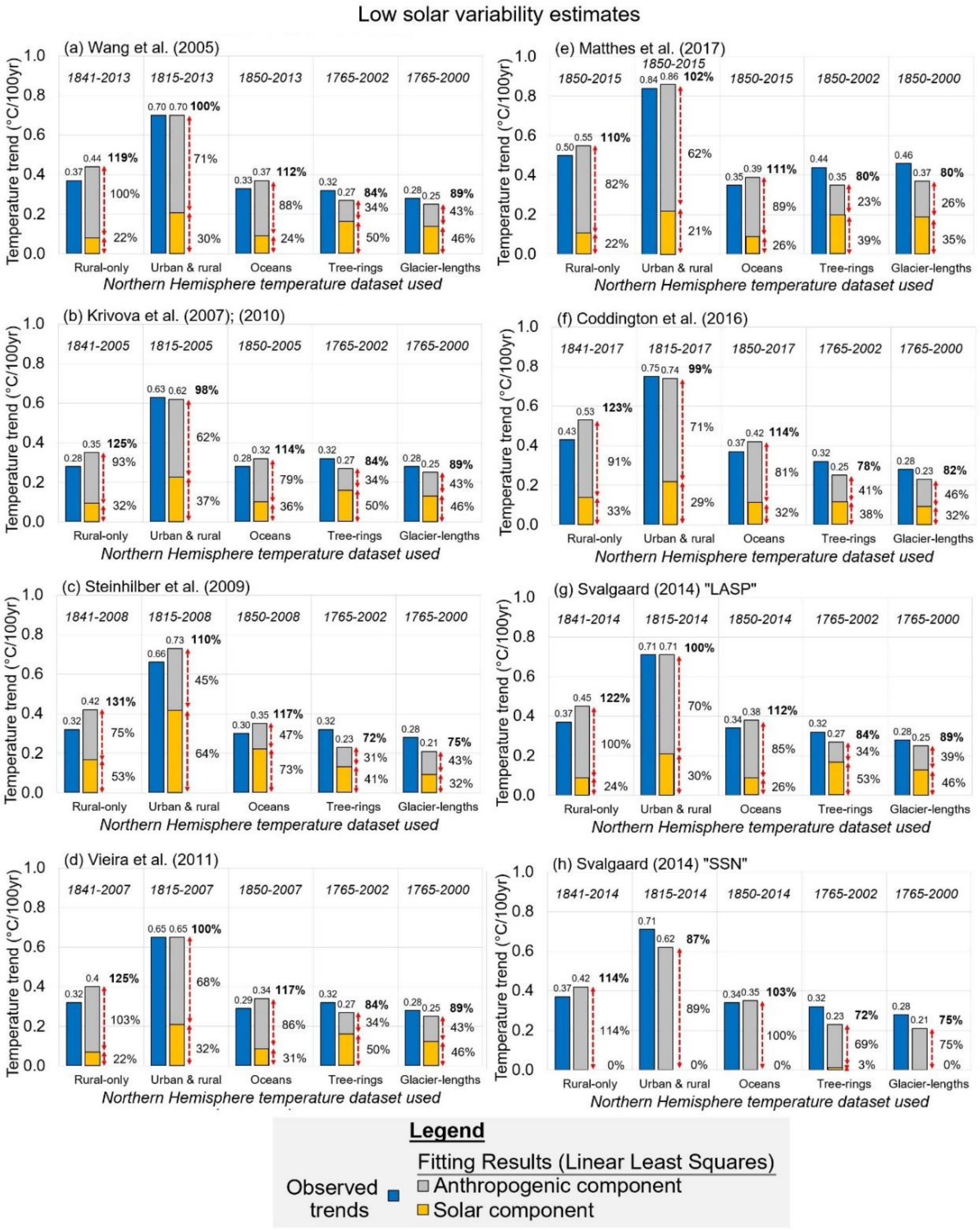

*Figure 15. The best linear least-squares fits to the five Northern Hemisphere temperature datasets of Figure 13 in terms of each of the eight "Low solar variability" TSI datasets from Figure 2 and the "anthropogenic forcings" time series of Figure 14. As described in the text, each of the temperature time series was first fit to the TSI time series over the maximum period of overlap of the two time series (starting in 1765 at the earliest). The residuals were then fit to the anthropogenic forcings time series. The linear temperature trends are plotted as the blue bars on the left. The linear temperature trends of the combined "solar plus anthropogenic" fits are plotted beside these bars, but split into the linear trends of the two components. The percentage of the observed linear trends that can be explained in terms of the solar component, the anthropogenic component and the combined fit are shown. Note that the percentage of the combined fits are often greater than or less than 100%, and that the sum of the two components does not always equal the combined fit.*

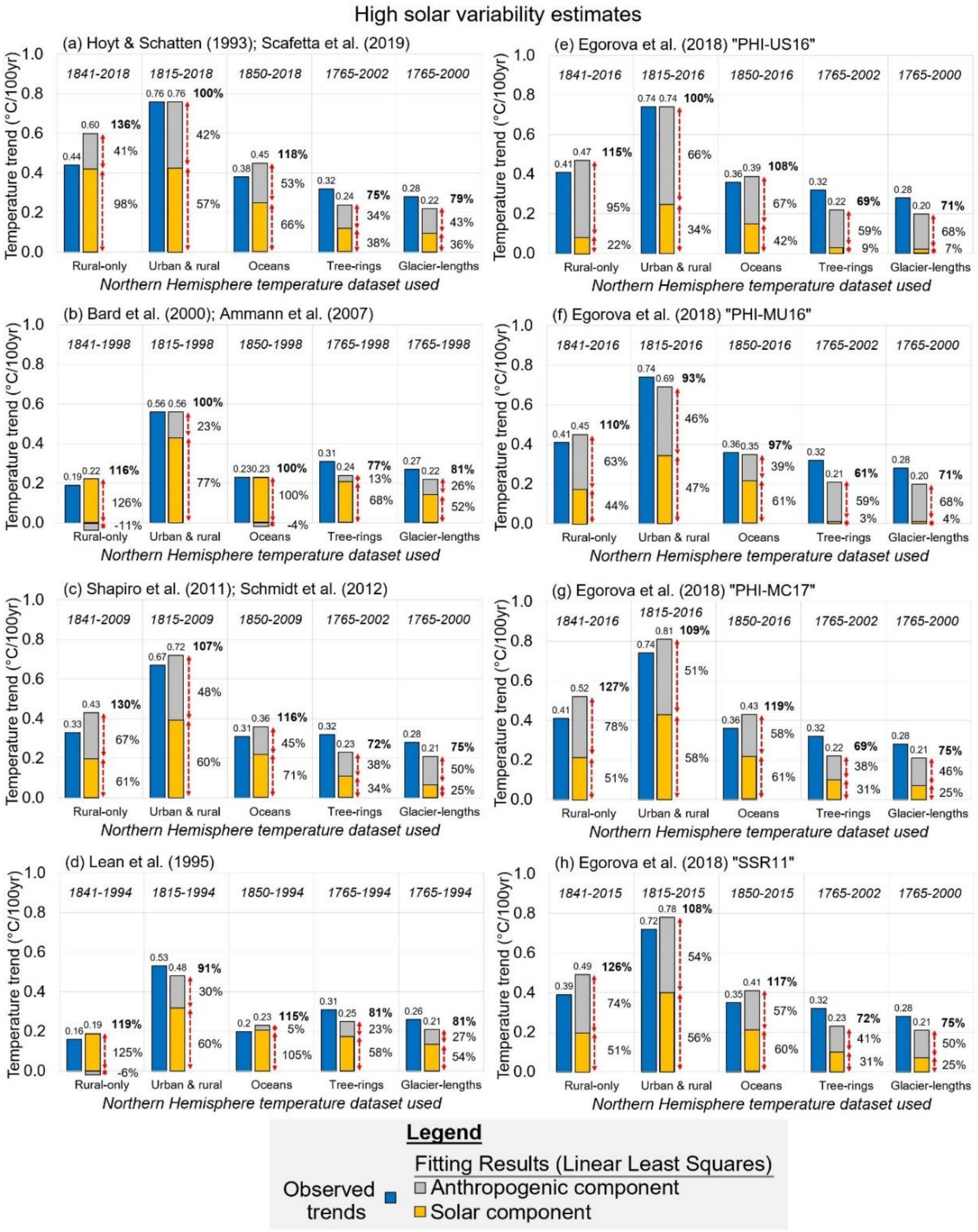

*Figure 16. The best linear least-squares fits to the five Northern Hemisphere temperature datasets of Figure 13 in terms of each of the eight "High solar variability" TSI datasets from Figure 3 and the "anthropogenic forcings" time series of Figure 14. As described in the text, each of the temperature time series was first fit to the TSI time series over the maximum period of overlap of the two time series (starting in 1765 at the earliest). The residuals were then fit to the anthropogenic forcings time series. The linear temperature trends are plotted as the blue bars on the left. The linear temperature trends of the combined "solar plus anthropogenic" fits are plotted beside these bars, but split into the linear trends of the two components. The percentage of the observed linear trends that can be explained in terms of the solar component, the anthropogenic component and the combined fit are shown. Note that the percentage of the combined fits are often greater than or less than 100%, and that the sum of the two components does not always equal the combined fit.*

Each of the TSI estimates covers a slightly different period (see Table 1), as do the Northern Hemisphere temperature datasets. So, the linear periods being analyzed varies a bit for each of the 80 combinations. The exact period is written above the corresponding bar plots. For each combination, the linear temperature trend over the relevant period (in °C/century) is indicated by a single bar (colored in blue). Beside it are the stacked double bars for the corresponding linear least-squares fits. The height of the stacked bars represents the linear temperature trend of the combined "solar plus anthropogenic" fits and can be directly compared to the observed trend. In some cases, this value is greater than the observed trend (i.e., >100%). I.e., this combined linear trend overestimates the observed linear trend, suggesting that either the anthropogenic and/or solar component is overestimated. In other cases, the value is less than the observed trend (i.e., <100%), suggesting there may have been other contributing factors than solar or anthropogenic forcing involved.

Meanwhile, the percentage of the observed temperature trend that can be explained in terms of the solar and anthropogenic components are listed on the right-hand side of the stacked bars. It should be noted that the sum of these percentages does not always equal the value of the combined fit. For example, the combined fit for the Matthes et al. (2017) TSI estimate to the urban and rural temperature series, i.e., Figure 15(e), explains 102% of the observed trend, but individually the solar component only explains 21% and the anthropogenic component only explains 62%. On the other hand, for the Wang et al. (2005) estimate, the combined fit to the rural-only series explains 119% of the observed trend, but the sum of the solar and anthropogenic fits adds to 122% of the observed trend. Some of these apparent inconsistencies are due to rounding errors. However, mostly this arises because neither the Northern Hemisphere temperature trends nor the modelled contributions are strictly linear. Hence, linear least-squares fits can only approximate the observed trends, and we caution that the results from this analysis should be explicitly treated with this caveat in mind. Nonetheless, the results can give us a reasonable estimate of which combinations imply the greatest and lowest role for solar variability in Northern Hemisphere temperature trends.

With regards to the smallest role for the Sun, Svalgaard (2014)'s "SSN" estimate, i.e., Figure 15(h), implies that 0% (or at most 3%) of the Northern Hemisphere temperature trends since at least the 19th century have been driven by solar variability. This TSI estimate is essentially a rescaled version of the Sunspot Number time series. This is worth noting for those readers who believe the Sunspot Number record is perfectly correlated with the TSI record. If this were the case, then the answer to the question posed in the title of our article would simply be, "nothing". But as discussed in Section 2, there seem to be many more factors to consider in estimating TSI than "sunspot numbers".

As mentioned in the introduction, the UN IPCC concluded in their 5th Assessment Report (AR5) [1] that, "*It is **extremely likely** that human influence has been the dominant cause of the observed warming since the mid-20th century*". The following analysis provides insights into the origin of this striking conclusion. As can be seen from Table 1 - and as discussed in Soon et al. (2015) [56] – all four of the estimates of TSI that were considered by the CMIP5 climate modelling groups for the hindcasts that were submitted to the IPCC AR5 are:

1. Wang et al. (2005) [172]
2. Krivova et al. (2007) [176]; (2010) [177]
3. Steinhilber et al. (2009) [252]
4. Vieira et al. (2011) [253]

The fitting results for these four estimates are plotted in Figure 15 (a) to (d), respectively. The long-term trends since the 17th century for each of these series are shown in Figure 2 (a) to (d) similarly. Except for the Steinhilber et al. (2009) series, all of the estimates do indeed imply that the long-term trends of the non-proxy based temperature series have been dominated by anthropogenic forcings. E.g., for the Wang et al. (2005) estimate, only 22% of the rural trends can be explained in terms of solar variability, and it is possible to explain 100% of the linear trend in terms of anthropogenic forcing – see Figure 15(a).

We can see that the new Matthes et al. (2017) estimate that has been recommended to the CMIP6 modelling groups [110] yields very similar results, and if anything implies even less of a solar contribution – Figure 15(e). Therefore, we anticipate that if the CMIP6 modelling groups have adopted the recommendations of Matthes et al. (2017) [110], then the IPCC 6th Assessment Report would probably come to a similar conclusion to the 5th Assessment Report. The same would occur if they were to use either the Coddington et al. (2016) [237] or the Svalgaard (2014) "LASP" estimates – Figure 15(f) and (g), respectively.

However, from Figure 16, we can see that the use of many of the "high solar variability" TSI estimates could imply a much greater role for the Sun than the IPCC AR5 had prematurely concluded. In particular, we note that either the Hoyt and Schatten (1993) [52,60,179] or Bard et al. (2000) [534,535] would imply the opposite of the IPCC AR5, i.e., that solar variability has been the dominant cause of the long-term warming – see Figure 16(a) and (b). For instance, the Hoyt and Schatten (1993) estimate suggests that 98% of the long-term trend (1841-2018) of the rural-only temperature estimates can be explained in terms of solar variability – see Figure 16(a). This is equivalent to what Soon et al. (2015) found [56]. Meanwhile, the Bard et al. (2000) [534,535] estimate suggests that all of the warming trend (although this estimate ends in 1998) observed for the rural-only and ocean temperature series can be explained in terms of solar variability, and the residuals imply if anything a slight cooling from anthropogenic forcings – see Figure 16(b).

It should already be apparent that the answer to our question, "How much has the Sun influenced Northern Hemisphere temperature trends?" depends substantially on which estimate of TSI we choose. To those readers who find this disquieting, we should stress that we share your discomfort, and we are surprised that neither the IPCC AR5 [1] nor Matthes et al. (2017) [110] seems to have been concerned about this.

There is an additional concern that deserves attention too. From Figure 15(c), we can see that the Steinhilber et al. (2009) estimate [252] implies that a substantial percentage of the long-term warming can be explained in terms of solar variability, e.g., up to 64% of the 1815-2008 trend of the urban & rural temperature estimate and 73% of the 1850-2008 ocean temperature estimate could be explained in terms of solar variability. This appears to contradict the IPCC AR5's conclusion, yet this was one of the four TSI estimates that were used by the CMIP5 climate modelling groups. One partial explanation is that most of the CMIP5 modelling groups used one of the other three estimates mentioned earlier. However, we suggest that a bigger factor is the fact that the long-term variability of the Steinhilber et al. (2009) estimate has been substantially dampened relative to the variability intrinsic in the underlying solar proxy data that were used to construct the estimates.

The Steinhilber et al. (2009) estimate is predominantly based on cosmogenic nucleotide solar proxies [252], similar to the Bard et al. (2000) estimate [534]. To convert this proxy series into absolute TSI values, Steinhilber et al. (2009) applied a linear relationship to the proxy values (see their Equation 4) [252]. Steinhilber et al. (2009) argued that their scaling was reasonable as it compared favourably with the variability implied by the PMOD satellite composite. However, as we discussed in Section 2.2, there is ongoing debate over which of the rival satellite composites are most reliable. Also, Soon (2014) has criticised the scaling of Steinhilber et al. (2009) by noting that it was effectively based on just four data points, and that two of those data points were problematic [536].

At any rate, it can be seen from Figure 2(c) that the absolute variability in W/m$^2$ of the Steinhilber et al. (2009) time series is very low compared to those of the "high solar variability" estimates of Figure 3, even though the relative variability of the underlying data is quite substantial. Since the current Global Climate Models essentially model the potential influence of solar variability on the climate through changes in the absolute TSI (in W/m$^2$), they are unlikely to detect as large a solar role as implied by Figure 15(c) when using the scaling applied by Steinhilber et al. (2009).

Another point to notice about the results in Figures 15 and 16 is that none of the fits were able to explain 100% of the long-term warming trends of the two proxy-based estimates in terms of just solar and anthropogenic factors. If the proxy-based estimates are reliable then this suggests that there are important additional climatic drivers that are not being considered by the current Global Climate Models. It may also indicate that the current models are underestimating the magnitude of internal climate variability [119,120].

A limitation of the above analysis is that it only compares and contrasts the linear trends and, as mentioned, the true multidecadal trends of both the observed data and the modelled fits are only crudely approximated in terms of linear trends. Also, the above analysis was only fitting the solar and anthropogenic components to the means of the five Northern Hemisphere temperature estimates. As discussed in Section 3, there are considerable uncertainties associated with each of the Northern Hemisphere temperature estimates, and for this reason we have provided upper and lower bounds for each estimate – see Figure 13.

Therefore, in Figures 17 and 18, we present a more detailed analysis for 10 of our 80 combinations. In this analysis, we plot the annual time series for each of the fits and compare them with the temperature estimates they were fit to. This allows us to compare and contrast the relative magnitudes and timings of the multidecadal variations in the TSI and Northern Hemisphere temperature estimates, rather than just the long-term linear trends.

In Figure 17, we present the results for the five fits to the Matthes et al. (2017) [110] TSI dataset, since this is the dataset that the CMIP6 modelling groups have been asked to use for their simulations that will be used in the IPCC's upcoming 6$^{th}$ Assessment Report (AR6), which is currently due to be published between 2021 and 2022. We also obtained qualitatively similar results to Matthes et al. (2017) for six of the remaining fifteen TSI datasets:

- Wang et al. (2005) [172]
- Krivova et al. (2007) [176]; (2010) [177]
- Vieira et al. (2011) [253]
- Coddington et al. (2016) [237]
- Svalgaard (2014) "LASP"
- Egorova et al. (2018) "PHI-US16" [254]

As described above, the first three of these TSI datasets – along with the Steinhilber et al. (2009) dataset – were the only ones considered by the CMIP5 modelling groups, i.e., the climate model results that the IPCC AR5 predominantly relied on. Therefore, the results in Figure 17 are also broadly applicable to the IPCC's 5$^{th}$ Assessment Report [1].

Meanwhile, in Figure 18, we present the results of the TSI datasets that provide the best statistical fit to solar variability for each of our five Northern Hemisphere temperature estimates. For the rural-only temperature series, Figure 18(a), this corresponds to the Hoyt and Schatten (1993) estimate [52,60,179]. This, therefore, represents an update of the final analysis of Soon et al. (2015) [56], which reached the opposite conclusion to the IPCC's AR5. However, for the other four Northern Hemisphere temperature estimates, a slightly better statistical fit was found for a different TSI dataset, i.e., Bard et al. (2000) [534,535].

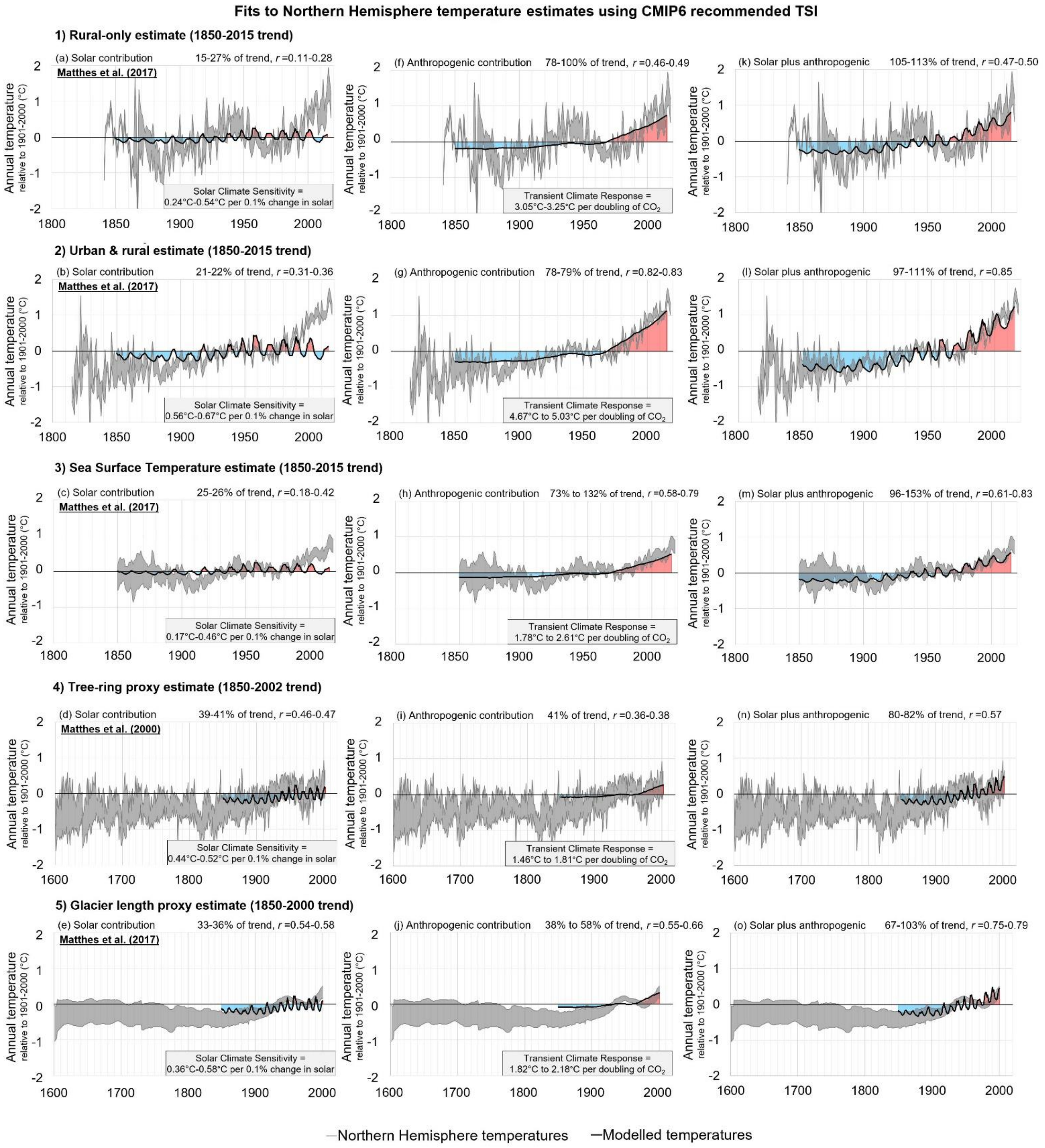

*Figure 17. The results of the fittings to the five different Northern Hemisphere temperature datasets using the Matthes et al. (2017) [110] Total Solar Irradiance (TSI) estimate, which has been recommended for use by the CMIP6 modelling groups in their simulations for the IPCC's upcoming 6th Assessment Report (currently due 2021-22). (a)-(e) provide the results for the maximum solar contribution implied by linear least-squares fitting. (f)-(j) provide the results for the best fits of the "anthropogenic forcings" dataset to the statistical residuals remaining after the solar fit. (k)-(o) provide the results of the combined "solar plus anthropogenic" fits. For ease of comparison, the y-axes in Figures 7-13, 17 and 18 are all plotted to the same scale, as are the x-axes except for some of the extended plots using paleoclimate estimates.*

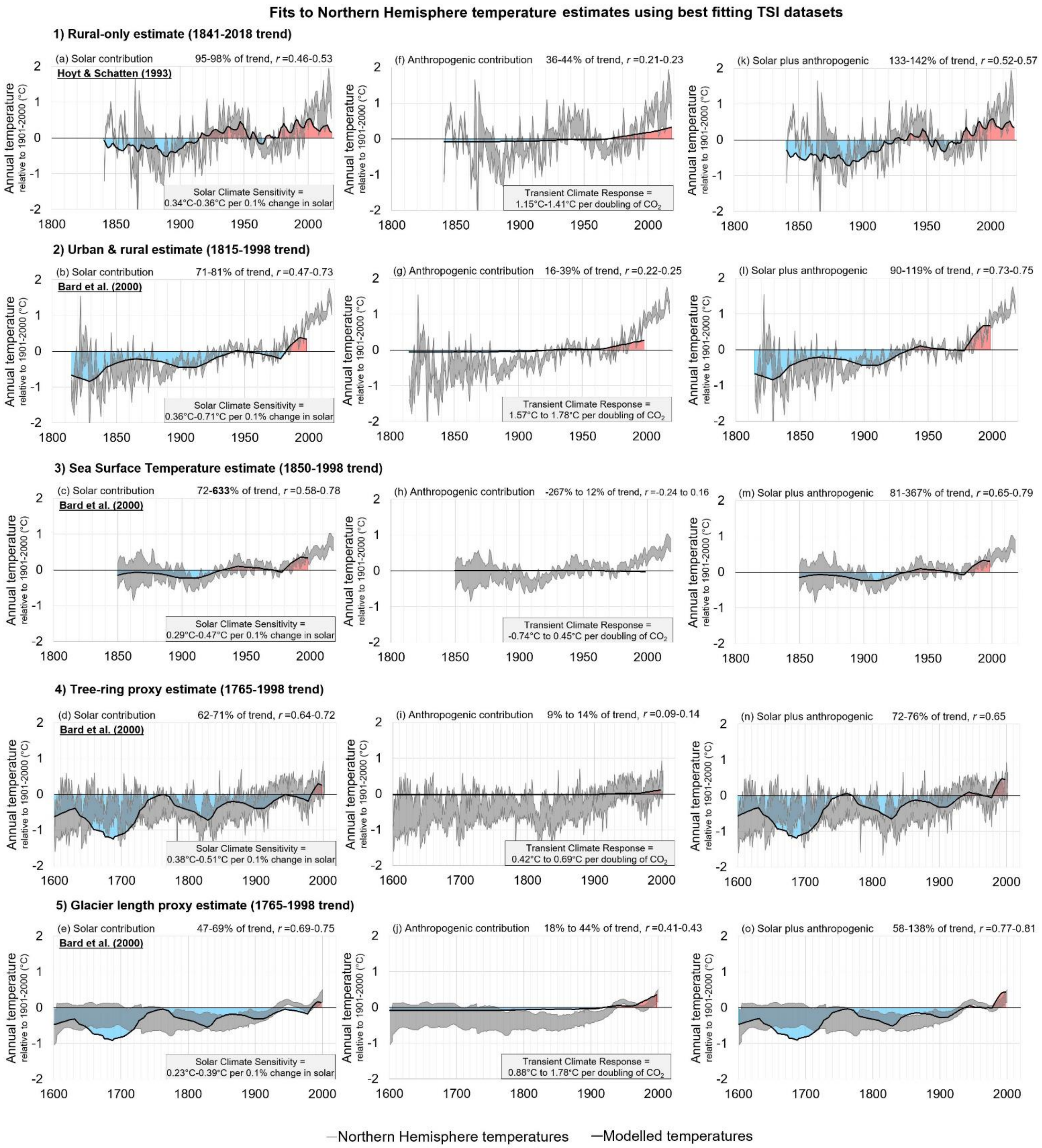

*Figure 18. The results of the fittings to the five different Northern Hemisphere temperature datasets using the Total Solar Irradiance (TSI) estimates that provide the largest role for solar variability of the 16. For the rural-only temperature series, this corresponds to the Hoyt and Schatten (1993) [52,60,179] estimate, while for the other four temperature series, this corresponds to the Bard et al. (2000) [534,535] estimate. (a)-(e) provide the results for the maximum solar contribution implied by linear least-squares fitting. (f)-(j) provide the results for the best fits of the "anthropogenic forcings" dataset to the statistical residuals remaining after the solar fit. (k)-(o) provide the results of the combined "solar plus anthropogenic" fits. For ease of comparison, the y-axes in Figures 7-13, 17 and 18 are all plotted to the same scale, as are the x-axes except for some of the extended plots using paleoclimate estimates.*

Therefore, by explicitly assuming that the relationship between TSI and Northern Hemisphere surface temperatures is linear, then the results of Figure 18 provide an upper bound for the solar contribution – according to the currently available datasets. Figure 17 is representative of the position adopted by the IPCC for their 5th Assessment Report [1], and it is likely to reflect that of their 6th Assessment Report if they repeat the approach of AR5 except adopting the TSI recommendations by Matthes et al. (2017) [110]. We re-emphasise that there is plenty of research to suggest that the sun/climate relationships may be non-linear, as discussed in Sections 2.5-2.6. Indeed, Scafetta (2009) notes that acknowledging a non-linear relationship can potentially increase the magnitude of the solar component of recent climate change [41].

Readers might wonder about what the corresponding "lower bound" for the role of Total Solar Irradiance on Northern Hemisphere temperature is. From Figure 15(h), we can see that the lower bound is described by the Svalgaard (2014) "SSN" dataset and is simply 0%. For brevity, we have not plotted the equivalent to Figures 17 or 18 for this time series as it is a rather trivial result – the Svalgaard (2014) "SSN" dataset essentially implies that there has been no detectable influence of Total Solar Irradiance on Northern Hemisphere temperatures since at least 1700 (when the time series begins). We invite interested readers to study the full set of results for all 80 combinations that we provide numerically as a Microsoft Excel datafile as Supplementary Materials.

From Figures 15 and 17, we can now see why the IPCC 5th Assessment Report was unable to find any role for solar variability in the observed warming since the mid-20th century. It is hard to find much of a role (if any) for solar variability on Northern Hemisphere temperature trends if you use any of the following TSI estimates: Wang et al. (2005) [172]; Krivova et al. (2007) [176]; (2010) [177]; Vieira et al. (2011) [253]; Matthes et al. (2017) [110]; Coddington et al. (2016) [237]; Svalgaard (2014) "LASP"; or the "PHI-US16" version of Egorova et al. (2018) [254]. As mentioned above, if you use the Svalgaard (2014) "SSN" estimate - which is essentially just a rescaled version of the Sunspot Number (SSN) time series – then this implies almost no role for solar variability since at least the 18th century.

On the other hand, from Figures 16 and 18, we can also see why Soon et al. (2015) [56] disputed the IPCC AR5's conclusion, and also why Scafetta et al. had disputed the earlier IPCC AR4's similar conclusion [34,41]. If you use the Hoyt and Schatten (1993) dataset [52,60,179] updated with the ACRIM record, you can explain 95-98% of the long-term warming trend (1841-2018) of the rural-only estimate of Northern Hemisphere temperature trends in terms of solar variability alone. Meanwhile, the Bard et al. (2000) estimate [534,535] can explain 100% of the observed Sea Surface Temperature warming trend over the maximum period of overlap (i.e., 1850-1998). It can also explain the majority of the warming trend for the urban and rural-based estimate as well as the tree-ring proxy series. Admittedly, the Bard et al. (2000) estimate can only explain 47-69% of the long-term warming trend from the glacier-length proxy series, but this is still greater than the 18-44% that can be explained in terms of the corresponding anthropogenic forcing – see Figure 18(e), (j) and (o).

In other words, both the Hoyt and Schatten (1993) dataset and the Bard et al. (2000) dataset imply that most (if not all) of the Northern Hemisphere warming trend since the 19th century (and earlier) has been due to solar variability.

Some readers might counter that the Bard et al. (2000) dataset ends in 1998 and the large solar role up to 1998 might have declined in more recent years. We agree that this is an unfortunate limitation of the available data (although the two proxy-based series also end around the same time). This is especially plausible for the urban & rural-based estimate which implies a much greater warming trend relative to the other four estimates (see Figure 13), although we suggest that the extra warming of this urban & rural-based estimate may be at least partially due to urbanization bias – see Section 3.1-3.2. However, rather than dismissing the fit for this dataset up to the end of the 20th century as out-of-date, to us this suggests that it is time to update the Bard et al. (2000) dataset. Possibly this could be done by calibrating and extending the time series with one of the rival satellite composites, but we urge researchers considering this approach to acknowledge the ongoing debates between the rival satellite groups (see Section 2.2), and we would recommend comparing and contrasting at least the ACRIM and the PMOD TSI satellite composites [60,162]. We would also recommend considering the effects of varying the solar proxies used and the debate over whether the high or low variability estimates are more reliable (see Section 2.3).

Of the remaining TSI estimates, five of them are high variability estimates: Shapiro et al. (2011) [255,256]; Lean et al. (1995) [185]; and the "PHI-MU16", "PHI-MC17" and "SSR11" versions of Egorova et al. (2018) [254]. The only other estimate, Steinhilber et al. (2009) [252], is one of the low variability estimates – see Figure 2. Broadly, these estimates imply a role for solar variability for Northern Hemisphere temperature trends that is intermediate between the "mostly human-caused" conclusion of IPCC AR5 [1] and the "mostly natural" conclusion of Soon et al. (2015) [56].

## 6. Conclusions and recommendations

By reviewing the literature and available data, we identified 16 different estimates of how the Total Solar Irradiance (TSI) has varied since the 19th century (and earlier) – see Table 1 and Figures 2 and 3. Although some of these estimates are very similar to each other, others imply quite different trends and hence can lead to different conclusions. The IPCC 5th Assessment Report (AR5) appears to have tried to overcome this problem by ignoring those datasets that give conflicting results. Worryingly, from reading Matthes et al. (2017), it appears that the CMIP6 modelling groups have been actively encouraged to consider only one estimate of TSI for the 1850-present period, i.e., the Matthes et al. (2017) dataset [110]. In terms of scientific objectivity, this seems to us to have been an approach that is not compatible with the results already published in the scientific literature and even unwise relative to the results highlighted by this paper and of other recently published works.

**Recommendation 1**. We urge researchers who are genuinely interested in trying to answer the question posed by the title of this paper to consider a wide range of TSI estimates and not just ones that agree with the researchers' prior beliefs or expectations. The 16 TSI estimates described in Figures 2 and 3, as well as the 4 additional estimates in Figure 1, are all provided in the Supplementary Materials.

—

Even among these 20 different estimates, it appears that many of the underlying challenges and uncertainties involved in estimating how solar activity has varied over recent decades, let alone centuries, have not been satisfactorily addressed.

**Recommendation 2**. We urge researchers to pay more attention to the scientific debate between the rival TSI satellite composites (see Section 2.2) and to consider the competing datasets when assessing solar trends during the satellite era. In particular, many researchers appear to have overlooked the ongoing scientific debate between the ACRIM and PMOD groups over the trends during the satellite era. For recent reviews of the current debate from different perspectives, we recommend reading/revisiting Zacharias (2014) [164]; de Wit et al. (2017) [156]; and Scafetta et al. (2019) [60] for instance.

For the pre-satellite era, many researchers appear to have become over-reliant on the use of simplistic TSI proxy models based on simple linear regression analysis between sunspots and faculae records or other proxies for describing solar activity during the pre-satellite era, while it is evident from multiple observations that solar luminosity variability is a much more complex phenomenon. As a starting point, we suggest readers read or revisit, e.g., Hoyt and Schatten (1993) [179]; Livingston (1994) [180]; Soon et al. (2015) [56].

—

Another ongoing problem is establishing what the true Northern Hemisphere temperature trends have been. In Section 3, we identified multiple different ways of calculating and estimating temperature trends since the 19th century (or earlier) – see Table 2. Most of these estimates have several common features, e.g., a warming from the 1900s to the 1940s; a cooling or plateau from the 1950s to the 1970s; a warming from the 1980s to the 2000s. However, as discussed in Section 3.6, there are important differences between the estimates on the exact timings and relative magnitudes of each of the warming and cooling periods.

Strikingly, it is only in the estimates that use both urban and rural station records in which the recent warming period appears particularly unusual. This suggests to us that urbanization bias does remain a significant problem for current temperature trend estimates [56,116–118]. However, we recognize that this disagrees with some researchers who have claimed that urbanization bias is only a small problem for global and hemispheric temperature trends, e.g., Jones et al. (1990) [487], Parker (2006) [488], Wickham et al. (2013) [489], as well as with a separate set of researchers who argue that after statistical homogenization techniques (usually automated) have been applied to the data, most of the non-climatic biases (including urbanization bias) are removed or substantially reduced, e.g., Peterson et al. (1999) [490], Menne and Williams (2009) [491], Hausfather et al. (2013) [492], Li and Yang (2019) [493], Li et al. (2020) [494].

**Recommendation 3**. Therefore, we urge researchers to look more closely at the differences between the various estimates of Northern Hemisphere temperature trends. In particular, we caution that despite many claims to the contrary in the literature, e.g., Refs. [487–494], the urbanization bias problem does not appear to have been satisfactorily resolved yet. Although our analysis was explicitly confined to the Northern Hemisphere because there are much less data available for the Southern Hemisphere, this recommendation is also relevant for those looking at global temperature trends.

—

**Recommendation 4**. In this review, we have mostly focused on the simple hypothesis that there is a direct linear relationship between TSI and Northern Hemisphere surface temperatures. However, in Sections 2.5 and 2.6, we showed that there is considerable evidence that the sun/climate relationships are more nuanced and complex. Therefore, we also encourage further research into the potential sun/climate relationships reviewed in Sections 2.5-2.6.

—

**Recommendation 5**. In this paper, we have focused on the role of the Sun in recent climate change and compared this with the role of anthropogenic factors. Therefore, other than in passing, we have not explicitly investigated the possible role of other non-solar driven natural factors such as internal changes in oceanic and/or atmospheric circulation. As discussed throughout Sections 2.5-2.6, such factors may actually have a solar component, e.g., Refs. [39,40,61,63,71,93,96,99,111–113,211,234,475]. However, we encourage further research into the role of other possible natural factors which do not necessarily have a solar component on recent climate change, e.g., Refs. [119–123].

—

**Conclusion**. In the title of this paper, we asked, “How much has the Sun influenced Northern Hemisphere temperature trends?” However, it should now be apparent that, despite the confidence with which many studies claim to have answered this question, it has not yet been satisfactorily answered. Given the many valid dissenting scientific opinions that remain on these issues, we argue that recent attempts to force an apparent scientific consensus (including the IPCC reports) on these scientific debates are premature and ultimately unhelpful for scientific progress. We hope that the analysis in this paper will encourage and stimulate further analysis and discussion. In the meantime, the debate is ongoing.

## Acknowledgements

The main analysis and first draft of the manuscript were carried out by the first three authors (RC, WS and MC). All other co-authors are listed in alphabetical order. As explained in the Introduction, the approach we have taken in this review is to explicitly acknowledge the many dissenting scientific perspectives on a lot of the issues reviewed. As a result, the co-authors have not reached a mutual consensus on all issues. Rather, we have endeavoured to present all competing scientific perspectives as fairly and open-mindedly as possible. With that in mind, all co-authors have approved of the text, even though most of us have definite opinions on many of the debates which have been described, and those opinions vary between co-authors.

NS also provided the updated Hoyt and Schatten (1993) time series in Figure 3(a), and VMVH carried out the interpolations of the missing daily values used for constructing Figure 5. We thank the providers of the datasets used in this paper for allowing access to their time series, either through publicly archiving them, or through personal communication. We thank Dr. Tatiana Egorova for providing the four Egorova et al. (2018) time series (Figure 3e-h). We also thank Prof. Vincent Courtillot, Dr. Ricky Egeland, Prof. Demetris Koutsoyiannis, Dr. Frank Stefani, Prof. Henrik Svensmark and Prof. HongRui Wang for useful comments and feedback on early drafts. RC and WS received financial support from the Center for Environmental Research and Earth Sciences (CERES), while carrying out the research for this paper. The aim of CERES is to promote open-minded and independent scientific inquiry. For this reason, donors to CERES are strictly required not to attempt to influence either the research directions or the findings of CERES. Readers interested in supporting CERES can find details at https://ceres-science.com/. GWH acknowledges long-term support from NASA, NSF, Tennessee State University, and the State of Tennessee through its Centers of Excellence program.